\documentclass[a4paper,11pt]{article}

\pdfoutput=1 

\usepackage{jheppub} 

\usepackage[utf8]{inputenc}
\usepackage{color} 

\usepackage{graphicx}
\graphicspath{{./img/}}

\newcommand{\tr}[1]{\text{Tr} \left[ #1 \right]}   
\renewcommand{\exp}[1]{\text{exp} \left\{ #1 \right\}}   
\renewcommand{\bf}[1]{\mathbf{#1}}  

\title{\boldmath Multiparticle production in proton-nucleus collisions beyond eikonal accuracy}

\author[a]{Pedro Agostini,}
\author[b]{Tolga Altinoluk,}
\author[a]{N\'estor Armesto,}
\author[a]{Fabio Dominguez,}
\author[c,d]{and Jos\'e Guilherme Milhano}

\affiliation[a]{Departamento de F\'{\i}sica de Part\'{\i}culas and IGFAE, Universidade de Santiago de Compostela, 15782 Santiago de Compostela,Galicia--Spain}
\affiliation[b]{Theoretical Physics Division, National Centre for Nuclear Research, Pasteura 7, Warsaw 02-093, Poland}
\affiliation[c]{Laborat\'orio de Instrumenta\c c\~ao e F\'{\i}sica Experimental de Part\'{\i}culas (LIP), Av.
Professor Gama Pinto 2, 1649-003 Lisboa, Portugal}
\affiliation[d]{Departamento de F\'{\i}sica, Instituto Superior T\'ecnico, Universidade de Lisboa, Av. Rovisco Pais 1, 1609-001,
Lisboa, Portugal}

\emailAdd{pedro.agostini@usc.es}
\emailAdd{tolga.altinoluk@ncbj.gov.pl}
\emailAdd{nestor.armesto@usc.es}
\emailAdd{fabio.dominguez@usc.es}
\emailAdd{gmilhano@lip.pt}

\abstract{We study the effects on multigluon production at mid-rapidity in the Color Glass Condensate of the non-eikonal corrections that stem from relaxing the shockwave approximation and giving the target a finite size. We extend previous works performed in the dilute-dilute approximation suitable for proton-proton collisions, to the dilute-dense one applicable to proton-nucleus. We employ the McLerran-Venugopalan model for the projectile averages. For the target averages, we use the Golec-Biernat--W\"usthoff model and restrict to the leading contributions in overlap area that allow a factorization of ensembles of Wilson lines into products of dipoles. We make the connection with the jet quenching formalism and compare with previous results in the literature, providing a parametrization of the so-called decorated dipoles. We show that the non-eikonal effects on single inclusive particle production, contrary to what happens in jet quenching, are only sizable, of a few percent, for modest energies $\sqrt{s_{NN}}\leq 100$ GeV and central rapidities. On the other hand, we find that the effects on double inclusive gluon production are larger for the same kinematics. We show that, as found previously in the dilute-dilute situation, non-eikonal corrections break the accidental symmetry in the CGC, allowing for the existence of non-vanishing odd azimuthal harmonics.}

\begin{document}
	
\maketitle	


\section{Introduction}
\label{sec6:introduction}

The understanding of multiparticle production is a central issue in high energy physics.
Multiparticle production is a direct consequence of the interaction and also acts as background for new phenomena that may eventually appear. It is of particular interest in collisions at high energies or involving extended objects like nuclei. There, the number of produced particles is so large that a microscopic treatment is defying and investigating the possibilities of simplified treatments becomes compulsory.

In heavy ion collisions evidence for the creation of a new state of matter, the Quark Gluon Plasma (QGP), is found. After an initial short time, smaller than 1 fm/c, such state is amenable to a macroscopic description in terms of relativistic hydrodynamics (see e.g.~\cite{Jeon:2016uym,Romatschke:2017ejr}). One of the key findings at the Large Hadron Collider (LHC) at CERN is the observation that many of the behaviors of observables that in heavy ion collisions are considered as evidence of QGP formation and as tools to analyse its properties, have also been observed in collisions of smaller systems, proton-proton ($pp$) and proton-nucleus ($pA$)~\cite{Schlichting:2016sqo,Schenke:2017bog,Loizides:2016tew,Citron:2018lsq,Nagle:2018nvi}. In these small collision systems the success of the hydrodynamic explanation based on final state interactions seems difficult to justify and alternatives based on the dynamics in the initial state have also been essayed.

In $pA$ collisions at high energies, particle production is commonly studied under the approximations that consider the projectile  as a dilute probe which scatters off a dense target described by an intense color field. This process can be computed within the Color Glass Condensate (CGC) effective theory~\cite{Gelis:2010nm,Kovchegov:2012mbw} where the  dilute projectile probes the small-$x$ glue inside the target. This framework relies on the eikonal approximation where the collision is described by infinitely boosted partons in the projectile that traverse a medium in the target with infinitesimal width -- a situation referred to as the shock-wave approximation. In this approximation, fluctuations in the projectile wave function are long-lived and can be considered as frozen throughout the interaction time with the target. Indeed, the eikonal approximation amounts to considering the target field of a left-moving nucleus $A^\mu(x^+,x^-,\mathbf{x})\propto \delta^{\mu -} \delta(x^+) A^-(\mathbf{x})$, where light-cone coordinates $x^\pm=(x^0\pm x^3)/\sqrt{2}$  are used and transverse coordinates  are denoted by boldface letters $\mathbf{x}=(x^1,x^2)$. The infinite boost is then present in three different ways in the expression for the target field: the $\delta$-function in $x^+$ which sets the target width to zero, the independence of $x^-$ which is equivalent to having recoilless interactions, and keeping only the minus component which is enhanced by the boost. Each of these features can be relaxed separately, thus leading to sub-eikonal corrections.

The CGC offers a framework where particle production at the very early phases of the collision~\cite{Gelis:2010nm,Kovchegov:2012mbw,Lappi:2006fp} and initial state correlations (see the review~\cite{Altinoluk:2020wpf}) can be analysed. The origin of such initial state correlations lies in the quantum effects (Bose enhancement in the case of gluons) in the wave function of the colliding hadrons~\cite{Dumitru:2008wn,Dumitru:2010iy,Kovchegov:2012nd,Kovchegov:2013ewa,Altinoluk:2015uaa,Altinoluk:2015eka,Altinoluk:2016vax}, but problems still exist to include subleading density effects (see recent efforts in~\cite{Li:2021zmf,Li:2021yiv,Li:2021ntt}), and to understand the behaviour of correlations with centrality or multiplicity in the event~\cite{Kovner:2018fxj,Altinoluk:2020psk}. Furthermore, odd coefficients of the azimuthal Fourier decomposition of the distribution of produced particles are absent in standard CGC calculations, and can only be generated by including density corrections~\cite{Kovner:2016jfp,Kovchegov:2018jun}, or a modification of the usual isotropic field ensembles in the target~\cite{Dumitru:2014vka,Dumitru:2014dra}, or non-eikonal corrections.

Sub-eikonal effects scale with the inverse of the beam energy. Therefore, the eikonal approximation is well justified at the kinematics of the Large Hadron Collider (LHC) at CERN where $pA$ collisions are performed at centre-of-mass energies per nucleon, $\sqrt{s_{\rm NN}}$, of order $10^3$ GeV. It is only approximate, with the corrections being around a few percent~\cite{Agostini:2019avp,Agostini:2019hkj},
at the top energies, $\sqrt{s_{\rm NN}}=200$ GeV, of the Relativistic Heavy Ion Collider (RHIC). Therefore, with the upcoming Electron-Ion Collider (EIC)~\cite{Accardi:2012qut,AbdulKhalek:2021gbh} where electron-nucleus collisions will be performed at  $\sqrt{s_{\rm NN}}\sim 20\div  100$ GeV, these sub-eikonal contributions will become relevant, as they already are for RHIC energies. For this reason studies aimed at including non-eikonal corrections, such as finite width effects, have experienced an intense growth in recent years.

Further, spin physics requires going beyond the eikonal approximation where spin-flip terms and sub-leading exchanges in the $t$-channel are absent. Sub-eikonal corrections are introduced in the quark and gluon propagators by first assuming a target with finite width and then relaxing the other mentioned approximations~\cite{Altinoluk:2014oxa,Altinoluk:2015gia,Altinoluk:2015xuy,Chirilli:2018kkw,Chirilli:2021lif,Altinoluk:2020oyd,Altinoluk:2021lvu}. In order to compute quark and anti-quark helicity transverse momentum distributions and parton distribution functions, modifications to the JIMWLK evolution equations to include helicity-dependent effects are introduced through polarized Wilson lines~\cite{Kovchegov:2018znm,Cougoulic:2019aja,Kovchegov:2015pbl,Kovchegov:2016zex,Kovchegov:2021iyc,Cougoulic:2020tbc,Cougoulic:2022gbk}, with several numerical analyses of the impact of small-$x$ evolution on the comparison with experimental data done recently~\cite{Adamiak:2021ppq,Kovchegov:2017jxc}. A different method in order to study the effects of the non-eikonal corrections was pursued in \cite{Jalilian-Marian:2017ttv, Jalilian-Marian:2018iui,Jalilian-Marian:2019kaf} by including the longitudinal momentum exchange between the projectile and the target during the interaction. Moreover, studies including the effects of an $x^-$-dependent, i.e., dynamical, target field in the quark propagator are being carried out~\cite{Sadofyev:2021ohn,Altinoluk:2021lvu,Andres:2022ndd}.

As mentioned before, it has been shown by the inclusion of finite width effects in double gluon production that the so-called accidental symmetry, i.e., an artificial azimuthal symmetry that appears in the leading order multigluon spectrum within the CGC and results in vanishing odd azimuthal harmonics~\cite{Altinoluk:2020wpf}, is broken~\cite{Agostini:2019avp,Agostini:2019hkj}. Thus, finite width effects open a new window for explaining, within the initial state framework provided by the CGC, the away- and near-side asymmetry seen in small  collision systems at RHIC where non-eikonal corrections are sizeable.
	
Finite width effects appear naturally in jet quenching where, in order to study the modification of jets traversing the QGP, one assumes that a high-energy parton traverses a finite colored medium and loses its energy through the induced emission of gluons~\cite{Casalderrey-Solana:2007knd,Mehtar-Tani:2013pia,Blaizot:2015lma}. One is usually interested in hard partons created in the initial collision which have not had time to radiate before interacting with the medium and therefore would not have any modification if one keeps the strict shock-wave approximation. In order to have non-trivial contributions from medium effects it is necessary to consider a finite longitudinal extent of the target, thus including sub-eikonal corrections which are resummed through in-medium propagators of gluons and quarks in a background field. Thus, inspired by jet quenching calculations, the approach of this manuscript will be to include the non-eikonal effects to the multigluon spectrum coming from considering a target with finite width through a modification of the eikonal gluon propagator. It is worth noting that, even though the techniques are the same and the calculations resemble each other to some degree, we are indeed calculating very different things in the jet quenching case from  particle production in $pA$. What appears as the main contribution for in-medium emission (transverse motion along a finite longitudinal extent) is only a subleading correction when the incoming particles are allowed to develop long-lived fluctuations before scattering with a background medium.
	
The main goal of this work is to generalize the results of~\cite{Agostini:2019avp,Agostini:2019hkj}, where some of us have relaxed the shock-wave approximation and analysed the effects of including such non-eikonal corrections in multigluon production in $pp$ collisions, to the case in which one of the participants is dense. We study the \emph{finite width effects} in multigluon production in the dilute-dense limit of the Color Glass Condensate framework which is suitable for $pA$ collisions, while still assuming that all momentum exchanges are only transverse and the only non-zero component of the background field is the minus component. We note that we work at a semiclassical level, and no rapidity evolution is considered. This manuscript is organised as follows. In Sec.~\ref{sec6:brief_review} we present a short introduction of the framework, similar to that employed for jet quenching, which will be used along this manuscript. In Sec.~\ref{sec6:NE_target_averaging} we generalize the results of~\cite{Agostini:2021xca}, where we have computed the target averages of Wilson lines within a model~\cite{Kovner:2017ssr,Kovner:2018vec,Altinoluk:2018ogz} -- named the area enhancement model -- that keeps those contributions that are leading in the product overlap area of the collision times the squared saturation momentum of the target, to the non-eikonal case in which the Wilson line has to be substituted by the scalar gluon propagator. In Sec.~\ref{sec6:single_gluon} we compute the single gluon spectrum beyond the eikonal accuracy -- in the sense of finite width effects. We analyze the effects of the non-eikonal corrections in Subsec.~\ref{sec6:single_numerics}. In Subsec.~\ref{sec6:comparison_altinoluk} we compare our results with the next-to-next-to-eikonal expansion performed in~\cite{Altinoluk:2014oxa,Altinoluk:2015gia} obtaining a parameterisation, within the Gaussian approximation for target averages, for the decorated dipole functions defined in those works. In Sec.~\ref{sec6:ne_multiparticle_production} we introduce a general framework for computing the multigluon spectrum for any number of gluons. In Subsec.~\ref{sec6:2gluon_production} we apply this framework to double inclusive gluon production and we analyse the dependence of the double gluon spectrum on the non-eikonal effects. Finally, in Sec.~\ref{sec6:conclusions} we conclude with a summary and outlook. Technical details are provided in Appendices~\ref{app6:path_integrals} and \ref{app:2}.

\section{Brief review of the theoretical framework}
\label{sec6:brief_review}
	
\begin{figure}
	\centering
	\includegraphics[scale=0.65]{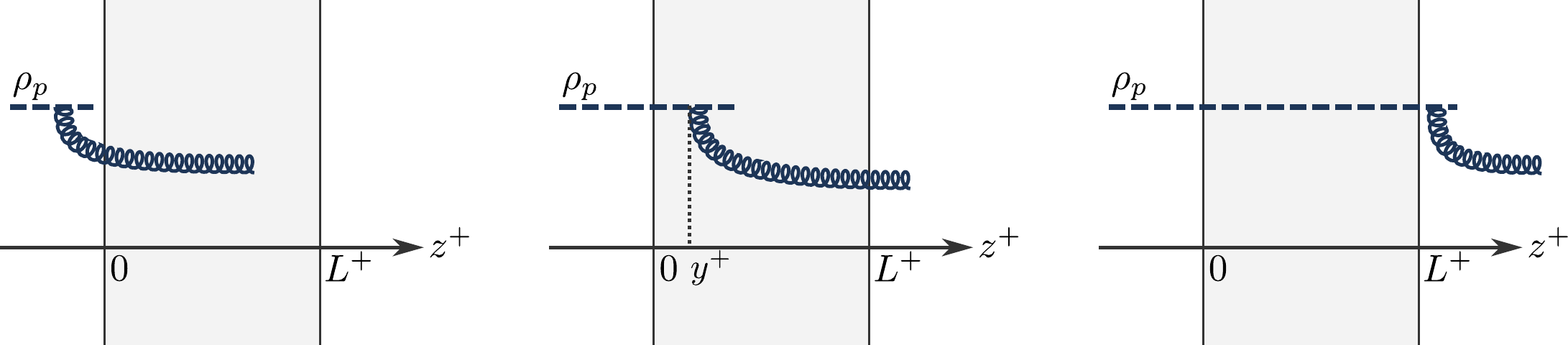}
	\caption{Diagrammatic representation of the three terms appearing in Eq.~\eqref{eq6:reduced_amplitude} where the gluon is emitted before, inside or after the medium.}
	\label{fig:amplitude}
\end{figure}

	
In this section, we briefly summarize the theoretical framework that we adopt to study multigluon production in $pA$ collisions beyond eikonal accuracy. In this framework, the dilute projectile, which is a highly boosted right-moving proton, is composed of large-$x$ partons that act as a color source with color charge $\rho_p^a({\bf x})$, with  index $a$ denoting  color and ${\bf x}$ being transverse position. The left-moving dense target is defined through the target field $A^-(x^+,\bf{x})$ which has a finite support from $0$ to $L^+$ along the $+$ direction, and thus goes beyond the eikonal description of the target where the target fields are localized around $x^+=0$ -- the shockwave limit. In this setup, by restricting ourselves to the light-cone gauge ${A^+=0}$, the production amplitude of a gluon with transverse momenta ${\bf k}$, longitudinal (plus) momentum $k^+$, color $a$ and polarization $\lambda$ can be obtained by using the LSZ reduction formula and reads\footnote{A detailed derivation of the gluon production amplitude can be found in Refs. \cite{Altinoluk:2014oxa, Altinoluk:2015gia, Mehtar-Tani:2006vpj}.}
\begin{align}\label{eq6:factorization_amplitude}
		\mathcal{M}^a_\lambda(k^+,\bf{k})=g\int \frac{d^2 \textbf{q}}{(2 \pi)^2} \overline{\mathcal{M}}^{ab}_\lambda(k^+,\bf{k},\textbf{q}) \rho_p^b(\textbf{q}),
	\end{align} 
where $\rho^b_p(\bf q)$ is the Fourier transform of the projectile color charge density and the reduced matrix amplitude is given by\footnote{Hereafter we employ the notation $\int_{\bf{x}}\equiv \int d^2 \bf{x}$, $\int_{\bf{q}}\equiv \int d^2 \bf{q}$.}
\begin{align}\label{eq6:reduced_amplitude}
		\overline{\mathcal{M}}^{ab}_\lambda(k^+,\bf{k},\textbf{q})&=\epsilon_\perp^{\lambda i*}  i e^{i k^- L^+} \Bigg\{ 	2\frac{\textbf{k}^i}{\textbf{k}^2} \int_\bf{y} e^{-i(\textbf{k}-\textbf{q}) \cdot \textbf{y}}   
		U^{ab}(L^+,0,\bf{y}) \Bigg. \nonumber \\
		\Bigg. & \hspace{0.5cm} -2\frac{\textbf{q}^i}{\textbf{q}^2} \int_{\textbf{x},\textbf{y}} e^{i \textbf{q} \cdot \textbf{y}-i \textbf{k} \cdot 	\textbf{x}}  \mathcal{G}^{ab}_{k^+}(L^+,\textbf{x};0,\textbf{y})
		\Bigg. \nonumber \\
		\Bigg. & \hspace{0.5cm}+ \int_{\textbf{x},\textbf{y}} e^{i \textbf{q} \cdot \textbf{y}-i\textbf{k} \cdot \textbf{x}} \frac{1}{k^+} 	\int_0^{L^+} dy^+ [\partial_{\textbf{y}^i} \mathcal{G}^{ac}_{k^+}(L^+,\textbf{x};y^+,\textbf{y})] U^{cb}(y^+,0,\bf{y}) \Bigg\}.
	\end{align}
The three terms in the reduced matrix amplitude, Eq.~\eqref{eq6:reduced_amplitude}, describe emission of the gluon from the projectile color charges after, before or inside the medium respectively and are illustrated in Fig.~\ref{fig:amplitude}. 

The propagation of the gluon inside the medium is described by the scalar retarded propagator 
\begin{align}\label{eq6:scalar_propagator}
		\mathcal{G}^{ab}_{k^+}(x^+,\bf{x} ; y^+,\bf{y}) = \int_{\bf{z}(y^+)=\bf{y}}^{\bf{z}(x^+)=\bf{x}} [\mathcal{D} \bf{z}(z^+)] 	\exp{\frac{i k^+}{2} \int_{y^+}^{x^+} dz^+ \dot{\bf{z}}^2(z^+)} U^{ab}(x^+,y^+; \bf{z}(z^+)).
	\end{align}
This path integral, from transverse position $\mathbf{y}$ at $y^+$ to $\mathbf{x}$ at $x^+$ along trajectory $\mathbf{z}(z^+)$, encodes both the color rotation and the motion of the gluon in the transverse plane while traversing the medium. The propagation of the gluon outside the medium is given by the free scalar propagator
\begin{align}\label{free_scalar_propagator}
\mathcal{G}^{ab}_{0,k^+}(x^+,\bf{x} ; y^+,\bf{y}) = \frac{-ik^+\delta^{ab}}{2\pi} \frac{\theta(x^+-y^+)}{x^+-y^+} {\rm exp}\bigg\{\frac{ik^+}{2(x^+-y^+)}({\bf x}-{\bf y})^2\bigg\}
\end{align}
(which can be obtained by setting $U^{ab}$ to $\delta^{ab}$ and computing the remaining integral).
Finally, $U^{ab}(x^+,y^+; {\bf x})$ is the Wilson line which accounts for multiple gluon exchanges between the projectile and the target  and it is given by the path-ordered exponential
\begin{align}\label{def:Wilson_line}
U^{ab}(x^+,y^+; {\bf x})={\cal P}_+{\rm exp}\bigg\{ ig\int_{y^+}^{x^+} dz^+ A^-(z^+,{\bf x})\bigg\}^{ab} \, .
\end{align}

In this approach the single inclusive gluon spectrum is given by 
\begin{align}
	2k^+(2\pi)^3 \frac{dN}{d k^+ d^2 \bf{k}} = 
	\Big\langle \mathcal{M}^a_\lambda(k^+,\bf{k}) \mathcal{M}^{a \dagger}_\lambda(k^+,\bf{k}) \Big\rangle_{p,T}\ ,
\end{align}
where $\langle \cdots \rangle_{p,T}$ accounts for the average over the color configurations of the projectile and the target respectively. The explicit expression of this function is derived later in this manuscript and can be found in \eqref{multi2} and \eqref{eq6:target_average}.
%
%

Within the same setup, single inclusive gluon production was studied in Refs. \cite{Altinoluk:2014oxa, Altinoluk:2015gia}  by developing a systematic method to compute higher order corrections to the eikonal approximation (or equivalently to the shockwave approximation) that are due to the non-zero longitudinal width of the target. Specifically, the method was developed to expand the scalar retarded propagator given in Eq. \eqref{eq6:scalar_propagator}. The expansion was performed at next-to-next-to-eikonal accuracy and the result was applied to the single inclusive gluon production process in $pA$ collisions. In the present work, we follow a different approach to account for the non-eikonal effects that stem from the finite longitudinal thickness of the target. Contrary to the studies performed in Refs. \cite{Altinoluk:2014oxa, Altinoluk:2015gia}, in the present work we discuss how to perform target averaging of non-eikonal objects that appear at the level of the squared amplitude without adopting the aforementioned eikonal expansion. 

As mentioned briefly in Section~\ref{sec6:introduction}, inclusive multigluon spectra is an essential observable for the study of particle correlations from the initial state point of view for small size systems such as $pp$ or $pA$ collisions. Within the CGC framework, these studies are usually performed in the ``glasma graph approximation"~\cite{Armesto:2006bv,Dumitru:2008wn,Dumitru:2010iy,Ozonder:2014sra,Ozonder:2017wmh} which amounts to allowing the emission of a gluon from a single color source in the projectile and neglecting the contributions where more than one gluon is emitted from the same color source in the projectile. One important aspect of the glasma graph calculations is that it is only valid for dilute-dilute scattering processes, i.e., $pp$ collisions, because it only considers two gluon exchanges with the target. However, its extension that accounts for the multiple interactions with the target which corresponds to dilute-dense scatterings, or equivalently $pA$ collisions, was performed in~\cite{Altinoluk:2018hcu,Altinoluk:2018ogz}. In a recent study~\cite{Agostini:2021xca}, the extension of the glasma graph calculations  to compute  $n$-gluon production in $pA$ collisions was developed and the four-gluon production spectrum was explicitly computed in the strict eikonal limit. 

Here, we extend the eikonal calculation of the $n$-gluon spectrum to the non-eikonal case where the effects of the finite longitudinal width of the target are accounted for. The non-eikonal generalization of the eikonal $n$-gluon spectrum computed in~\cite{Agostini:2021xca} can be written as 
	\begin{align}
		\label{eq6:multi_particle_pA}
		2^n (2\pi)^{3n} \frac{d^n N}{\prod_{i=1}^n d k_i^+/k_i^+ d^2 \textbf{k}_i}
		= & \,g^{2n}\int_{\textbf{q}_1,\dots,\textbf{q}_{2n}} 
		\Big \langle 
		\rho_p^{b_1}(\textbf{q}_1) \rho_p^{*b_2}(\textbf{q}_2) \cdots \rho_p^{*b_{2n}}(\textbf{q}_{2n})
		\Big \rangle_p
		\nonumber \\ & \hskip -5cm \times
		\Big \langle 
		\overline{\mathcal{M}}^{a_1 b_1}_{\lambda_{1}}(\underline{k}_1,\textbf{q}_1)
		\overline{\mathcal{M}}^{\dagger b_2 a_1}_{\lambda_{1}}(\underline{k}_1,\textbf{q}_2)
		\cdots
		\overline{\mathcal{M}}^{a_{n} b_{2n-1}}_{\lambda_{n}}(\underline{k}_n,\textbf{q}_{2n-1})
		\overline{\mathcal{M}}^{\dagger b_{2n} a_{n}}_{\lambda_{n}}(\underline{k}_n,\textbf{q}_{2n})
		\Big \rangle_T,
	\end{align}
where a shorthand notation $\underline{k} \equiv (k^+,\bf{k})$ is introduced.
Due to the finite longitudinal extend of the target, the reduced matrix amplitude is a more complicated object compared to its eikonal limit. It includes not only the standard Wilson lines as in the case of strict eikonal limit but also the retarded scalar propagator $\mathcal{G}_{k^+}(x^+,\bf{x} ; y^+,\bf{y}) $ (see Eq. \eqref{eq6:reduced_amplitude}). Therefore, at the squared amplitude level one gets not only eikonal multipoles (dipoles, quadrupoles, etc.) but their non-eikonal generalizations which include scalar propagators. In the next section, we discuss how to evaluate the target averaging of such new objects that appear beyond eikonal accuracy. 

	
\section{Non-eikonal target averaging}
\label{sec6:NE_target_averaging}
	
In order to obtain an analytical result for the inclusive $n$-gluon spectra given in Eq. \eqref{eq6:multi_particle_pA}, the target average of the $2n$ reduced matrix amplitudes has to be computed. It is well known that in the strict eikonal limit of the single inclusive gluon spectrum, one ends up with the eikonal dipole function which is usually evaluated by using some model like McLerran-Venugopalan (MV)~\cite{McLerran:1993ni,McLerran:1994vd} or Golec-Biernat--W\"usthoff (GBW)~\cite{GolecBiernat:1998js,GolecBiernat:1999qd}. However, for two or more gluons multipole functions such as quadrupoles, sextuples, etc., appear. In order to easily compute the target averaging of these multipole functions, the Area Enhancement (AE) model\footnote{This model has been called dipole approximation in~\cite{Li:2021ntt}, where its validity is also discussed.} is introduced in Refs.~\cite{Kovner:2017ssr,Kovner:2018vec,Altinoluk:2018ogz}. This model is based on the following ansatz: Any multipole can be written in terms of dipole functions through a Wick expansion for those configurations of multipole functions that maximize the phase space integration, up to  terms that are suppressed by the collision area\footnote{See Appendix A of Ref.~\cite{Agostini:2021xca} for a detailed discussion of the validity of the AE model and its numerical comparison with the MV model for double dipole and quadrupole operators in the fundamental representation. Specifically, the difference between the Fourier transform for the double dipole in both models is just a few \% and clearly decreasing with increasing collision area. But for the Fourier transform of the quadrupole such difference is considerably larger, up to 30 \%, and the convergence with increasing collision area much slower than for the double dipole. Further numerical checks and discussions can be found in~\cite{Li:2021ntt}.}. In other words, in the AE model, after the integration over the phase space, the Wilson lines can approximately be described by a Gaussian distribution up to the corrections that are of the order of the inverse of the phase space area. In this section, we will generalize the target averaging of eikonal two point functions (eikonal dipole functions) to compute the non-eikonal two point functions (i.e., non-eikonal dipole functions) that appear in the single inclusive gluon production when one includes the target finite longitudinal width effects. Then, these results will be used together with the AE argument to study the non-eikonal multigluon spectra. 
	
	
	Non-eikonal single inclusive gluon spectrum is given by setting $n=1$ in Eq. \eqref{eq6:multi_particle_pA}, which requires the evaluation of the following three objects: 
	
	\begin{align}
		\frac{1}{N_c^2-1}\Big\langle \tr{U_{\bf y}(x^+,y^+)U^\dagger_{\bar{\bf y}}(x^+,y^+)} \Big\rangle_T
		& \equiv d^{(0)}(x^+,y^+|{\bf y},\bar{\bf y}),
		\label{eq6:dipole_uu}
		\\
		\frac{1}{N_c^2-1} \Big \langle \tr{\mathcal{G}_{k^+}(x^+,\textbf{x};y^+,\textbf{y}) U_{\bar{\textbf{y}}}^\dagger(x^+,y^+)}  \Big \rangle_T
		& \equiv d^{(1)}(x^+,y^+|{\bf x},{\bf y},k^+;\bar{\bf y}),
		\label{eq6:dipole_gu}
		\\
		\frac{1}{N_c^2-1}\Big\langle  \tr{{\cal G}_{k_1^+}(x^+,{\bf x};y^+,{\bf y}){\cal G}_{k_2^+}^\dagger(x^+,\bar{\bf x}; y^+,\bar{\bf y})} \Big\rangle_T
		& \equiv d^{(2)}(x^+,y^+|{\bf x},{\bf y},k_1^+; \bar{\bf x},\bar{\bf y},k_2^+),
		\label{eq6:dipole_gg}
	\end{align}
	where we have written $U_{\bf x}(x^+,y^+) \equiv U(x^+,y^+;\bf{x})$ for simplicity. It is straightforward to note that $d^{(0)}(x^+,y^+|{\bf y},\bar{\bf y})$ given in Eq.~\eqref{eq6:dipole_uu} can be identified with the eikonal dipole function evaluated over a finite longitudinal extent $z^+\in[y^+,x^+]$ of the target. The functions defined in Eqs.~\eqref{eq6:dipole_gu} and \eqref{eq6:dipole_gg} are non-eikonal generalizations of the dipole function.
	
	For simplicity, we assume that the eikonal dipole function is described by the GBW model which is valid as long as the dipole size is much smaller than $\Lambda_{\rm QCD}^{-1}$. In this case, the eikonal dipole function reads
	\begin{align}\label{eq6:gbw_local}
		d^{(0)}(x^+,y^+|{\bf y},\bar{\bf y})=\exp{- \frac{q_s^2(x^+,y^+)}{4} (\bf{x}-\bf{y})^2},
	\end{align}
	where $q_s(x^+,y^+)$ can be identified as the effective saturation momentum in a longitudinal slice $[y^+,x^+]$ of the target. Within the MV model, saturation momentum can be defined as\footnote{This definition of the saturation momentum is arbitrary and depends on the representation of the target parton that interacts with the projectile. However, it is convenient since it makes the dipole function independent of the value of the Casimir of the representation.}
	\begin{align}\label{eq4:saturation_scale}
	Q_s^2=\frac{g^4\, C_R\, \mu^2}{4\pi}\ ,
	\end{align}
	with $\mu^2$  the color density of the target and $C_R$  the Casimir of the projectile parton interacting with the target. With this definition of the saturation momentum, the effective saturation momentum $q_s(x^+,y^+)$ can be defined as 
	\begin{align}
		q_s^2(x^+,y^+) = \frac{g^4 C_R}{4 \pi} \int_{y^+}^{x^+} dz^+ \tilde{\mu}^2(z^+).
	\end{align}
	Thus, assuming that the color density $\tilde{\mu}^2(z^+)$ is constant along the target longitudinal extent, i.e., $\tilde{\mu}^2(z^+) = \mu^2/L^+$, the effective saturation momentum takes the following form:
	\begin{align}
	\label{eq:effqs}
		q_s^2(x^+,y^+) = \frac{x^+-y^+}{L^+} Q_s^2.
	\end{align}

The object defined in Eq.~\eqref{eq6:dipole_gu} is one of the generalizations of the eikonal dipole function which stems from the finite longitudinal extent of the target and we refer to it as the  \textit{$1^{\rm st}$ order NE dipole function}. It requires averaging over the scalar propagator $\mathcal{G}_{k^+}(x^+,\bf{x} ; y^+,\bf{y})$ defined in Eq. \eqref{eq6:scalar_propagator} and therefore implies solving a path integral. In order to do so, we discretize the longitudinal space into $N$ slices where each discretized point is labeled as  $z_i^+$. In this case, the scalar propagator can be written as~\cite{Altinoluk:2014oxa,Altinoluk:2015gia,Mehtar-Tani:2006vpj}
	\begin{align}\label{eq6:scalar_propagator_discretized}
		&\mathcal{G}^{ab}_{k^+}(\underline{x},\underline{y})= \lim_{N \rightarrow \infty} \Theta(x^+-y^+) 
		\int \left(\prod_{n=1}^{N-1} d^2 \textbf{z}_n \right) 
		\left( \frac{-i k^+ N}{2(x^+-y^+)} \right)^N  
		\nonumber \\ & \hskip4cm \times 
		\exp{\frac{i k^+ N}{2(x^+-y^+)} \sum_{n=1}^{N}(\textbf{z}_{n}-\textbf{z}_{n-1})^2 }
		\prod_{n=1}^{N} 
		U^{ab}_{\bf{z}_n}(z_{n-1}^+,z_n^+),
	\end{align}
	where we have defined ${\underline x} \equiv (x^+,\bf{x})$. This equation describes the discrete random walk of the emitted gluon through the transverse points $\bf{z}_n(z_n^+)$ inside the target. The emitted gluon is propagated from $\bf{z}_0(z_0^+)=\bf{y}$ with $z_0^+=y^+$ to $\bf{z}_N(z_N^+)=\bf{x}$ with $z_N^+=x^+$. The only dependence on the target configuration in Eq.~\eqref{eq6:scalar_propagator_discretized} appears through the discretized Wilson lines, $U^{ab}_{\bf{z}_n}(z_{n-1}^+,z_n^+)$, that account for the eikonal propagation of the gluon in each longitudinal slice. Thus, the $1^{\rm st}$ order NE dipole function defined in Eq.~\eqref{eq6:dipole_gu} can be written as  
	\begin{align}\label{eq6:dip_aux}
		&d^{(1)}(x^+,y^+|{\bf x},{\bf y},k^+;\bar{\bf y}) = \lim_{N \rightarrow \infty}
		\int \left(\prod_{n=1}^{N-1} d^2 \textbf{z}_n \right)   
		\exp{\frac{i k^+ N}{2(x^+-y^+)} \sum_{n=1}^{N}(\textbf{z}_{n}-\textbf{z}_{n-1})^2 }
		\nonumber \\ & \hskip3cm \times
		\left( \frac{-i k^+ N}{2(x^+-y^+)} \right)^N
		\Bigg \langle  \tr{\left(\prod_{n=1}^{N}U_{\bf{z}_n}(z_{n-1}^+,z_n^+)\right) U^\dagger_{\bf{\bar y}}(x^+,y^+)}\Bigg\rangle_T.
	\end{align}
	This expression can be further simplified by realizing that, due to the properties of path-ordered exponentials, the Wilson line evaluated over some longitudinal extent $[y^+,x^+]$ can be factorized into a product of independent contributions at each slice of the discretized axis:
	\begin{align}
		U_\bf{x}(x^+,y^+) = \prod_{i=1}^n U_{\bf{x}}(x_{i-1}^+,x_i^+),
	\end{align}
	where $x_0^+=y^+$ and $x_n^+=x^+$. Moreover, noting that the MV model is local in the longitudinal direction which implies that the average of Wilson lines evaluated at different points on the longitudinal axis factorizes into independent averages, we simplify the target average in Eq.~\eqref{eq6:dip_aux} to
	\begin{align}
		&\Bigg \langle  \tr{\left(\prod_{n=1}^{N}U_{\bf{z}_n}(z_{n-1}^+,z_n^+)\right) U^\dagger_{\bf{\bar y}}(x^+,y^+)}\Bigg\rangle_T\nonumber \\
		=
		&\prod_{n=1}^{N}\Bigg \langle  \tr{U_{\bf{z}_n}(z_{n-1}^+,z_n^+) U^\dagger_{\bf{\bar y}}(z_{n-1}^+,z_n^+)}\Bigg\rangle_T
		.
	\end{align}
	Finally, by using the local GBW model given in Eq.~\eqref{eq6:gbw_local} and noting that $z_i^{+}-z_{i-1}^{+}=L^+/N$, the $1^{\rm st}$ order NE dipole function takes the following form: 
	\begin{align}\label{eq:def_1st_type_1}
		&d^{(1)}(x^+,y^+|{\bf x},{\bf y},k^+;\bar{\bf y}) = \lim_{N \rightarrow \infty}
		\int \left(\prod_{n=1}^{N-1} d^2 \textbf{z}_n \right)  
		\left( \frac{-i k^+ N}{2(x^+-y^+)} \right)^N 
		\nonumber \\ & \hskip3cm \times
		\exp{\frac{i k^+ N}{2(x^+-y^+)} \sum_{n=1}^{N}(\textbf{z}_{n}-\textbf{z}_{n-1})^2 - \frac{Q_s^2}{4N} \sum_{n=1}^N(\bf{z}_n-\bf{\bar{y}})^2}.
	\end{align}
	In the continuum limit, by defining $\bf{r}=\bf{z}-\bf{\bar{y}}$, Eq.~\eqref{eq:def_1st_type_1} can be written as
	\begin{align}\label{eq6:gu_harmonic_oscillator}
		d^{(1)}(x^+,y^+|{\bf x},{\bf y},k^+;\bar{\bf y}) = \int [\mathcal{D} \bf{r}] 	\exp{ \int_{y^+}^{x^+} dz^+ \left[\frac{i k^+}{2}  \dot{\bf{r}}^2 - \frac{Q_s^2}{4 L^+} \bf{r}^2\right]}.
	\end{align}
	
	Two comments are in order here. First, the result given in Eq.~\eqref{eq6:gu_harmonic_oscillator} can be obtained in a straightforward manner by using the expression for the scalar propagator $\mathcal{G}_{k^+}(x^+,\bf{x} ; y^+,\bf{y})$ given in Eq.~\eqref{eq6:scalar_propagator} in the definition of the $1^{\rm st}$ order NE dipole function and adopting the GBW model directly without introducing the discretization along the longitudinal axis. However, the way that Eq.~\eqref{eq6:gu_harmonic_oscillator} is obtained above justifies the use of the locality argument which plays a crucial role in our discussion.
Second, Eq.~\eqref{eq6:gu_harmonic_oscillator} is the well known path integral of the harmonic oscillator with ``mass" $k^+/2$ and imaginary ``frequency" $\sqrt{-iQ_s^2/2L^+k^+}$ that is frequently used in jet quenching calculations (see for example~\cite{Mehtar-Tani:2013pia}). For completeness, we compute this integral in Appendix~\ref{app6:path_integrals} with the result given in~\eqref{eq6:path_int_1ho_sol}. Thus, after performing the path integral, the $1^{\rm st}$ order NE dipole function reads
	\begin{align}\label{eq6:gu_function}
		d^{(1)}(x^+,y^+|{\bf x},{\bf y},k_i^+;\bar{\bf y}) =	\frac{- Q_s^2}{4 \pi \epsilon_i \sin\frac{\epsilon_i \Delta^+}{L^+}} \exp{\frac{Q_s^2}{4\epsilon_i} \left[ \frac{\bf{a}_0^2+\bf{a}_N^2}{\tan\frac{\epsilon_i \Delta^+}{L^+}} -2 \frac{ \bf{a}_0 \cdot \bf{a}_N}{\sin\frac{\epsilon_i \Delta^+}{L^+}} \right]},
	\end{align}
	where we have defined $\Delta^+=x^+-y^+$, $\bf{a}_0=\bf{y}-\bf{\bar y}$, $\bf{a}_N=\bf{x}-\bf{\bar y}$ and
	\begin{align}\label{eq6:epsilon}
		\epsilon_i^2 =\frac{Q_s^2 L^+}{2 i k_i^+}.
	\end{align}
	Note that $\epsilon_i^2$ is a dimensionless parameter that vanishes in  the eikonal limit ($k_i^+ \to \infty$ and $L^+ \to 0$)\footnote{The meaning of the non-eikonal parameter in \eqref{eq6:epsilon} is the following: $Q_s$ is the typical transverse momentum of the fluctuations of the projectile that are resolved by the target, and $k^+$ their plus momentum. Then $\epsilon$ becomes the ratio of $L^+$ to the formation time of such fluctuations, the latter being infinite (very large) in the strict eikonal limit. The deviations from eikonality are then given by $\epsilon$ becoming larger and larger, which happens when the formation time becomes of the order or even smaller that the length of the target.}. Its role is to weight the strength of the finite longitudinal width effects and we will refer to it as the \textit{non-eikonal parameter}. 
	
	The next two-point function that we consider is defined in Eq.~\eqref{eq6:dipole_gg}. It is referred to as the \textit{$2^{\rm nd}$ order NE dipole} function which can be computed in the same manner as the $1^{\rm st}$ order one. By using the GBW model and the locality of the MV model we can write it as
	\begin{align}\label{eq6:gg_path_integral}
		&d^{(2)}(x^+,y^+|{\bf x},{\bf y},k_1^+; \bar{\bf x},\bar{\bf y},k_2^+) 
		\nonumber \\ & \hskip3cm 
		= \int [\mathcal{D} \bf{r}] [\mathcal{D} \bf{\bar{r}}]	
		\exp{ \int_{y^+}^{x^+} dz^+ \left[\frac{i k_1^+}{2}  \bf{\dot r}^2 - \frac{i k_2^+}{2}  \bf{\dot{\bar{r}}}^2 - \frac{Q_s^2}{4 L^+} (\bf{r}-\bf{\bar{r}})^2\right]},
	\end{align}
	where $\bf{r}(y^+)=\bf{y}$, $\bf{r}(x^+)=\bf{x}$, $\bf{\bar r}(y^+)=\bf{\bar y}$ and $\bf{\bar r}(x^+)=\bf{\bar x}$. This equation can be identified with the path integral of two coupled harmonic oscillators which can be decoupled. The solution is also derived in Appendix~\ref{app6:path_integrals} and the final result is given in Eq.~\eqref{eq6:path_int_2ho_sol}. Thus, the $2^{\rm nd}$ order NE dipole function reads
	\begin{align}\label{eq6:gg_function}
		&d^{(2)}(x^+,y^+|{\bf x},{\bf y},k_1^+; \bar{\bf x},\bar{\bf y},k_2^+) 
		= \frac{-Q_s^4}{(4\pi)^2}
		\frac{\epsilon_{-} L^+ }{\Delta^+ \epsilon_1^2 \epsilon_2^2 \sin \frac{\Delta^+ \epsilon_{-}}{L^+}}
		{\rm exp} \Bigg\{\frac{Q_s^2}{4 \epsilon_{-}^2} \Bigg(
		\frac{\epsilon_{-} (\bf{r}_0^2+\bf{r}_N^2)}{\tan\frac{\Delta^+ \epsilon_{-}}{L^+}}
		\nonumber \\ & \hskip2.5cm
		-2\frac{\epsilon_{-} \bf{r}_0 \cdot \bf{r_N}}{\sin\frac{\Delta^+ \epsilon_{-}}{L^+}}
		-
		\frac{L^+}{\Delta^+ \epsilon_1^2 \epsilon_2^2 \epsilon_{+}^4} \left[
		2\epsilon_1^2\epsilon_2^2 (\bf{r}_N-\bf{r}_0)
		-\epsilon_{+}^2 \epsilon_{-}^2
		(\bf{b}_N-\bf{b}_0)
		\right]^2
		\Bigg)	
		\Bigg\},
	\end{align}
	where we have defined for simplicity $\epsilon_{\pm}^2=\epsilon_1^2\pm\epsilon_2^2$, $\bf{r}_0=\bf{y}-\bf{\bar{y}}$, $\bf{r}_N=\bf{x}-\bf{\bar{x}}$, $\bf{b}_0=(k_1^+ \bf{y}+k_2^+ \bf{\bar{y}})/(k_1^++k_2^+)$ and $\bf{b}_N=(k_1^+ \bf{x}+k_2^+ \bf{\bar{x}})/(k_1^++k_2^+)$. 
	
	We would like to emphasize that this solution is novel and only in the limit $k_2^+=k_1^+$ it simplifies and leads to the known results~\cite{Casalderrey-Solana:2007knd,Zakharov:1998sv,Blaizot:2012fh,Apolinario:2014csa}.  
	\begin{align}\label{eq6:gg_eq_function} 
		&d^{(2)}(x^+,y^+|{\bf x},{\bf y},k_1^+; \bar{\bf x},\bar{\bf y},k_1^+) =
		-\left(\frac{L^+ Q_s^2}{4 \pi \Delta^+ \epsilon_1^2}\right)^2
		\nonumber \\ & \hskip1cm \times
		{\rm exp}\Bigg\{  
		-\frac{Q_s^2}{12 \epsilon_1^2} \Bigg[
		\epsilon_1^2 \frac{\Delta^+}{L^+}  (\bf{r}_0^2+\bf{r}_N^2+\bf{r}_0 \cdot \bf{r}_N)
		-6 \frac{L^+}{\Delta^+}(\bf{r}_N-\bf{r}_0)\cdot (\bf{b}_N-\bf{b}_0)
		\Bigg]
		\Bigg\}
	\end{align}
which is required for single inclusive gluon production or medium-induced gluon radiation. But the more general result for $k_2^+\ne k_1^+$ is required for double inclusive gluon production, that we analyse below (see also Appendix~\ref{app:2}).
	
	With Eqs.~\eqref{eq6:gbw_local}, \eqref{eq6:gu_function} and \eqref{eq6:gg_function}  we have determined the non-eikonal dipole functions evaluated within a longitudinal extent $\Delta^+$ of the target medium. We note that these functions are extensively used by the jet quenching community where the effects of the parton propagation within a dense medium is computed. In that case, the GBW model is usually referred to as the \textit{harmonic oscillator approximation} and the effective saturation scale defined in  Eq.~\eqref{eq6:gbw_local} is written  
	\begin{align}
	q_s^2(x^+,y^+)=\int_{y^+}^{x^+} \hat{q}(z^+)dz^+,
	\end{align}
	where $\hat{q}(z^+)$ is the medium transport coefficient. For a static homogeneous medium in which $\hat{q}(z^+)$ is constant, we get $\hat{q}=Q_s^2/L^+$.
	
	Before we conclude this section, we would like to comment on the computation of the higher order functions, i.e., multipoles, where one has to evaluate the non-eikonal target average of multiple Wilson lines and scalar propagators. As stated shortly at the beginning of this section, in order to compute the NE multipoles we use the AE model. This model is based on the chromo-electric domain structure in the target transverse plane and leads to a similar result when compared with the MV model. The advantage of this approach is that, upon integration over the transverse phase space, the Wilson lines follow approximately a Gaussian distribution -- with the non-Gaussian corrections suppressed by the collision area -- and therefore allows one to adopt Wick's theorem. However, in the case of a target with a finite longitudinal extent, one should be more careful since
the chromo-electric domains may decohere in the longitudinal direction. 

To justify the application of the AE model in target ensembles with non-zero width, $L^+$, we examine the space-time kinematics of the interaction in the center-of-mass (CoM) frame. In this frame the Lorentz contracted target width is $L^+ \sim A^{1/3}/\sqrt{s_{\rm NN}}$. On the other hand, the chromo-electric fields are defined by the low-$x$ gluons which have the following spread in the longitudinal direction: $\Delta x^+ \sim 1/q^- = A/(xQ_T^-)$, where $Q_T^- = A \sqrt{s_{\rm NN}}/\sqrt{2}$ and $q^-$ are the longitudinal momentum (in the CoM frame) of the target and the gluon respectively. Thus, as long as $\Delta x^+ \gg L^+$ the domain structure of the target will not decohere within its extent and the AE model is justified. Therefore, for $L^+/\Delta x^+ \sim x A^{1/3} \ll 1$ the Area Enhancement model should be a good approximation in the non-eikonal case.\footnote{Note that its application in jet quenching calculations, where the length of the medium is not Lorentz contracted in this frame but also the longitudinal spread of the target gluons may be larger, seems more delicate.}
	
	
\section{Non-Eikonal single inclusive gluon production}
\label{sec6:single_gluon}
	
	Before computing the non-eikonal multi gluon production, we first consider the single inclusive case as a warm up. In this section, we first derive the analytical solutions of the non-eikonal single inclusive spectrum, then perform the numerical analysis of these results and finally compare our results with the ones that exist in the literature (specifically with the results of Ref.~\cite{Altinoluk:2014oxa,Altinoluk:2015gia}).
	
	\subsection{Analytical results}\label{sec:analytical}

	The non-eikonal single inclusive gluon spectrum is given by Eq.~\eqref{eq6:multi_particle_pA} for $n=1$, which reads 
	\begin{align}\label{multi2}
		(2 \pi)^3(2 k^+) \frac{d N}{d k^+ d^2 \textbf{k}}=g^2\int_{\textbf{q}_1 ,\textbf{q}_2} \big\langle \rho^{b_1}(\textbf{q}_1) \rho^{*b_2}(\textbf{q}_2) \big\rangle_p \big\langle \overline{\mathcal{M}}^{b_1 a_1}_\lambda(\underline{k},\textbf{q}_1) \overline{\mathcal{M}}^{\dagger a_1 b_2}_\lambda(\underline{k},\textbf{q}_2) \big\rangle_T.
	\end{align}
	We employ the MV model to compute the averaging over the projectile color sources:
	\begin{align}
		\big\langle \rho^{b_1}(\textbf{q}_1) \rho^{*b_2}(\textbf{q}_2) \big\rangle_p = \frac{\delta^{b_1 b_2}}{N_c^2-1} \mu^2(\bf{q}_1,-\bf{q}_2),
	\end{align}
	where the factor $1/(N_c^2-1)$ is introduced by convenience but will not be relevant in the analysis performed in this manuscript\footnote{Compared to other definitions in the literature, this color factor goes into the definition of $\mu^2$.}. Since this contribution is proportional to $\delta^{b_1 b_2}$, it leads to a color trace of the reduced amplitudes in Eq.~\eqref{multi2}. By using the expression for the reduced amplitude given in Eq.~\eqref{eq6:reduced_amplitude}, the target average of the reduced amplitudes for single inclusive gluon spectrum can be written as\footnote{As explained before, this equation is very similar to the medium-induced radiation spectrum used in jet quenching calculations~\cite{Baier:1996sk,Baier:1996kr,Zakharov:1997uu}. The main difference between this equation and those used in jet quenching calculations is that the reduced amplitude given in Eq.~\eqref{eq6:reduced_amplitude} and therefore the trace given in Eq.~\eqref{eq6:target_average} include interference terms with the contribution where the gluon is emitted before the interaction of the source with the target, which are absent in the jet quenching framework~\cite{Mehtar-Tani:2013pia,Wiedemann:2000za}.}
	\begin{align}\label{eq6:target_average}
		&
		\frac{1}{N_c^2-1}\Big\langle \tr{\overline{\mathcal{M}}_\lambda(\underline{k},\textbf{q}_1) \overline{\mathcal{M}}^{\dagger}_\lambda(\underline{k},\textbf{q}_2)} \Big\rangle_T
		=
		\int_{\textbf{y},\bar{\textbf{y}}} e^{i{\bf q}_1\cdot{\bf y}-i{\bf q}_2\cdot\bar{\bf y}}\bigg\{
		\frac{2}{{\bf k}^2}e^{-i{\bf k}\cdot({\bf y}-\bar{\bf y})} d^{(0)}(L^+,0|{\bf y};\bar{\bf y})
		\nonumber \\
		& \hskip6cm
		+2 \frac{{\bf q}_1\cdot{\bf q}_2}{{\bf q}_1^2{\bf q}_2^2}
		\int_{{\bf x},\bar{\bf x}}e^{-i{\bf k}\cdot({\bf x}-\bar{\bf x})}d^{(2)}(L^+,0|,{\bf x},{\bf y},k^+;\bar{\bf x},\bar{\bf y},k^+)
		\nonumber \\
		&
		-4 \frac{{\bf k}\cdot{\bf q}_1}{{\bf k}^2{\bf q}_1^2}\int_{\bf x}e^{-i{\bf k}\cdot({\bf x}-\bar{\bf y})}d^{(1)}(L^+,0|{\bf x},{\bf y},k^+; \bar{\bf y})
		\nonumber\\
		&
		+2\frac{{\bf k}^i}{{\bf k}^2}\frac{1}{k^+}\int_0^{L^+}dy^+\int_{{\bf x}}e^{-i{\bf k}\cdot({\bf x}-\bar{\bf y})}
		d^{(0)}(y^+,0|{\bf y};\bar{\bf y}) \big[ \partial_{{\bf y}^i}d^{(1)}(L^+,0|{\bf x},{\bf y},k^+; \bar{\bf y})\big]
		\nonumber \\
		&
		-2\frac{{\bf q}_1^i}{{\bf q}_1^2}\frac{1}{k^+}\int_0^{L^+}d y^+\int_{{\bf x},\bar{\bf x},{\bf u}} e^{-i{\bf k}\cdot({\bf x}-\bar{\bf x})}
		\big[\partial_{\bar{\bf y}^i}d^{(2)}(L^+,0|,{\bf x},{\bf y},k^+;\bar{\bf x},\bar{\bf y},k^+)\big]
		\nonumber \\ & \hskip9cm \times
		d^{(1)}(L^+,0|{\bf x},{\bf y},k^+; \bar{\bf y})
		\nonumber \\
		&
		+\frac{1}{(k^+)^2}\int_0^{L^+}dy^+\int_{y^+}^{L^+}d\bar y^+\int_{{\bf x},\bar{\bf x},{\bf u}}e^{-{\bf k}\cdot({\bf x}-\bar{\bf x})}
		\big[ \partial_{\bar{\bf y}^i}d^{(2)}(L^+,0|,{\bf x},{\bf y},k^+;\bar{\bf x},\bar{\bf y},k^+)\big]
		\nonumber \\ & \hskip5cm \times
		\big[\partial_{{\bf y}^i}d^{(1)}(L^+,0|{\bf x},{\bf y},k^+; \bar{\bf y})\big]
		d^{(0)}(y^+,0|{\bf y};\bar{\bf y}) 
		\bigg\} + \bf{c.c.},
	\end{align}
	where we have used the convolution property of the scalar propagator
		\begin{align}
		{\cal G}_{k^+}(x^+,\bf{x};y^+,\bf{y}) = \int_\bf{u} {\cal G}_{k^+}(x^+,\bf{x};z^+,\bf{u}) {\cal G}_{k^+}(z^+,\bf{u};y^+,\bf{y}),\ \ y^+<z^+<x^+,
	\end{align}
	in order to break the target average of the scalar propagator into its local pieces where the number of objects is fixed.

\begin{figure}[h!]
		\centering
		\includegraphics[scale=0.6]{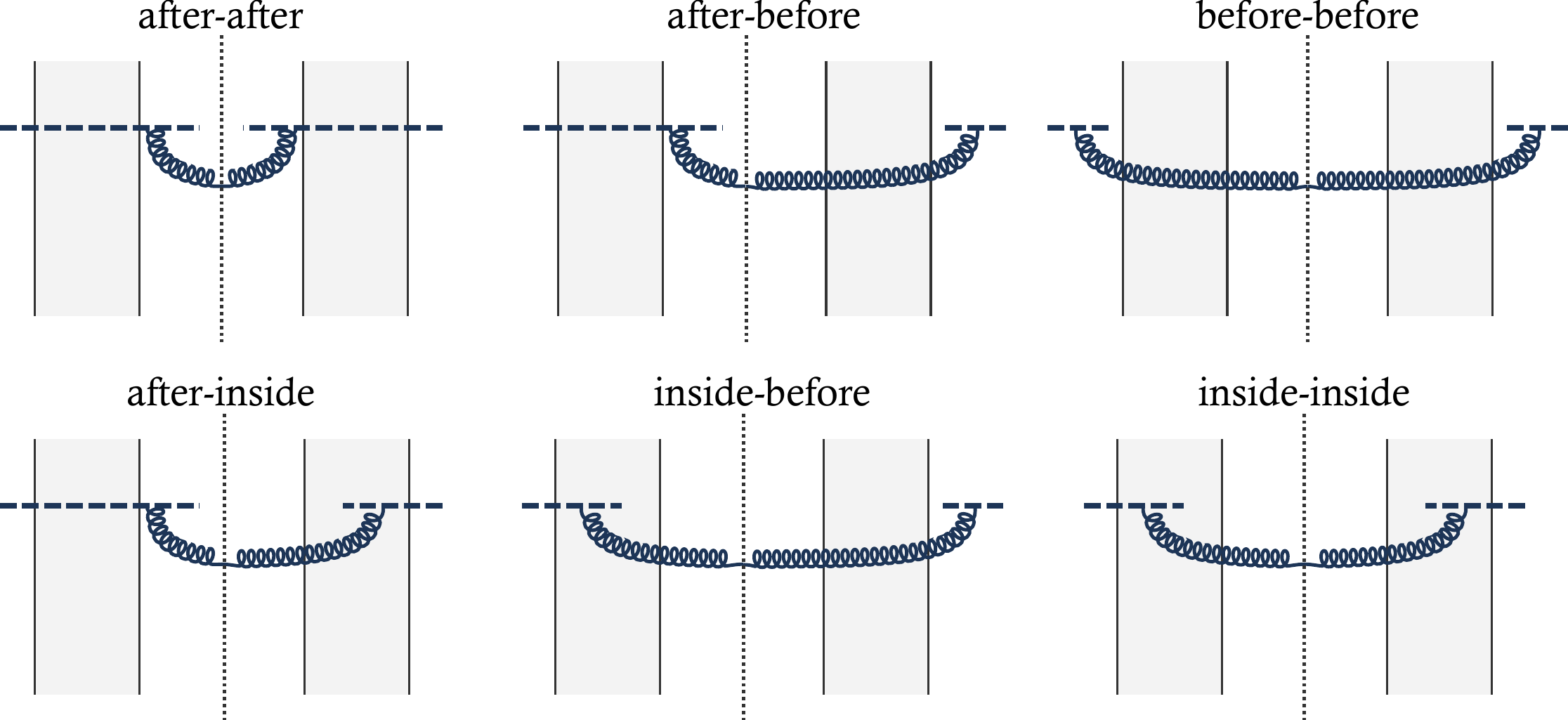}
		\caption{All possible contributions to the single gluon spectrum given in Eq.~\eqref{eq6:single_spectrum_NE}, which result from squaring the amplitude of a gluon being emitted before, during (inside) or after the interaction of the source with the target.}
		\label{fig6:contributions}
	\end{figure}

	
	It is worth mentioning that since we are using the GBW model (i.e., a Gaussian form) for the non-eikonal dipole functions, all the integrals over the transverse coordinates or transverse momenta appearing in Eq.~\eqref{eq6:target_average}, or in general in this manuscript, are of the form
	\begin{align}\label{eq6:gauss_int}
		\int_{\bf{x}} (a+\bf{b}^i \bf{x}^i + \bf{c}^i \bf{d}^j \bf{x}^i\bf{x}^j) e^{-A \bf{x}^2 + \bf{B}\cdot \bf{x}}
		=
		\frac{\pi}{A} e^{\frac{\bf{B}^2}{4A}} \Bigg[ a+ \bf{b}^i \frac{\bf{B}^i}{2A} + \frac{\bf{c}^i \bf{d}^j}{2A} \left(\delta^{ij}+\frac{\bf{B}^i \bf{B}^j}{2A} \right) \Bigg],
	\end{align}
	and, therefore, although very tedious in some cases, it is straightforward to compute them. Moreover, the argument of the integrals over $\bf{y}$ and $\bf{\bar{y}}$ is translational invariant under these coordinates and thus the integral is proportional to $\delta^{(2)}(\bf{q}_1-\bf{q}_2)$. Then, the single inclusive gluon spectrum given in Eq.~\eqref{multi2} can explicitly be written as 
	\begin{align}\label{eq6:single_spectrum_NE}
		&\frac{d N}{d \eta d^2 \textbf{k}}=\frac{g^2 \pi B_p }{(2 \pi)^3}
		\int_{\textbf{q}} \operatorname{Re} \Big\{ \mathcal{I}_{\text{aft-aft}}+\mathcal{I}_{\text{bef-bef}}+\mathcal{I}_{\text{aft-bef}}+\mathcal{I}_{\text{aft-in}}+\mathcal{I}_{\text{in-bef}}+\mathcal{I}_{\text{in-in}} \Big\},
	\end{align}
	where we have used the fact that $\mu^2(\bf{q},-\bf{q})=\pi B_p$, with $B_p$  the gluonic transverse area of the projectile. We have also changed our notation from the longitudinal momentum to rapidty: $dk^+/k^+=d\eta$. The terms that appear in the argument of the integral in Eq.~\eqref{eq6:single_spectrum_NE} result from squaring the contributions in Eq.~\eqref{eq6:reduced_amplitude} where the gluon is emitted before, after or inside the target medium (see Fig.~\ref{fig6:contributions}).
	
	By using the expressions for the eikonal dipole function and the two non-eikonal dipole functions, explicit expressions of each contribution in Eq.~\eqref{eq6:single_spectrum_NE} can be computed. Let us start with the  $\mathcal{I}_{\text{aft-aft}}$ term. This term is the contribution where the gluon is emitted after the interaction of the source with the target on both sides of the cut. It reads
	\begin{align}
		\label{eq6:single_afaf}
		\mathcal{I}_{\text{aft-aft}} = \frac{2}{{\bf k}^2} \int_{\bf{r}} e^{-i (\bf{k}-\bf{q}) \cdot \bf{r}} 
		d^{(0)}\left(L^+,0\Big|\frac{{\bf r}}{2};-\frac{{\bf r}}{2}\right) = \frac{8 \pi}{Q_s^2} \frac{1}{\textbf{k}^2}   \exp{-\frac{(\textbf{k}-\textbf{q})^2}{Q_s^2}}.
	\end{align}
	Note that we have dropped the dependence on the impact parameter $\bf{b}$ because the eikonal dipole function is translational invariant.
	
	The  next contribution in Eq.~\eqref{eq6:single_spectrum_NE} is the $\mathcal{I}_{\text{bef-bef}}$ term which corresponds to the emission of the gluon before the interaction of the source with the target on both sides of the cut. It reads 
	\begin{align}\label{bef-bef_additional}
		\mathcal{I}_{\text{bef-bef}} =
		\frac{2}{\bf{q}^2} \int_{{\bf r}} e^{-i\bf{q} \cdot \bf{r}}
		\int_{{\bf x},\bar{\bf x}}e^{-i{\bf k}\cdot({\bf x}-\bar{\bf x})}
		d^{(2)}\left(L^+,0\Big|,{\bf x},\bf{b}+\frac{{\bf r}}{2},k^+;\bar{\bf x},\bf{b}-\frac{{\bf r}}{2},k^+\right).	
	\end{align}
	We would like to mention that we keep the impact parameter $\bf{b}$ dependence in Eq. \eqref{bef-bef_additional} for consistency of the notation but, after performing the Fourier transforms, the final result is translationally invariant, i.e., independent of $\bf{b}$ (see also the comment before Eq.~\eqref{eq6:single_spectrum_NE}). Moreover, the Fourier transform of the $2^{\rm nd}$ order NE dipole when the two longitudinal momenta are equal (as in the case of Eq.~\eqref{eq6:gg_eq_function}) reads  
	\begin{align}\label{eq6:gg_eq_fourier}
		\hskip-0.5cm
		\int_{\textbf{x}, \bar{\textbf{x}}} e^{-i \textbf{k} (\textbf{x}- \bar{\textbf{x}})} d^{(2)}(x^+,y^+|{\bf x},{\bf y},k^+; \bar{\bf x},\bar{\bf y},k^+)
		=\exp{-\frac{\Delta^+ Q_s^2}{4 L^+} (\bf{y}-\bf{\bar{y}})^2-i \bf{k} \cdot (\bf{y}-\bf{\bar{y}})},
	\end{align}
	which shows that in this case the $2^{\rm nd}$ order NE dipole function does not provide any non-eikonal correction. As we shall see later, this is not the case when the two longitudinal momenta are different. Thus, the result for the $\mathcal{I}_{\text{bef-bef}}$ contribution reads 
	\begin{align}
		\label{eq6:single_bebe}
		\mathcal{I}_{\text{bef-bef}} =
		\frac{8 \pi}{Q_s^2} \frac{1}{\textbf{q}^2}   \exp{- \frac{(\textbf{k}-\textbf{q})^2}{Q_s^2}}.
	\end{align}
	
	The $\mathcal{I}_{\text{aft-bef}}$ term corresponds to the case where the gluon is emitted before the interaction of the source with target on one side of the cut and after on the other side. This term can be written as 
	\begin{align}
		\mathcal{I}_{\text{aft-bef}}=-4 \frac{{\bf k}\cdot{\bf q}}{{\bf k}^2{\bf q}^2}
		\int_{{\bf r}} e^{-i\bf{q} \cdot \bf{r}} \int_{\bf x}e^{-i{\bf k}\cdot({\bf x}-\bf{b}+\bf{r}/2)}
		d^{(1)}\left(L^+,0\Big|{\bf x},\bf{b}+\frac{{\bf r}}{2},k^+;\bf{b}-\frac{{\bf r}}{2}\right).
	\end{align}
	Noting that the Fourier transform of the $1^{\rm st}$ order NE dipole given in Eq.~\eqref{eq6:gu_function} reads 
	\begin{align}\label{eq6:gu_fourier}
		&\int_\textbf{x} e^{-i\textbf{k}\textbf{x}} d_{k^+}^{(1)}(x^+,y^+|{\bf x},{\bf y};\bar{\bf y})
		\nonumber \\ & \hskip1cm 
		=\frac{1}{\sin \frac{\Delta^+ \epsilon}{L^+}}
		e^{-i\bf{k}\cdot \bf{\bar{y}}}
		\exp{
			\left(-\frac{Q_s^2}{4 \epsilon} (\bf{y}-\bf{\bar{y}})^2 + \epsilon \frac{\bf{k}^2}{Q_s^2}\right) \tan\frac{\Delta^+ \epsilon}{L^+}
			+i \frac{\bf{k} \cdot (\bf{y}-\bf{\bar{y}})}{\sin\frac{\Delta^+ \epsilon}{L^+}}
		} ,
	\end{align}
$\mathcal{I}_{\text{aft-bef}}$ takes the following form:
	\begin{align}
		\label{eq6:single_beaf}
		\mathcal{I}_{\text{aft-bef}}=
		-\frac{16 \pi \epsilon}{ Q_s^2 \sin \epsilon} \frac{\textbf{k}\cdot\textbf{q}}{\textbf{k}^2 \textbf{q}^2}
		\exp{ 
			- \frac{\epsilon}{Q_s^2} \left[
			\frac{\textbf{k}^2+\textbf{q}^2}{\tan \epsilon}-2 \frac{\textbf{k} \cdot \textbf{q}}{\sin \epsilon} \right]
		}.
	\end{align}
	It is straightforward to realize that from this equation one can recover the results coming from the GBW model by considering the eikonal limit $|\epsilon^2| \to 0$ (see the dependence on $\textbf{k}$ and \textbf{q} in Eq.~\eqref{eq6:single_afaf}). 
	
	$\mathcal{I}_{\text{aft-in}}$  corresponds to the case where the gluon is emitted after the interaction of the source with the target on one side of the cut and inside the target on the other side. This contribution can be written  
	\begin{align}
		\mathcal{I}_{\text{aft-in}}=
		\frac{{\bf k}^i}{{\bf k}^2}\frac{2}{k^+}\int_0^{L^+}dy^+\int_{{\bf x}, \bf{r}}
		e^{-i\bf{q} \cdot \bf{r}-i{\bf k}\cdot({\bf x}-\bar{\bf y})}
		d^{(0)}(y^+,0|{\bf y};\bar{\bf y}) \big[ \partial_{{\bf y}^i}d^{(1)}(L^+,y^+|{\bf x},{\bf y},k^+;\bar{\bf y})\big],
	\end{align}
	where $\bf{y}=\bf{b}+\frac{\bf{r}}{2}$, $\bf{\bar{y}}=\bf{b}-\frac{\bf{r}}{2}$ and $y^+$ is the longitudinal coordinate where the gluon is emitted inside the target. Using Eq.~\eqref{eq6:gu_fourier} and defining $\alpha=\frac{L^+-y^+}{L^+}$ as the fraction of the target that the gluon traverses, we can write this contribution as
	\begin{align}
		\label{eq6:single_afin}
		&\mathcal{I}_{\text{aft-in}}=
		\frac{16 \pi \epsilon^3}{Q_s^4}\int_0^{1}d\alpha
		\frac{\sec^2 (\alpha \epsilon)}{(\epsilon \tilde{\alpha}+\tan(\alpha \epsilon))^2} \frac{\textbf{k}^2 \tilde{\alpha} \epsilon+\textbf{k}\cdot \textbf{q} \sin(\alpha \epsilon)}{\textbf{k}^2}
		\nonumber \\ & \hskip1cm \times
		\exp{- \frac{\epsilon}{Q_s^2(\epsilon \tilde{\alpha} +\tan(\alpha \epsilon))} 
			\Big[\textbf{k}^2\big(1-\tilde{\alpha} \epsilon \tan (\alpha \epsilon)\big)+\textbf{q}^2-2\textbf{k} \cdot \textbf{q} \sec(\alpha \epsilon)\Big]
		}.
	\end{align}
	Here, we also introduced the notation $\tilde{\alpha}=1-\alpha$ that corresponds to the longitudinal extent of the target  traversed by the source before emitting the gluon. It is worth mentioning that when the trigonometric functions are expanded to the leading order in  $\epsilon$, $\mathcal{I}_{\text{aft-in}}$ is $\mathcal{O}(\epsilon^2)$ while the first three contributions ($\mathcal{I}_{\text{aft-aft}}$, $\mathcal{I}_{\text{bef-bef}}$, $\mathcal{I}_{\text{aft-bef}}$) are  $\mathcal{O}(\epsilon^0)$. This implies that $\mathcal{I}_{\text{aft-in}}$ term is the first genuine non-eikonal contribution which is absent in the shockwave approximation. 
	
	
	$\mathcal{I}_{\text{in-bef}}$ term is another interference contribution that corresponds to emission of the gluon inside the target on one side of the cut and before on the other side and it can be written 
	\begin{align}
		&\mathcal{I}_{\text{in-bef}}=
		-\frac{{\bf q}^i}{{\bf q}^2}\frac{2}{k^+}\int_0^{L^+}d y^+
		\int_{{\bf x},\bar{\bf x},{\bf u},\bf{r}} e^{-i\bf{q} \cdot \bf{r}-i{\bf k}\cdot({\bf x}-\bar{\bf x})}
		\big[\partial_{\bar{\bf y}^i}d^{(2)}(L^+, y^+|{\bf x},{\bf u},k^+; \bar{\bf x},\bar{\bf y},k^+)\big]
		\nonumber \\ & \hskip9cm \times
		d^{(1)}( y^+,0|{\bf u},{\bf y},k^+;\bar{\bf y}).
	\end{align}
	This integral can be performed in the same manner as before, using Eq.~\eqref{eq6:gg_eq_fourier} and  the same definition of $\alpha$. The result reads
	\begin{align}
		\label{eq6:single_bein}
		&\mathcal{I}_{\text{in-bef}}=
		-\frac{16 \pi \epsilon^3}{Q_s^4}
		\int_0^{1}d\alpha
		\frac{\sec^2 (\alpha \epsilon)}{(\epsilon \tilde{\alpha}+\tan(\alpha \epsilon))^2} \frac{\textbf{q}^2 \tilde{\alpha} \epsilon+\textbf{k}\cdot \textbf{q} \sin(\alpha \epsilon)}{\textbf{q}^2}
		\nonumber \\ & \hskip1cm \times
		\exp{
			- \frac{\epsilon}{Q_s^2(\epsilon \tilde{\alpha} +\tan(\alpha \epsilon))}
			\Big[\textbf{k}^2+\textbf{q}^2\left[1-\tilde{\alpha} \epsilon \tan (\alpha \epsilon)\right]-2\textbf{k} \cdot \textbf{q} \sec(\alpha \epsilon)\Big]	
		}.
	\end{align}
	
	Finally, $\mathcal{I}_{\text{in-in}}$  is the last contribution to the single inclusive gluon spectrum given in Eq.~\eqref{eq6:single_spectrum_NE} and it accounts for the case in which the gluon is emitted inside the target on both sides of the cut. This contribution can be written 
	\begin{align}
		&\mathcal{I}_{\text{in-in}}=
		\frac{1}{(k^+)^2}\int_0^{L^+}dy^+\int_{y^+}^{L^+}d\bar y^+\int_{{\bf x},\bar{\bf x},{\bf u},\bf{r}}
		e^{-i\bf{q} \cdot \bf{r}-{\bf k}\cdot({\bf x}-\bar{\bf x})}
		\big[ \partial_{\bar{\bf y}^i}d^{(2)}(L^+,\bar{y}^+|{\bf x},{\bf u},k^+;\bar{\bf x},\bar{\bf y},k^+)\big]
		\nonumber \\ & \hskip6cm \times
		\big[\partial_{{\bf y}^i}d^{(1)}(\bar y^+,y^+|{\bf u},{\bf y},k^+;\bar{\bf y})\big]
		d^{(0)}(y^+,0|{\bf y};\bar{\bf y}).
	\end{align}
	Here the gluon on the left side of the cut is emitted at longitudinal position $y^+$ while the one on the right side of the cut is emitted at position $\bar{y}^+>y^+$. By defining $\xi=(\bar{y}^+-y^+)/L^+$ as the fractional longitudinal length traveled by the first gluon before the second one is emitted, $\tilde{\alpha}=y^+/L^+$ and using Eqs.~\eqref{eq6:gg_eq_fourier}, \eqref{eq6:gu_function} and \eqref{eq6:gbw_local}, we can write
	\begin{align}
		\label{eq6:single_inin}
		&\mathcal{I}_{\text{in-in}}=
		-\frac{16 \pi \epsilon^5}{Q_s^6}
		\int_0^1 d\tilde{\alpha} \int_0^{1-\tilde{\alpha}} d\xi
		\frac{\csc^2(\xi \epsilon)}{\left[1-\epsilon^2 \tilde{\alpha} \gamma+(1-\xi) \epsilon \cot(\xi \epsilon) \right]^3} 
		\nonumber \\ & \hskip0cm \times
		\Bigg[ 
		\epsilon \tilde{\alpha} \gamma Q_s^2  \left(1-\epsilon^2 \tilde{\alpha} \gamma+(1-\xi) \epsilon \cot(\xi \epsilon) \right)
		+ \epsilon \tilde{\alpha} \textbf{k}^2  \Big(1+\epsilon \tilde{\alpha} \cot(\xi \epsilon)\Big)
		\nonumber \\ & \hskip0cm 
		+\epsilon \gamma  \textbf{q}^2 \Big(1+\epsilon \gamma \cot(\xi \epsilon)\Big)
		+ \frac{\epsilon^2 \tilde{\alpha} \gamma+ \sin^2 (\xi \epsilon) \Big(1+\epsilon \tilde{\alpha} \cot(\xi \epsilon)\Big) \Big(1+\epsilon \gamma \cot(\xi \epsilon)\Big)}{\sin(\xi \epsilon)} \textbf{k} \cdot \textbf{q} \Bigg]
		\nonumber \\ & \hskip2cm \times
		\exp{ -\frac{\epsilon  \Big(\epsilon \tilde{\alpha}- \cot(\xi \epsilon)\Big) \textbf{k}^2 +\Big(\epsilon \gamma- \cot(\xi \epsilon)\Big) \textbf{q}^2+2 \csc(\xi \epsilon) \textbf{k} \cdot \textbf{q} }{Q_s^2\left[1-\epsilon^2 \tilde{\alpha} \gamma+(1-\xi) \epsilon \cot(\xi \epsilon) \right]} }.
	\end{align}
	In Eq.~\eqref{eq6:single_inin} we introduced a new variable $\gamma=1-\xi-\tilde{\alpha}$ that corresponds to the longitudinal fraction of the target  which the two gluons traverse at the same time. We  would also like to mention that by expanding the trigonometric functions at leading order in powers of $\epsilon$, $\mathcal{I}_{\text{in-in}}$ term is  ${\cal O}(\epsilon^4)$ and therefore this non-eikonal correction is sub-leading with respect to the other contributions.
	
	Summing up, the non-eikonal single inclusive gluon spectrum is given by Eq.~\eqref{eq6:single_spectrum_NE}, with the explicit expressions for the contributions inside the integral  provided in  Eqs.~\eqref{eq6:single_afaf},  \eqref{eq6:single_bebe}, \eqref{eq6:single_beaf} \eqref{eq6:single_afin}, \eqref{eq6:single_bein} and \eqref{eq6:single_inin} respectively.  In the two following subsections, we investigate these results further.


\subsection{Numerical results}
\label{sec6:single_numerics}
	
	This subsection is devoted to a numerical analysis of the results derived in Subsection~\ref{sec:analytical} to illustrate the effects of the non-eikonal corrections on single inclusive gluon production in $pA$ collisions. These non-eikonal effects, that stem from the finite longitudinal width of the target, are encoded in the so-called {\it non-eikonal parameter} $\epsilon$ defined in Eq. \eqref{eq6:epsilon}. This non-eikonal parameter depends on the saturation momentum $Q_s$, the gluon longitudinal momentum $k^+$ and the longitudinal width of the target $L^+$. For convenience of the analysis, the longitudinal momentum of the gluon $k^+$ and the width of the target $L^+$ can be written as (see Ref.~\cite{Agostini:2019hkj} for a detailed discussion) 
	\begin{align}
	k^+ = \frac{1}{\sqrt{2}}|\bf{k}| e^\eta, \\
	L^+=\frac{20 A^{1/3}}{\sqrt{s_{\rm NN}}}\ .
	\end{align}
	Thus, non-eikonal corrections depend on the values of the mass number $A$ of the target, pseudorapidity $\eta$ and  transverse momentum $p_\perp \equiv |\bf{k}|$ of the produced gluon, and center of mass energy $\sqrt{s_{\rm NN}}$.  We analyze the dependence of the single inclusive gluon spectrum on these variables. 
	
	
	In our analysis, in order to regulate the infrared divergences that appear, we introduce a mass in the denominators that lead to these divergences\footnote{A comment is in order here. In the eikonal limit, the reduced matrix amplitude defined in Eq. \eqref{eq6:reduced_amplitude} can be further factorized into the Lipatov vertex and the standard Wilson lines that describe the interaction between the projectile and the target (see for example \cite{Altinoluk:2018ogz}). In the eikonal limit, the mass term introduced here regulates the infrared divergence appearing in the Lipatov vertex. In the case of a non-eikonal treatment of the target one can not factorize the Lipatov vertex from the reduced matrix amplitude. However, the mass term introduced here regulates the infrared divergence appearing the kinematic factor that is the analog of the Lipatov vertex of the eikonal limit.}, i.e., $1/\bf{q}^2 \to 1/(\bf{q}^2 + m^2)$ in Eqs. ~\eqref{eq6:single_bebe},\eqref{eq6:single_beaf} and \eqref{eq6:single_bein}.  We fix the value of the regulator to $m = 0.6$ GeV and check the dependence of our results on this value. Our analysis shows a mild dependence on the value of the regulator: non-eikonal corrections become slightly smaller for smaller values of the infrared regulator. Moreover, we fix the value of the saturation scale to $Q_s=\sqrt{2}$ GeV and take $A=197$ (gold nucleus as target). Note that the non-eikonal parameter depends mildly,  $\propto A^{1/3}$,  on $A$.
	
\begin{figure}[h!]
		\centering
		\includegraphics[scale=0.8]{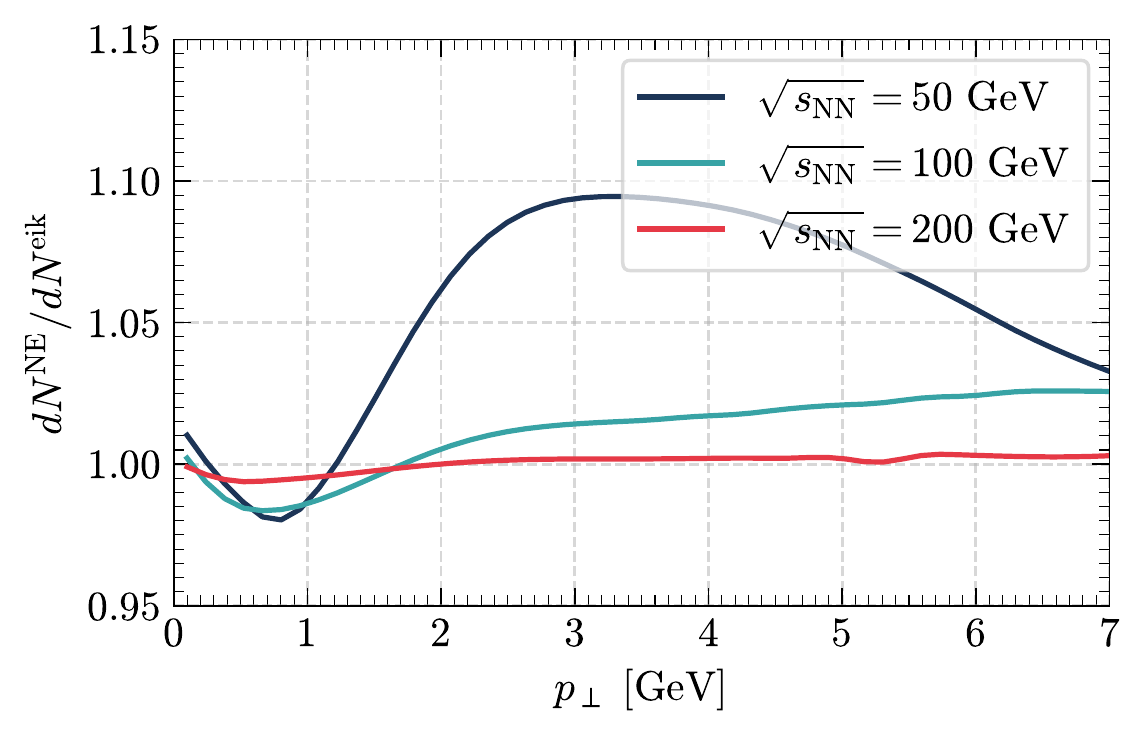}
		\caption{Ratio of the non-eikonal single inclusive gluon spectrum, Eq.~\eqref{eq6:single_spectrum_NE}, to the  eikonal result ($\epsilon=0$) as a function of gluon transverse momentum, $p_\perp$, for three values of the center-of-mass energy per nucleon and for $\eta=0$.}
		\label{fig6:NE_vs_pt}
	\end{figure}

	In Fig.~\ref{fig6:NE_vs_pt}, the ratio of the non-eikonal yield given in Eq.~\eqref{eq6:single_spectrum_NE} to the strict eikonal result (where $\epsilon=0$) is plotted as a function of  transverse momentum, $p_\perp$, for three  values of the center-of-mass energy per nucleon and for $\eta=0$. The results show that while non-eikonal corrections decrease the yield at low $p_\perp$, they cause an enhancement at $p_\perp \gtrsim 1$ GeV.  Moreover, the effect of non-eikonal contributions are about $\sim 10$ \% when $\sqrt{s_{\rm NN}} = 50$ GeV and $p_\perp \sim 3$ GeV, but they are almost negligible for the highest center-of-mass energy $\sqrt{s_{\rm NN}} = 200$ GeV at RHIC. The results suggest that the impact of the finite width effects on phenomenology studies in single inclusive gluon production at relatively high energies is almost negligible.

\begin{figure}[h!]
		\centering
		\includegraphics[scale=0.8]{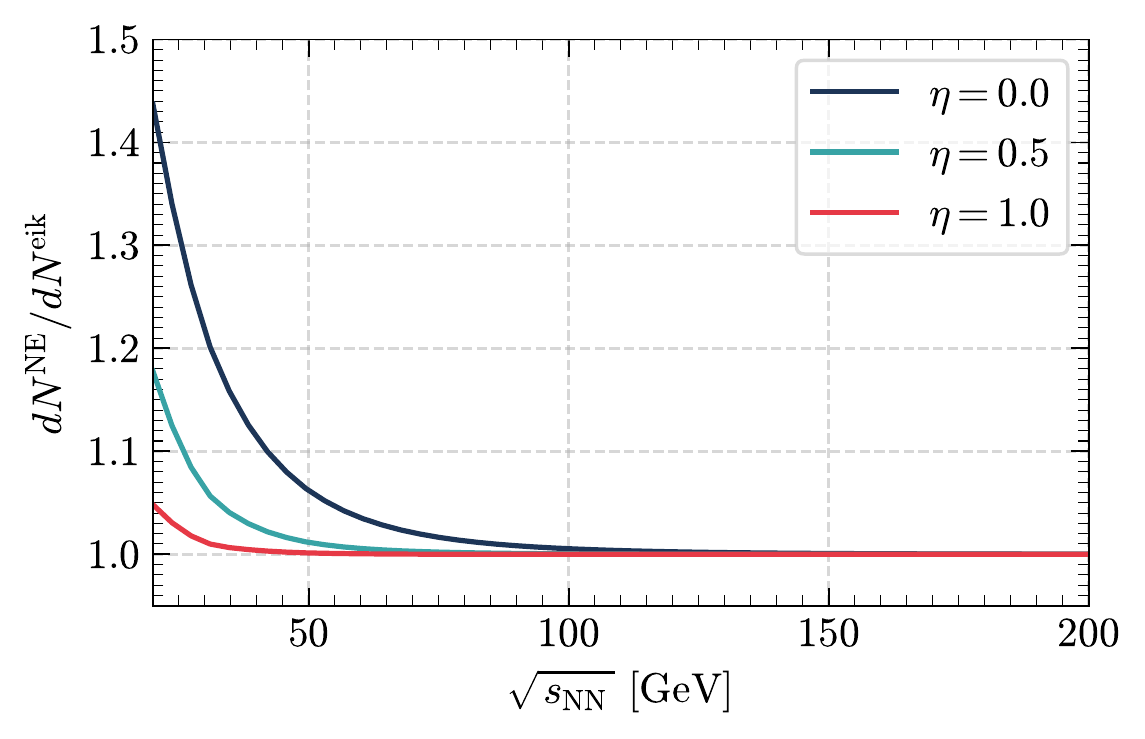}
		\caption{Ratio of the non-eikonal single gluon spectrum, given in Eq.~\eqref{eq6:single_spectrum_NE}, with respect to the eikonal one as a function of the center-of-mass energy per nucleon for three values of  pseudorapidity and for $p_\perp=2$ GeV.}
		\label{fig6:NE_vs_snn}
	\end{figure}

	In Fig. \ref{fig6:NE_vs_snn}, we plot the same ratio as a function of the center-of-mass energy per nucleon for three values of  pseudorapidity and  for $p_\perp=2$ GeV. The results indicate that for $\eta=0$ the finite width effects are changing roughly between $10$  \%  to $3$ \%  up to between $\sqrt{s_{\rm NN}}\sim 40$  GeV to $\sqrt{s_{\rm NN}}\sim 100$ GeV, for $\eta=0.5$ they lead to up  to $4$ \% effect between $\sqrt{s_{\rm NN}}\sim 40$ GeV and $\sqrt{s_{\rm NN}}\sim 50$ GeV and for $\eta=1$ they are almost negligible. Note that even though the results are plotted starting from $\sqrt{s_{\rm NN}} = 20$ GeV, at this low energy  Bjorken-$x$ of the target parton probed by the projectile is $x \sim \frac{1 \ {\rm GeV}}{\sqrt{s_{\rm NN}}} e^{-\eta}$, i.e., not small, and therefore our approach cannot not be considered reliable. Therefore, we consider our results starting from $\sqrt{s_{\rm NN}}\sim 40$ GeV where our approach is still valid.

\begin{figure}[h!]
		\centering
		\includegraphics[scale=0.8]{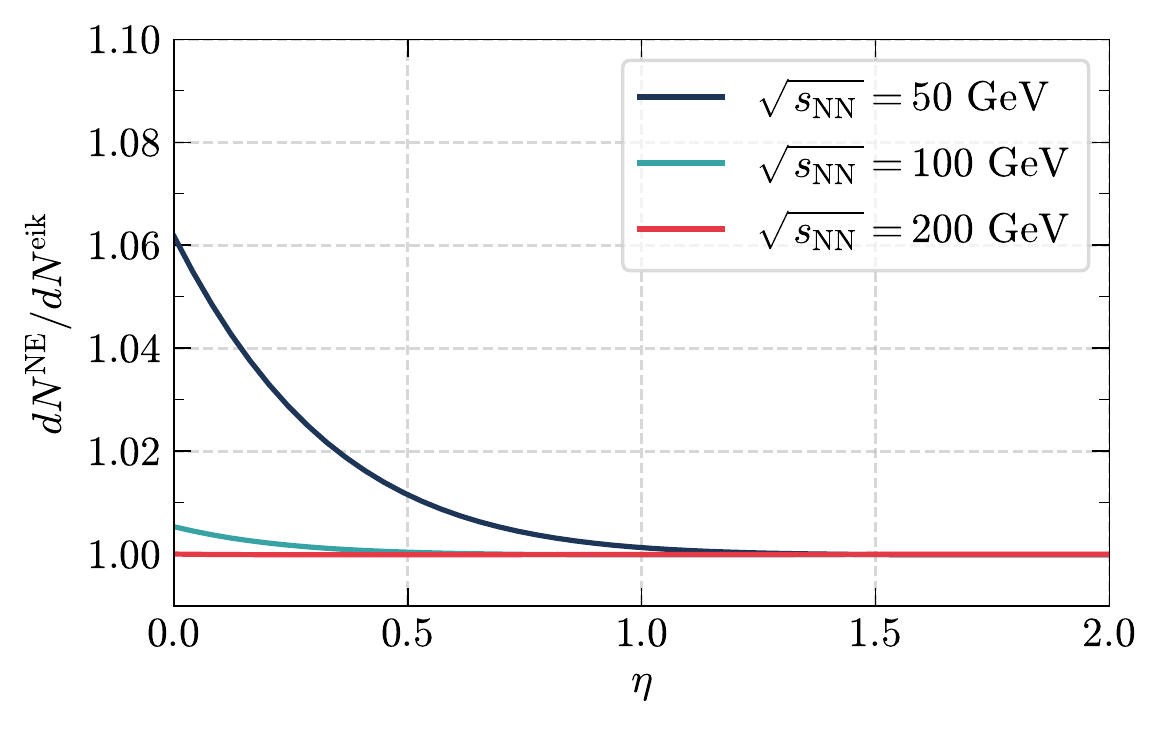}
		\caption{Ratio of the non-eikonal single gluon spectrum, given in Eq.~\eqref{eq6:single_spectrum_NE}, with respect to the eikonal one as a function of the pseudorapidity of the produced gluon for three values of $\sqrt{s_{\rm NN}}$ and for $p_\perp=2$ GeV.}
		\label{fig6:NE_vs_eta}
	\end{figure}

	Finally, in Fig. \ref{fig6:NE_vs_eta}, we plot the ratio as a function of the pseudorapidity of the produced gluon for three values of $\sqrt{s_{\rm NN}}$ and for $p_\perp=2$ GeV. The conclusions are analogous to the ones extracted from previous figures: the  non-eikonal corrections yield to up to $\sim 6$ \% effect for $\eta<1$ and when $\sqrt{s_{\rm NN}} = 50$ GeV but it is almost negligible at higher energies at all values of the pseudorapidity.

\subsection{Connection with earlier results in the literature}
\label{sec6:comparison_altinoluk}
	
	To finalize our study on the non-eikonal single inclusive gluon production, we make the explicit connection between our results and  the ones obtained in Refs.~\cite{Altinoluk:2014oxa,Altinoluk:2015gia}. We would like to emphasize that the results derived in this manuscript do not adopt any kind of expansion in the non-eikonal parameter $\epsilon^2$, and are therefore valid to all orders. On the other hand,  in Refs.~\cite{Altinoluk:2014oxa,Altinoluk:2015gia} a systematic expansion of the scalar propagator is proposed and the corrections are obtained up to next-to-next-to-eikonal accuracy. The expansion proposed in Refs.~\cite{Altinoluk:2014oxa,Altinoluk:2015gia} for the scalar propagator is performed in two steps. First, an expansion of the scalar propagator around its classical trajectory, i.e., the one that minimizes the path integral and leads to a linear path instead of the Brownian motion. Then, a second geometric expansion by assuming that the angle between the classical trajectory and the longitudinal axis with constant transverse coordinate is small since it is proportional to  $L^+/k^+$, and that the target field is weak far away from the classical solution. The corresponding result is an expansion in powers of $\frac{L^+}{k^+}=\frac{2 i \epsilon^2}{Q_s^2}$. It is effectively performed for the function $\mathcal{R}_{\underline{k}}(L^+,0;\textbf{y})$  which encodes the finite width target effects and is related to the scalar propagator via 
	\begin{align}\label{rkdef}
		\int_{\textbf{x}}e^{-i \textbf{k}\textbf{x}} \mathcal{G}_{k^+}(L^+,\textbf{x};0,\textbf{y})=e^{-ik^- L^+} e^{-i\textbf{k} \textbf{y}}  \mathcal{R}_{\underline{k}}(L^+,0;\textbf{y}).
	\end{align}
The result of the expansion at next-to-next-to-eikonal accuracy reads~\cite{Altinoluk:2015gia}
	%
	\begin{align}\label{rk}
		&\mathcal{R}_{\underline{k}}(L^+,0;\textbf{y})=U + \frac{2 i \epsilon^2}{Q_s^2} \left[ \textbf{k}^i U_{[0,1]}^i + \frac{i}{2} U_{[1,0]} \right]
		\nonumber \\ & \hskip4cm
		+\left( \frac{2 i \epsilon^2}{Q_s^2}\right)^2 \left[ \textbf{k}^i \textbf{k}^j U_{[0,2]}^{ij}+\frac{i}{2} \textbf{k}^i U_{[1,1]}^i-\frac{1}{4}U_{[2,0]} \right]+\mathcal{O}\left(\epsilon^6\right),
	\end{align}
	where the color objects $U^{i \cdots j}_{[\alpha,\beta]}(L^+,0; \textbf{y})$\footnote{For simplicity of the expressions we have dropped the arguments of these functions. Moreover, since the definition of these color objects is lengthy we refer to  Ref.~\cite{Altinoluk:2015gia} for their explicit expressions.} are referred to as \textit{decorated Wilson lines}. They can be identified with the non-eikonal partners of the standard Wilson lines. 
	
	In order to make the connection between the two results, one can immediately realize that the Fourier transform of the non-eikonal dipole functions that we have introduced in Eqs.~\eqref{eq6:dipole_gg} and \eqref{eq6:dipole_gu} can be written in terms of $\mathcal{R}_{\underline{k}}(L^+,0;\textbf{y})$ as  
	\begin{align}
		&\int_\textbf{x} e^{-i\textbf{k} \cdot \textbf{x}} d^{(1)}(L^+,0|{\bf x},{\bf y},k^+;\bar{\bf y})
		=\frac{e^{-i k^- L^+} e^{-i\textbf{k} \cdot \bf{y}}}{(N_c^2-1)}
		\Big \langle {\rm Tr} \left[\mathcal{R}_{\underline{k}}(L^+,0;\textbf{y}) U_{\bar{\textbf{y}}}^\dagger(L^+,0) \right] \Big \rangle_T , \label{FTd12}
		\\
		&\int_{\textbf{x}, \bar{\textbf{x}}} e^{-i \textbf{k} \cdot (\textbf{x}- \bar{\textbf{x}})} 
		d^{(2)}(L^+,0|{\bf x},{\bf y},k^+; \bar{\bf x},\bar{\bf y},k^+)
		\nonumber \\ & \hskip5cm 
		= \frac{e^{-i\textbf{k} \cdot (\bf{y}-\bf{\bar{y}})}}{(N_c^2-1)} \Big \langle {\rm Tr} \left[\mathcal{R}_{\underline{k}}(L^+,0;\textbf{y}) \mathcal{R}^{\dagger}_{\underline{k}}(L^+,0;\bar{\textbf{y}})\right] \Big \rangle_T.
		\label{FTd22}
	\end{align}
	
	Therefore, by using the non-eikonal expansion given in Eq.~\eqref{rk} we can write the Fourier transform of the $1^{\rm st}$ order NE dipole function as
	\begin{align}\label{RU}
		&e^{i k^- L^++i\textbf{k} \cdot \bf{y}}
		\int_\textbf{x} e^{-i\textbf{k} \cdot \textbf{x}} d^{(1)}(L^+,0|{\bf x},{\bf y},k^+;\bar{\bf y})
		= \mathcal{O}(\textbf{r}) + 
		\frac{2 i \epsilon^2}{Q_s^2} \left[ \textbf{k}^i \mathcal{O}_{[0,1]}^i(\textbf{r}) + \frac{i}{2} \mathcal{O}_{[1,0]}(\textbf{r}) \right]
		\nonumber \\ & \hskip3cm 
		+\left( \frac{2 i \epsilon^2}{Q_s^2} \right)^2\left[ \textbf{k}^i \textbf{k}^j \mathcal{O}_{[0,2]}^{ij}(\textbf{r})+\frac{i}{2} \textbf{k}^i \mathcal{O}_{[1,1]}^i(\textbf{r})-\frac{1}{4}\mathcal{O}_{[2,0]}(\textbf{r}) \right]+\mathcal{O}\left( \epsilon^6\right),
	\end{align}
	where the tensors $\mathcal{O}^{i \cdots j;l\cdots m}_{[\alpha,\beta];[\gamma,\delta]}(\textbf{r})$ are referred to as  \textit{decorated dipoles} and are the leading eikonal corrections to the eikonal dipole function $\mathcal{O}(\textbf{r})=d^{(0)}(\bf{r})$. Furthermore, we note that in order to write Eq.~\eqref{RU}, we assume translational invariance of the decorated dipoles and write $\bf{r}=\bf{y}-\bf{\bar{y}}$. The definitions of these dipole functions are:
	\begin{align}
		\mathcal{O}(\textbf{r})&=\frac{1}{N_c^2-1}
		\Big \langle {\rm Tr} \left[ U_{{\textbf{y}}}(L^+,0) U_{\bar{\textbf{y}}}^\dagger(L^+,0) \right] \Big \rangle_T,
		\\ 
		\mathcal{O}^{i \cdots j}_{[\alpha,\beta]}(\textbf{r})&=\frac{1}{N_c^2-1}
		\Big \langle {\rm Tr} \left[ U^{i \cdots j}_{[\alpha,\beta]}(L^+,0;\textbf{y}) U_{\bar{\textbf{y}}}^\dagger(L^+,0) \right] \Big \rangle_T,
		\\
		\mathcal{O}^{i \cdots j;l\cdots m}_{[\alpha,\beta];[\gamma,\delta]}(\textbf{r})&=\frac{1}{N_c^2-1}
		\Big \langle {\rm Tr} \left[ U^{i \cdots j}_{[\alpha,\beta]}(L^+,0;\textbf{y}) 
		U^{l \cdots m \dagger}_{[\gamma,\delta]}(L^+,0;\bar{\textbf{y}}) \right] \Big \rangle_T.
	\end{align}
	
	On the other hand, the Fourier transform of the $2^{\rm nd}$ order NE dipole function can be written in the same fashion, reading
	\begin{align}\label{RR}
		&e^{i\textbf{k} \cdot (\bf{y}-\bf{\bar{y}})}\int_{\textbf{x}, \bar{\textbf{x}}} e^{-i \textbf{k} \cdot (\textbf{x}- \bar{\textbf{x}})} 
		d^{(2)}(L^+,0|{\bf x},{\bf y},k^+; \bar{\bf x},\bar{\bf y},k^+)
		\nonumber \\ & \hskip1cm 
		=\mathcal{O}(\textbf{r})
		+\frac{2 i \epsilon^2}{Q_s^2} \left[ \textbf{k}^i\left( \mathcal{O}^i_{[0,1]}(\textbf{r})+\mathcal{O}^i_{[0,1]}(-\textbf{r}) \right)+\frac{i}{2} \left( \mathcal{O}_{[1,0]}(\textbf{r}) -\mathcal{O}_{[1,0]}(-\textbf{r})\right)\right]
		\nonumber \\ & \hskip1cm 
		+\left( \frac{2 i \epsilon^2}{Q_s^2} \right)^2 
		\Bigg[ \textbf{k}^i\textbf{k}^j \Big( \mathcal{O}^{ij}_{[0,2]}(\textbf{r})+\mathcal{O}^{ij}_{[0,2]}(-\textbf{r})+\mathcal{O}^{ij}_{[0,1];[0,1]}(\textbf{r}) \Big)
		\nonumber \\ & \hskip3cm 
		+\textbf{k}^i \frac{i}{2}\Big( \mathcal{O}^{i}_{[1,1]}(\textbf{r})-\mathcal{O}^{i}_{[1,1]}(-\textbf{r})+\mathcal{O}^{i}_{[1,0];[0,1]}(\textbf{r})-\mathcal{O}^{i}_{[1,0];[0,1]}(-\textbf{r})  \Big)
		\nonumber \\ & \hskip4.5cm 
		-\frac{1}{4}\Big( \mathcal{O}_{[2,0]}(\textbf{r})-\mathcal{O}_{[2,0]}(-\textbf{r})+\mathcal{O}_{[1,0];[1,0]}(\textbf{r}) \Big) \Bigg]
		+\mathcal{O}\left( \epsilon^6 \right).
	\end{align}

We can now make a one-to-one comparison between our result Eqs.~~\eqref{eq6:gu_fourier} and~\eqref{RU} (and also between Eqs.~\eqref{eq6:gg_eq_fourier} and~\eqref{RR}) by expanding our results to the appropriate order in order to obtain a parametrization of the decorated dipoles within the GBW model, i.e., assuming a Gaussian form. 
Expanding the Fourier transform of the $1^{\rm st}$ order NE dipole function given in Eq.~\eqref{eq6:gu_fourier} at next-to-next-to-eikonal accuracy, i.e., in terms of $\epsilon^2$ up to order $\epsilon^4$, we get 
	%
	\begin{align}\label{RU2}
		&e^{i k^- L^++i\textbf{k} \cdot \bf{y}}
		\int_\textbf{x} e^{-i\textbf{k} \cdot \textbf{x}} d^{(1)}(L^+,0|{\bf x},{\bf y},k^+;\bar{\bf y})
		\nonumber \\ & \hskip-0.5cm 
		= e^{-\frac{Q_s^2}{4} \bf{r}^2} \Bigg\{ 1 
		- 
		\frac{2 i \epsilon^2}{Q_s^2} \frac{Q_s^2}{4} \Bigg[
		\bf{k}^i \bf{r}^i - i \left(\frac{Q_s^2}{6} \bf{r}^2-1\right)
		\Bigg]
		+
		\left(\frac{2 i \epsilon^2}{Q_s^2}\right)^2 
		\Bigg[ \textbf{k}^i \textbf{k}^j \left( \frac{Q_s^4}{32} \textbf{r}^i\textbf{r}^j -\frac{Q_s^2}{12} \delta^{ij} \right)
		\nonumber \\ & \hskip1cm 
		- i \textbf{k}^i\textbf{r}^i Q_s^4\left( \frac{11}{96}+\frac{Q_s^2}{96} \textbf{r}^2 \right)
		+ Q_s^4\left( -\frac{5}{96}+\frac{3}{160} Q_s^2 \textbf{r}^2- \frac{Q_s^4}{1152} \textbf{r}^4 \right) \Bigg]
		\Bigg\}
		+
		{\cal O}(\epsilon^6)
		.
	\end{align}	
	
	A term by term comparison of Eqs.~\eqref{RU} and \eqref{RU2} leads to the following parametrization of the decorated dipoles in our Gaussian ansatz\footnote{These parametrizations can also be obtained in a straightforward way from the definition of the decorated Wilson lines, assuming a local version of the GBW model with Eq.~\eqref{eq:effqs}. We thank Carlota Andr\'es and Alexis Moscoso for this observation.}:
	\begin{align}
		\mathcal{O}(\textbf{r})&=e^{-\frac{Q_s^2}{4}\textbf{r}^2},
		\\
		\mathcal{O}_{[0,1]}^i(\textbf{r}) &=-\frac{Q_s^2}{4}\textbf{r}^i e^{-\frac{Q_s^2}{4}\textbf{r}^2},
		\\
		\mathcal{O}_{[1,0]}(\textbf{r}) &=\frac{Q_s^2}{12}\left(-6+Q_s^2 \textbf{r}^2\right) e^{-\frac{Q_s^2}{4}\textbf{r}^2},
		\\
		\mathcal{O}_{[0,2]}^{ij}(\textbf{r}) &=  \frac{Q_s^2}{4} \left( -\frac{1}{3} \delta^{ij} +\frac{Q_s^2}{8} \textbf{r}^i\textbf{r}^j\right) e^{-\frac{Q_s^2}{4}\textbf{r}^2},
		\\
		\mathcal{O}_{[1,1]}^i(\textbf{r})   &= -\frac{Q_s^4}{48} \textbf{r}^i \left( 11 + Q_s^2 \textbf{r}^2 \right) e^{-\frac{Q_s^2}{4}\textbf{r}^2},
		\\
		\mathcal{O}_{[2,0]}(\textbf{r}) &= \frac{Q_s^4}{8} \left( \frac{5}{3}-\frac{3}{5} Q_s^2\textbf{r}^2+\frac{1}{36} Q_s^4 \textbf{r}^4 \right) e^{-\frac{Q_s^2}{4}\textbf{r}^2}.
	\end{align}
	
	On the other hand, by using Eq.~\eqref{eq6:gg_eq_fourier}, the Fourier transform of the $2^{\rm nd}$ order NE dipole function  can be written as 
	\begin{align}\label{RR2}
		e^{i\textbf{k} \cdot (\bf{y}-\bf{\bar{y}})}\int_{\textbf{x}, \bar{\textbf{x}}} e^{-i \textbf{k} \cdot (\textbf{x}- \bar{\textbf{x}})} d^{(2)}(L^+,0|{\bf x},{\bf y},k^+; \bar{\bf x},\bar{\bf y},k^+)=e^{-\frac{1}{4}Q_s^2 \textbf{r}^2}.
	\end{align}
	As mentioned earlier, the $2^{\rm nd}$ order NE dipole function does not encode any finite width effect in the case in where both longitudinal momenta are equal and therefore it is of $\mathcal{O}(\epsilon^0)$. Thus, comparing Eqs.~\eqref{RR}  and \eqref{RR2} and keeping in mind that the non-eikonal corrections in Eq.~\eqref{RR} vanish in the Gaussian approximation, we obtain the following constraints for the decorated dipoles:
	\begin{align}
		\mathcal{O}^{ij}_{[0,1];[0,1]}(\textbf{r})&=-2\mathcal{O}^{ij}_{[0,2]}(\textbf{r}),
		\\
		\mathcal{O}^{i}_{[1,0];[0,1]}(\textbf{r})-\mathcal{O}^{i}_{[1,0];[0,1]}(-\textbf{r}) &=-2\mathcal{O}^{i}_{[1,1]}(\textbf{r}),
		\\
		\mathcal{O}_{[1,0];[1,0]}(\textbf{r})&=2\mathcal{O}_{[2,0]}(\textbf{r}).
	\end{align}
	
	We conclude that by comparing the model independent non-eikonal expansion given in Refs.~\cite{Altinoluk:2014oxa,Altinoluk:2015gia} with our approach, which relies on the Gaussian form of the dipole functions, we are able to obtain a parameterization of the decorated dipoles in the Gaussian ansatz that is valid for small size dipoles. This parameterization may be useful for including finite target width effects in future phenomenological studies.

\section{Non-eikonal multigluon production}
\label{sec6:ne_multiparticle_production}
	
	In this section we study the general case that corresponds to the non-eikonal  $n$-gluon production whose generic expression is given in Eq.~\eqref{eq6:multi_particle_pA}.  Throughout this study, we adopt the Area Enhancement approximation in order to write the multipoles as non-eikonal two point functions (or dipoles). The validity of the AE argument when the finite width target effects are included in the computation is discussed at the end of Section~\ref{sec6:NE_target_averaging}. We would like to also mention that  in Ref.~\cite{Agostini:2021xca} we investigated  multigluon production in the strict eikonal limit where we introduced a diagramatic approach to study the $2n$-point correlators within the AE argument.  Our aim, in this section, is to generalize that approach to the non-eikonal case where the finite longitudinal  width of the target is accounted for.

	Let us start by writing the $n$-gluon spectrum given in Eq.~\eqref{eq6:multi_particle_pA} in terms of the simplified notations introduced in Ref.~\cite{Agostini:2021xca}. Within the AE model, the non-eikonal $n$-gluon spectra can be written as 
	\begin{align}\label{eq6:multigluon_wick}
		\frac{d^n N}{\prod_{i=1}^{n} d\eta_{2i-1} d^2 \textbf{k}_{2i-1}}= \frac{1}{2^n (2 \pi)^{3n} }&\int \left(\prod_{i=1}^{2n} \frac{d^2 \textbf{q}_i}{(2 \pi)^2}\right)
		\left(\sum_{\sigma \in \Pi(\chi)} \prod_{\{i,j\}\in \sigma}  \Big \langle \Omega_i \Omega_j \Big \rangle_{p}\right)
		\nonumber \\ & \hskip3cm \times
		\left(\sum_{\omega \in \Pi(\chi)} \prod_{\{\alpha,\beta\}\in \omega}  \Big \langle \Lambda_\alpha \Lambda_\beta \Big \rangle_{T}\right),
	\end{align}
	where $\chi=\{1,2,\dots,2 n\}$  represents the total number of color charges in the amplitude and in the complex conjugate amplitude (i.e., on both sides of the cut) and $\Pi(\chi)$ represents the set of partitions of $\chi$ with disjoint pairs. The projectile is represented by function $\Omega_i$ which is given by
	\begin{align} \label{modfied_omega}
	\Omega_i= g \rho^{b_i}(\bf {q}_i),
	\end{align}
	with the property that for odd $i$ the color source sits on the left side of the cut and for even $i$ it sits on the right side. The two point correlator of the function $\Omega_i$ is defined as\footnote{We would like to note that both the definition of the function $\Omega_i$ (Eq. \eqref{modfied_omega}) and the definition of its 2-point correlator (Eq. \eqref{eq6:projectile_correlator}) differ from the definitions introduced in Ref.~\cite{Agostini:2021xca} for the following reason. In that reference, the transverse momenta of each projectile color source was denoted as $\bf{k}-\bf{q}$. In this manuscript, transverse momenta of each color source is denoted as $\bf {q}$. Therefore, compared to Ref.~\cite{Agostini:2021xca}, we perform a change of variable $\bf{k}-\bf{q}\to \bf{q}$. The factor $(-1)^{i+1}$ in the definition of the 2-point correlator is introduced in order to account for the change of the sign in the transverse momentum when the source is on the right side of the cut, i.e., for even $i$.} 
	%
	%
	\begin{align}\label{eq6:projectile_correlator}
		\Big \langle \Omega_i \Omega_j \Big \rangle_{p}=\frac{\delta^{b_i b_j}}{N_c^2-1} \mu_p^2\big((-1)^{i+1}\textbf{q}_i,(-1)^{j+1}\textbf{q}_j\big), \quad \mu_p^2(\bf{k},\bf{q})= \pi B_p e^{-\frac{(\bf{k}+\bf{q})^2}{4}B_p}.
	\end{align}
	
	On the other hand, the object $\Lambda_\alpha$ represents the target and its interaction with the projectile. It is defined as\footnote{The definition of $\Lambda_{\alpha}$ introduced in this manuscript differs from its definition introduced in Ref.~\cite{Agostini:2021xca}. Here, since we account for the finite longitudinal width effects of the target, the reduced amplitude $\overline{{\cal M}}^{a_\alpha b_\alpha}$ and therefore the object $\Lambda_{\alpha}$ is more involved compared to its eikonal limit which was used in Ref.~\cite{Agostini:2021xca}.} 
	\begin{align}
		\Lambda_\alpha = \overline{{\cal M}}^{a_\alpha b_\alpha}_{\lambda_\alpha}({\underline k}_\alpha,\bf{q}_\alpha),
	\end{align}
	with odd values of $\alpha$ corresponding to the left side of the cut and even values of $\alpha$  to the right side of the cut. Therefore, for even values of $\alpha$ one takes the complex conjugate of the reduced amplitude. 
	
	Similar to the strict eikonal case, we impose the following constraints on the produced gluons:
	\begin{align}
		{\underline k}_{2n} &= {\underline k}_{2n-1},
		\label{eq6:constraint_momenta}
		\\
		a_{2n} &= a_{2n-1},
		\\
		\lambda_{2n} &= \lambda_{2n-1},
	\end{align}
	namely  the color, polarization and both transverse and longitudinal momenta of the produced gluons have to be the same on both sides of the cut. 
We would like to remind that the arguments introduced at the end of Section \ref{sec6:NE_target_averaging} to justify the use of AE model in the non-eikonal case where the finite longitudinal width effects are accounted for, are based on the fact that the fluctuation time of the target fields is much larger than its longitudinal extent. This implies that  that chromo-electric domains do not decohere during the interaction of the projectile and the target. Moreover, it also allows one to argue that each domain has to be color neutral and the target average has to be a singlet. Therefore, we can write the two point correlators of  functions $\Lambda_\alpha$ as
	%
	%
	\begin{align}\label{def_simp_lambda}
		\Big \langle \Lambda_\alpha \Lambda_\beta \Big \rangle_{T} = \frac{\delta^{a_\alpha a_\beta} \delta^{b_\alpha b_\beta}}{(N_c^2-1)^2} \Big \langle \tr{ \overline{{\cal M}}_{\lambda_\alpha}({\underline k}_\alpha,\bf{q}_\alpha) \overline{{\cal M}}_{\lambda_\beta}({\underline k}_\beta,\bf{q}_\beta) } \Big \rangle_{T}.
	\end{align}
	Using the definition of the reduced amplitude given in Eq.~\eqref{eq6:reduced_amplitude}, and the definitions of the eikonal and non-eikonal dipole functions introduced in Eqs.~\eqref{eq6:dipole_gg}, \eqref{eq6:dipole_gu} and \eqref{eq6:dipole_uu}, the 2-point correlator becomes 
	\begin{align}\label{eq6:2point_target}
		&\Big \langle \Lambda_\alpha \Lambda_\beta \Big \rangle_T =
		\frac{\delta^{a_\alpha a_\beta} \delta^{b_\alpha b_\beta}}{N_c^2-1} (2\pi)^2 \delta^{(2)}\big[ \pm(\bf{k}_\alpha-\bf{q}_\alpha)+ \pm (\bf{k}_\beta-\bf{q}_\beta) \big]
		\nonumber \\ & \hskip0cm \times
		\Bigg\{
		2\frac{\bf{k}_\alpha^{\lambda_\alpha}\bf{k}_\beta^{\lambda_\beta}}{\bf{k}_\alpha^{2}\bf{k}_\beta^{2}} 
		I_{\rm aft-aft} \big(\bf{k}_\beta,\bf{q}_\beta \big)
		-
		4\frac{\bf{k}_\alpha^{\lambda_\alpha}\bf{q}_\beta^{\lambda_\beta}}{\bf{k}_\alpha^{2}\bf{q}_\beta^{2}}
		I_{\rm aft-bef} \Big(\bf{k}_\beta,\bf{q}_\beta,\pm k_\beta^+ \Big)
		\nonumber \\ & \hskip0cm
		+
		2\frac{\bf{q}_\alpha^{\lambda_\alpha}\bf{q}_\beta^{\lambda_\beta}}{\bf{q}_\alpha^{2}\bf{q}_\beta^{2}}
		I_{\rm bef-bef} \Big(\pm\bf{k}_\alpha,\pm\bf{q}_\alpha,\pm\bf{k}_\beta,\pm\bf{q}_\beta;\pm k_\alpha^+,\pm k_\beta^+ \Big)
		\nonumber \\ & \hskip0cm
		+
		2\frac{L^+}{k_\beta^+} \frac{\bf{k}_\alpha^{\lambda_\alpha}}{\bf{k}_\alpha^{2}} \int_0^1 df_\beta
		I^{\lambda_\beta}_{\rm aft-in} \Big(\pm \bf{k}_\beta,\pm \bf{q}_\beta,\pm k_\beta^+ ; f_\beta \Big)
		\nonumber \\ & \hskip0cm
		-
		2\frac{L^+}{k_\beta^+} \frac{\bf{q}_\alpha^{\lambda_\alpha}}{\bf{q}_\alpha^{2}} \int_0^1 df_\beta
		I^{\lambda_\beta}_{\rm in-bef} \Big(\pm\bf{k}_\alpha,\pm \bf{q}_\alpha,\pm\bf{k}_\beta,\pm\bf{q}_\beta;\pm k_\alpha^+,\pm k_\beta^+ ; f_\beta \Big)
		\nonumber \\ & \hskip0cm
		+
		\frac{L^+}{k_\alpha^+} \frac{L^+}{k_\beta^+} 
		\int_0^1 df_\beta \int_0^{f_\beta} df_\alpha
		I^{\lambda_\alpha \lambda_\beta}_{\rm in-in} \Big(\pm\bf{k}_\alpha,\pm\bf{k}_\beta,\pm \bf{q}_\beta;\pm k_\alpha^+,\pm k_\beta^+ ; f_\alpha, f_\beta \Big)
		\nonumber \\ & \hskip0cm
		+
		(\alpha \leftrightarrow \beta)
		\Bigg\},
	\end{align}
	where we have introduced the notation $\pm k_{a} \equiv (-1)^{a+1} k_a$. It corresponds to stating that when the respective momentum (either longitudinal or transverse) is on the left side of the cut, i.e., with $a$ odd, it is multiplied by $(+1)$ and when it is on the right side it is multiplied by $(-1)$, changing its sign due to the Fourier transform. The explicit expressions for the functions $I$ are given in Appendix~\ref{sec:2point}.

	Thus, in order to compute the $n$-gluon spectra, one needs to evaluate the Wick expansions given in Eq.~\eqref{eq6:multigluon_wick}. On the projectile side, since we are adopting the MV model for the projectile configurations, projectile averages factorize into 2-point correlators defined in Eq.~\eqref{eq6:projectile_correlator}. On the target side, we are using the AE model to perform the averages over the target fields which also factorize into 2-point correlators of the reduced matrix amplitudes, given in Eq.~\eqref{eq6:2point_target}. Expanding the $\left( \frac{(2n)!}{2^n n!} \right)^2$ terms in Eq.~\eqref{eq6:multigluon_wick} and evaluating the 2-point correlators (on both the projectile and the target sides),  we are able to compute the $n$-gluon spectra for any value of $n$ neglecting, of course, contributions that are subleading in powers of $\rho_p$. Since multigluon spectra contain a very large number of terms, it is convenient to use the diagrammatic approach introduced in Ref.~\cite{Agostini:2021xca}. 
	
	Finally, let us note that the single inclusive gluon spectrum can be computed in the approach presented in this section  by just writing 
	\begin{align}
		\frac{dN}{d\eta d^2\bf{k}} = \frac{g^2}{2 (2\pi)^3} \int_{\bf{q}_1,\bf{q}_2} \Big \langle \Omega_1 \Omega_2 \Big \rangle_{p} \Big \langle \Lambda_1 \Lambda_2 \Big \rangle_{T}.
	\end{align}
	It is straightforward to see that if we use Eqs.~\eqref{eq6:projectile_correlator} and \eqref{eq6:2point_target}, this equation reduces to Eq.~\eqref{eq6:single_spectrum_NE}. In the next subsection, we study the double inclusive gluon production by using the setup introduced here.

	\subsection{Non-eikonal double inclusive gluon production}
	\label{sec6:2gluon_production}
	
	Now we focus on the non-eikonal double inclusive gluon production -- the $n=2$ case in the multigluon spectrum in Eq.~\eqref{eq6:multigluon_wick}. In our set up, the double inclusive spectrum reads 
	\begin{align}
		\label{eq6:double_gluon_production}
		&\frac{d^2N}{d\eta_1 d^2\bf{k}_1 d\eta_3 d^2 \bf{k}_3} = \frac{g^4}{4(2\pi)^6} \int_{\bf{q}_1,\bf{q}_2,\bf{q}_3,\bf{q}_4}
		\nonumber \\ & \hskip2cm \times
		\Bigg( \Big \langle \Omega_1 \Omega_2 \Big \rangle_{p} \Big \langle \Omega_3 \Omega_4 \Big \rangle_{p}
		+ \Big \langle \Omega_1 \Omega_3 \Big \rangle_{p} \Big \langle \Omega_2 \Omega_4 \Big \rangle_{p}
		+ \Big \langle \Omega_1 \Omega_4 \Big \rangle_{p} \Big \langle \Omega_2 \Omega_3 \Big \rangle_{p}   \Bigg)
		\nonumber \\ & \hskip2cm \times
		\Bigg(
		\Big \langle \Lambda_1 \Lambda_2 \Big \rangle_{T} \Big \langle \Lambda_3 \Lambda_4 \Big \rangle_{T}
		+ \Big \langle \Lambda_1 \Lambda_3 \Big \rangle_{T} \Big \langle \Lambda_2 \Lambda_4 \Big \rangle_{T}
		+ \Big \langle \Lambda_1 \Lambda_4 \Big \rangle_{T} \Big \langle \Lambda_2 \Lambda_3 \Big \rangle_{T}
		\Bigg),
	\end{align}
	where the 2-point functions for the projectile and the target sides are given by Eqs.~\eqref{eq6:projectile_correlator} and \eqref{eq6:2point_target} respectively.
	
	Double inclusive gluon production has been studied extensively in the eikonal approximation (see Refs.~\cite{Dumitru:2008wn,Dumitru:2010iy,Kovchegov:2012nd,Kovchegov:2013ewa,Altinoluk:2015uaa,Altinoluk:2015eka,Altinoluk:2016vax}). In the eikonal limit, double inclusive gluon production spectrum contains not only a fully uncorrelated piece, but also  Bose Enhancement and Hanbury-Brown-Twiss-like (HBT) correlations encoded in the terms that are suppressed by powers of $(N_c^2-1)$ with respect to the fully uncorrelated one.  Moreover, it has been  shown that the eikonal results give non-zero even harmonics but vanishing odd harmonics. This is due to an accidental symmetry of the CGC which is encoded in the double inclusive spectrum through its $\bf{k}_3\to -\bf{k}_3$ symmetry, with ${\bf k}_3$ being the transverse momenta of the second gluon. Over the last decade, there have been many suggested ways to break this symmetry (see Ref.~\cite{Altinoluk:2020wpf} for a review). One of the ways to break this symmetry and generate non-zero odd harmonics through CGC calculations is to include non-eikonal corrections, as has been shown in Refs.~\cite{Agostini:2019avp,Agostini:2019hkj} for proton-proton collisions.  
	
	The non-eikonal double inclusive spectrum given in Eq.~\eqref{eq6:double_gluon_production} resembles   its eikonal limit in many features.  First, the term proportional to $\big \langle \Omega_1 \Omega_2 \big \rangle_{p} \big \langle \Omega_3 \Omega_4 \big \rangle_{p} \big \langle \Lambda_1 \Lambda_2 \big \rangle_{T} \big \langle \Lambda_3 \Lambda_4 \big \rangle_{T}$ is the fully uncorrelated piece and it is the leading term in the $N_c$ counting. Bose Enhancement and HBT-like correlations are encoded in the other terms in Eq.~\eqref{eq6:double_gluon_production} and they are subleading in powers of $(N_c^2-1)$ and collision area (actually in $B_p Q_s^2$) just as in its eikonal limit. The main difference between the non-eikonal double inclusive spectrum given in Eq.~\eqref{eq6:double_gluon_production} and its eikonal limit is that the non-eikonal corrections break the accidental symmetry of the CGC. Note that Eq.~\eqref{eq6:double_gluon_production} is symmetric under the exchange of $({\underline k}_3 \to - {\underline k}_3)$ where ${\underline k}_3\equiv(k_3^+, {\bf k}_3)$\footnote{Two comments are in order here. First, in the notation that is adopted to write Eq.~\eqref{eq6:double_gluon_production} the momenta of the second gluon is represented by $k_3$. Second, the exchange of $({\underline k}_3 \to - {\underline k}_3)$ in Eq.~\eqref{sec6:2gluon_production} is equivalent to exchanging indices 3 and 4, which obviously leaves the spectrum invariant.}, but not under $\bf{k}_3 \to -\bf{k}_3$. Therefore, non-eikonal corrections introduce an asymmetry  in the spectrum with respect to the azimuthal angle.   
	
	
	In order to see this asymmetry explicitly, we compute the non-eikonal double inclusive gluon spectrum given in Eq.~\eqref{eq6:double_gluon_production} numerically. For that purpose, we need to expand the terms inside the parenthesis in Eq.~\eqref{eq6:double_gluon_production} and use Eqs.~\eqref{eq6:projectile_correlator} and \eqref{eq6:2point_target} as the definitions of the 2-point correlators for the projectile and the target averages respectively. Similar to the treatment in the single inclusive case, the infrared divergences that appear in the limit $\bf{q}_i \to 0$ in Eq.~\eqref{eq6:2point_target} are regulated by introducing a mass term $m^2$ in the denominators, i.e., we perform the change  $1/\bf{q}_i^2 \to 1/(\bf{q}_i^2+m^2)$, where the value of the regulator is fixed to $m = 0.4$ GeV. We also fix in the numerical evaluations  $N_c=3$, $A^{1/3}=6$, $B_p=6$ GeV$^{-2}$ and $Q_s^2 = 2$ GeV$^{2}$. 
	
\begin{figure}[h!]
		\centering
		\includegraphics[scale=0.8]{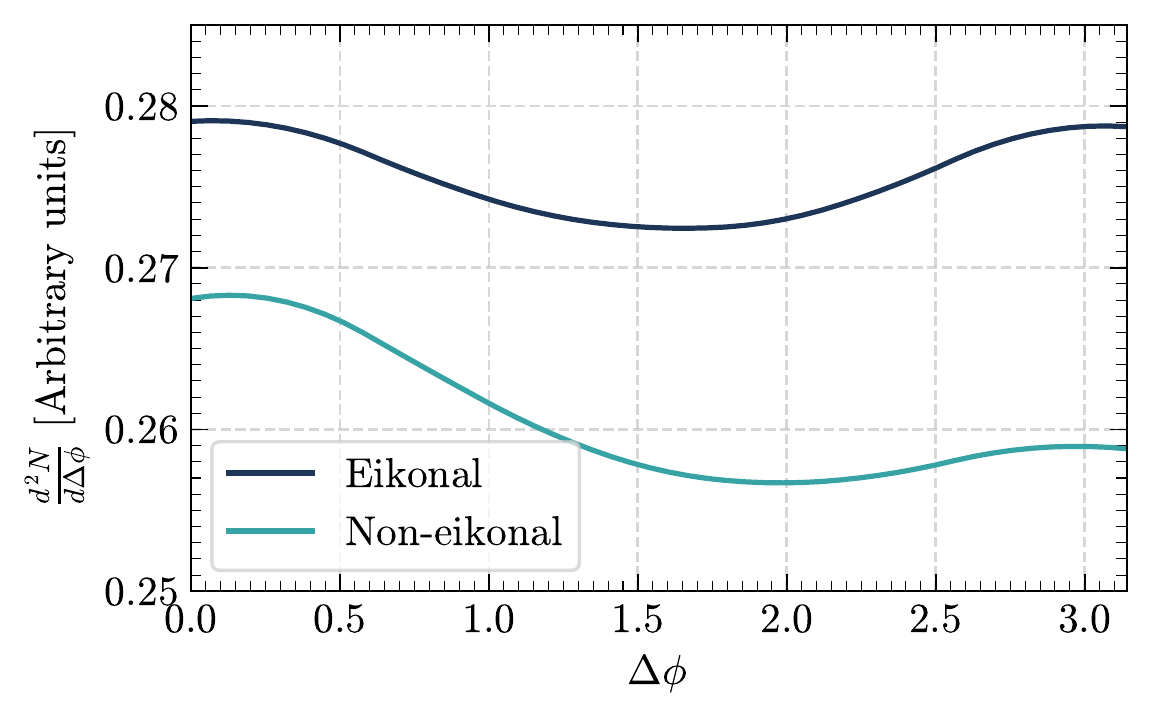}
		\caption{The double inclusive gluon spectrum given in Eq. \eqref{eq6:double_gluon_production} and its eikonal limit as a function of the azimuthal angle $\Delta \phi$. In this plot, we have fixed $\sqrt{s_{\rm NN}}= 100$ GeV, $\eta_1=\eta_3=0.2$ and $|\bf{k}_1|=|\bf{k}_3|=1$ GeV.}
		\label{fig6:azimuthal}
	\end{figure}

	
	The results are presented in Fig.~\ref{fig6:azimuthal}, where we plot the solution of Eq.~\eqref{eq6:double_gluon_production} and its respective eikonal limit as a function of the azimuthal angle $\Delta \phi = \arccos \frac{\bf{k}_1 \cdot \bf{k}_3}{|\bf{k}_1||\bf{k}_3|}$. In this plot, we used $\sqrt{s_{\rm NN}}= 100$ GeV, $\eta_1=\eta_3=0.2$ and $|\bf{k}_1|=|\bf{k}_3|=1$ GeV. As discussed above, a clear asymmetry between the near- and away-side peaks appears in the non-eikonal double gluon spectrum. This observation confirms the breaking of the accidental symmetry of the CGC by introducing the non-eikonal corrections that here we check in the dilute-dense ($pA$) situation. Moreover, Fig.~\ref{fig6:azimuthal} shows that the non-eikonal spectrum differs from the eikonal one by $4 \%$ in the near-side and $8 \%$ in the away-side peak. In this kinematics, this is a {\color{blue} slightly larger }correction than the one in the single inclusive spectrum discussed in Section~\ref{sec6:single_numerics}.

	These results clearly indicate that non-eikonal corrections provide non-vanishing odd harmonics since the accidental symmetry of the CGC is broken by their inclusion. We leave a detailed analysis of the azimuthal harmonics for a future work.

\section{Conclusions and Outlook}
\label{sec6:conclusions}
	
	In this work we generalized to proton-nucleus collisions the framework proposed in Refs.~\cite{Agostini:2019avp,Agostini:2019hkj} for proton-proton collisions. We computed multigluon spectra including the non-eikonal corrections that stem from the finite longitudinal width of the target. We use the Area Enhancement model that allows for an expansion of multipoles in terms of dipoles by neglecting contributions suppressed by powers of the overlap area of projectile and target. The formalism adopted in this work is inspired by the one used in jet quenching studies~\cite{Casalderrey-Solana:2007knd,Mehtar-Tani:2013pia,Blaizot:2015lma} where a non-zero longitudinal extent of the dense medium is essential to study parton propagation in the Quark Gluon Plasma. We adopt and extend such formalism to compute particle production and correlations in the Color Glass Condensate for the first time, where
the effect of considering a finite medium width is small as compared to the jet quenching
case where this is the main contribution to emission processes.
	
	
	We have shown that the non-eikonal corrections in the target 2-point correlators can be computed analytically using the GBW approximation that assumes that the eikonal dipole function has a Gaussian form. This approximation is justified as long as the dipole size is  $\ll \Lambda_{\rm QCD}^{-1}$. The results for the non-eikonal dipole functions are given in Eqs.~\eqref{eq6:gbw_local}, \eqref{eq6:gu_function} and \eqref{eq6:gg_function}, where the finite width target effects are accounted for, encoded in  the dimensionless non-eikonal parameter $\epsilon^2=\frac{Q_s^2 L^+}{2 i k^+}$. Using these results, we computed the single inclusive gluon spectrum beyond eikonal accuracy. The results of the single inclusive gluon production are summarized in Figs.~\ref{fig6:NE_vs_pt}, \ref{fig6:NE_vs_snn} and \ref{fig6:NE_vs_eta}. They show that including the non-eikonal corrections to the single inclusive gluon spectrum results in between $2$ \% to $10$ \% effect for center-of-mass energies per nucleon smaller than 100 GeV and for pseudorapidities of the produced particle smaller than 1. Thus, we conclude that  the non-eikonal corrections are not sizable for phenomenological studies in single inclusive gluon production at top RHIC or LHC energies. In Section~\ref{sec6:comparison_altinoluk}, we compare the next-to-next-to-eikonal approximation of our results, i.e., an expansion up to order $\epsilon^4$, with the one presented in~\cite{Altinoluk:2014oxa,Altinoluk:2015gia}. We obtained a parameterization of the so-called decorated dipoles within the Gaussian ansatz that we have used here.
	
	
	In Section~\ref{sec6:ne_multiparticle_production} we generalize the framework introduced in~\cite{Agostini:2021xca} to compute multigluon production, to the non-eikonal case where the corrections to the eikonal approximation stem from the finite longitudinal with of the target. Like in its eikonal limit, our framework is based on the Area Enhancement model that we argue should be valid for a target with finite longitudinal size since  the coherence length of the chromo-electric domains is much larger than the longitudinal extent of the target. We then focused on double inclusive gluon production and performed its numerical analysis (see Fig.~\ref{fig6:azimuthal}). The results suggest two important outcomes. First, for not very high energies and rapidities, the non-eikonal corrections to the double inclusive gluon spectrum are slightly larger than those to the single inclusive case. Second, the non-eikonal corrections induce an azimuthal asymmetry in the double inclusive spectrum which demonstrates that the inclusion of such corrections breaks the accidental symmetry of the CGC, as in $pp$~\cite{Agostini:2019avp,Agostini:2019hkj}. Therefore, we expect to obtain non-vanishing odd harmonics from the non-eikonal double inclusive gluon spectrum in $pA$ collisions. A detailed study of the effects of non-eikonal corrections to both even and odd azimuthal harmonics demands a dedicated numerical effort that we leave for the future. 
	

\section*{Acknowledgements}

PA, NA and FD have received financial support from Xunta de Galicia (Centro singular de investigaci\'on de Galicia accreditation 2019-2022), by European Union ERDF, by  the ``Mar\'{\i}a  de Maeztu" Units  of  Excellence program  MDM-2016-0692, and by the Spanish Research State Agency under project PID2020-119632GB-I00.
JGM acknowledges financial support from Funda\c c$\tilde{\mathrm{a}}$o para a Ci$\hat{\mathrm{e}}$ncia e a Tecnologia (Portugal) under projects CERN/FIS- PAR/0024/2019 and CERN/FIS-PAR/0032/2021, and he gratefully acknowledges the hospitality of the CERN theory group.
This work has been performed in the framework of the European Research Council project ERC-2018-ADG-835105 YoctoLHC and the MSCA RISE 823947 ``Heavy ion collisions: collectivity and precision in saturation physics''  (HI\-EIC), and has received funding from the European Un\-ion's Horizon 2020 research and innovation programme under grant agreement No. 824093. 	

\appendix

\section{Harmonic oscillator path integrals}
\label{app6:path_integrals}
	
	In this section we compute solve the path integrals that appear in Section~\ref{sec6:NE_target_averaging}. Before starting our discussion we define the following matrix:
	\begin{align}\label{eq6:matrix_fn}
		[F_{N}(x)]_{ij}= \begin{cases}
			x & \text{if $i=j$},\\
			-1 & \text{if $|i-j|=1$},\\
			0 & \text{otherwise},
		\end{cases}
	\end{align}
	with $x$ a positive real scalar and $i,j=1,\dots, N$. Therefore $F_N$ is a $N \times N$ tridiagonal matrix with $x$ in the main diagonal and $-1$ in the neighbour diagonals. Since this matrix will be a key ingredient in the calculation of this section we will study some of its properties.
	
	By performing the Laplace expansion of this matrix it is easy to see that it follows the following recursive equation:
	\begin{align}
		\det F_N(x)=x \det F_{N-1}(x)-\det F_{N-2}(x).
	\end{align}
	The solution of this equation is
	\begin{align}\label{eq6:matrix_determinant}
		\det F_N(x)=\frac{2^{-(N+1)}}{\sqrt{x^2-4}} \left(  x_+^{N+1}-x_{-}^{N+1} \right),
	\end{align}
	where $x_{\pm}=x \pm \sqrt{x^2-4}$.
	
	The inverse of $F_N(x)$ can be computed by using the expression of the inverse of a block matrix:
	\begin{align}\label{eq6:block_matrix_inv}
		\begin{pmatrix}
			\mathbf {A} &\mathbf {B} \\\mathbf {C} &\mathbf {D}
		\end{pmatrix}^{-1}
		&= 
		\begin{pmatrix}
			\left(\mathbf {A} -\mathbf {BD} ^{-1}\mathbf {C} \right)^{-1}&-\left(\mathbf {A} -\mathbf {BD} ^{-1}\mathbf {C} \right)^{-1}\mathbf {BD} ^{-1}\\-\mathbf {D} ^{-1}\mathbf {C} \left(\mathbf {A} -\mathbf {BD} ^{-1}\mathbf {C} \right)^{-1}&\quad \mathbf {D} ^{-1}+\mathbf {D} ^{-1}\mathbf {C} \left(\mathbf {A} -\mathbf {BD} ^{-1}\mathbf {C} \right)^{-1}\mathbf {BD} ^{-1}
		\end{pmatrix}
		\nonumber \\
		& \hskip-1cm = 
		\begin{pmatrix}
			\mathbf {A} ^{-1}+\mathbf {A} ^{-1}\mathbf {B} \left(\mathbf {D} -\mathbf {CA} ^{-1}\mathbf {B} \right)^{-1}\mathbf {CA} ^{-1}&-\mathbf {A} ^{-1}\mathbf {B} \left(\mathbf {D} -\mathbf {CA} ^{-1}\mathbf {B} \right)^{-1}\\-\left(\mathbf {D} -\mathbf {CA} ^{-1}\mathbf {B} \right)^{-1}\mathbf {CA} ^{-1}&\left(\mathbf {D} -\mathbf {CA} ^{-1}\mathbf {B} \right)^{-1}
		\end{pmatrix}.
	\end{align}
	Noting that $F_N(x)$ can be written as a block matrix with $\bf{A}=x$, $\bf{D}=F_{N-1}(x)$, $\bf{C}_i=-\delta_{i, 2}$ and $\bf{B}_j=-\delta_{j, 2}$, we can compute the inverse of its component $(1,1)$ by using the first equality of \eqref{eq6:block_matrix_inv}:
	\begin{align}\label{eq6:aux_matrix}
		[F^{-1}_N(x)]_{1,1} = \Big( x -[F^{-1}_{N-1}(x)]_{1,1} \Big)^{-1}.
	\end{align}
	Analogously, we could use the second equality of \eqref{eq6:block_matrix_inv} with $\bf{A}=F_{N-1}(x)$, $\bf{D}=x$, $\bf{C}_j=-\delta_{j, N-1}$ and $\bf{B}_i=-\delta_{i, N-1}$ to obtain the same recursive equation for $[F_N(x)^{-1}]_{N,N}$. Solving \eqref{eq6:aux_matrix} we obtain
	\begin{align}\label{eq6:matrix_inverse_nn}
		[F_N^{-1}(x)]_{1,1}=[F_N^{-1}(x)]_{N,N}=\frac{2^{-N}}{\sqrt{x^2-4} \det F_N(x)}\left( x_+^{N}-x_{-}^{N} \right).
	\end{align}
	
	Inspecting \eqref{eq6:block_matrix_inv}, we realize that the rest of the components of $F_N^{-1}(x)$ can be computed just as a function of its $(1,1)$ and $(N,N)$ components. In this manuscript we will only need the $(1,N)$ and $(N,1)$ components of this matrix which can be written
		\begin{align}\label{eq6:matrix_inverse_1n}
		[F_N^{-1}(x)]_{1,N}=[F_N^{-1}(x)]_{N,1}=\frac{1}{\det F_N(x)}\ .
	\end{align}
	
	We now proceed to solve the path integrals\footnote{While in the standard case the exponent is written as $i$ times the action, in our case we absorb this $i$ in the definition of the corresponding parameters, masses and couplings, in order to simplify the notation. On the other hand, we assume convergence of the integrals which can be checked in the final results, but a more rigorous treatment would require a Wick rotation.}. The first path integral that we encounter in Section~\ref{sec6:NE_target_averaging} is the one of a single harmonic oscillator of mass $m$ and coupling $\kappa$:
	\begin{align}
		\int_{\textbf{y}}^\textbf{x} \left[\mathcal{D} \textbf{z}\right] {\rm exp} \left\{ \int_{y^+}^{x^+}dz^+ \left[ m \dot{\textbf{z}}^2-\kappa \textbf{z}^2 \right]  \right\},
	\end{align}
	where, from \eqref{eq6:gu_harmonic_oscillator}, $m=\frac{i k^+}{2}$ and $\kappa =\frac{Q_s^2}{4 L^+}$. By discretizing the longitudinal direction into $N$ slices we can write this equation as
	\begin{align}\label{eq6:path_aux2}
		&\lim_{N \rightarrow \infty} \int \left( \prod_{n=1}^{N-1} d^2 \textbf{z}_n \right) \left( \frac{- m N}{\pi \Delta^+} \right)^N
		{\rm exp} \left\{ \sum_{n=1}^{N} \left[\frac{m N}{\Delta^+} (\textbf{z}_n-\textbf{z}_{n-1})^2-\frac{\kappa \Delta^+}{N} \textbf{z}_n^2 \right]\right\}
		\nonumber \\ & \hskip0cm =
		\lim_{N \rightarrow \infty} \int \left( \prod_{n=1}^{N-1} d^2 \textbf{z}_n \right) \left( \frac{a}{2 \pi} \right)^N
		{\rm exp} \left\{ \sum_{n=1}^{N} \left[-\frac{a}{2} (\textbf{z}_n-\textbf{z}_{n-1})^2-b \textbf{z}_n^2 \right]\right\},
	\end{align}
	where we have defined $\Delta^+=x^+-y^+$ and
	\begin{align}
		a= -\frac{2mN}{\Delta^+}\ , \qquad b= \frac{\kappa \Delta^+}{N}\ .
	\end{align}
	
	In order to solve this integral it is convenient to write it in a matrix form. In order to do so we define the vector $\vec{x}=(\textbf{z}_1,\dots,\textbf{z}_{N-1})$ and we can make the following simplifications:
	\begin{align}
		\sum_{n=1}^{N} (\textbf{z}_n-\textbf{z}_{n-1})^2=\textbf{z}_0^2+\textbf{z}_N^2 -2 \textbf{z}_0 \cdot \textbf{z}_1-2\textbf{z}_{N-1} \cdot \textbf{z}_N + \vec{x}^T F_{N-1}(2) \vec{x}
	\end{align}
	and
	\begin{align}
		\sum_{n=1}^{N} \textbf{z}_n^2=\textbf{z}_N^2+\vec{x}^T \mathbb{I} \vec{x},
	\end{align}
	where the matrix $F_N(x)$ is defined in \eqref{eq6:matrix_fn}.
	
	Thus, by introducing the vector $\vec{x}$, we can write \eqref{eq6:path_aux2} as a multidimensional Gaussian integral:
	\begin{align}\label{eq6:path_aux3}
		\lim_{N \rightarrow \infty} \int  d^{2(N-1)} \vec{x} \left( \frac{a}{2 \pi} \right)^N
		{\rm exp} \left\{ -\frac{a}{2} (\textbf{z}_0^2+\textbf{z}_N^2)-\frac{a}{2} \vec{x}^T F_{N-1}\left(2+\frac{2b}{a}\right) \vec{x}+\vec{J} \cdot \vec{x} \right\},
	\end{align}
	where we have defined the vector
	\begin{align}
		J^i=a \textbf{z}_0 \delta^{1,i}+a \textbf{z}_N \delta^{N-1,i}.
	\end{align}
	
	Since the solution of a multidimensional Gaussian integral is
	\begin{align}\label{eq6:multi_gauss_int}
		\int d^Dx \, \exp{-\frac{a}{2} x^T M x + B \cdot x} = \frac{(2\pi)^{D/2}}{\det M} \exp{ \frac{1}{2} B^T M^{-1} B},
	\end{align}
	where $M$ is a constant matrix $D \times D$ and $B$ is a constant $D$-dimensional vector, we can write \eqref{eq6:path_aux3} as
	\begin{align}
		\hskip-0.6cm
		\lim_{N \rightarrow \infty} \left( \frac{a}{2 \pi} \right)^N \frac{(2\pi)^{N-1}}{a^{N-1} \det F_{N-1}\left(2+\frac{2b}{a}\right)}
		{\rm exp} \left\{ -\frac{a}{2} (\textbf{z}_0^2+\textbf{z}_N^2)+\frac{1}{2 a} \vec{J}^T F_{N-1}^{-1}\left(2+\frac{2b}{a}\right) \vec{J} \right\}.
	\end{align}
	
	Using \eqref{eq6:matrix_determinant} and taking the $N\to \infty$ limit we obtain 
	\begin{align}
		\lim_{N \rightarrow \infty} \left( \frac{a}{2 \pi} \right)^N \frac{(2\pi)^{N-1}}{a^{N-1} \det F_{N-1}\left(2+\frac{2b}{a}\right)}
		= \frac{-m \omega}{\pi} \frac{1}{\sin \omega \Delta^+}\ ,
	\end{align}
	where we define
	\begin{align}
		\omega^2 =- \frac{2b}{a} \frac{N^2}{(\Delta^+)^2} = \frac{\kappa}{m}
	\end{align}
	and we have used the fact that
	\begin{align}\label{eq6:determinant_limit}
		\lim_{N\to\infty} \frac{N}{\det F_{N-1}\left( 2-\frac{x^2}{N^2} \right)} = \frac{x}{\sin x}\ .
	\end{align}
	
	On the other hand, using \eqref{eq6:matrix_inverse_nn} and \eqref{eq6:matrix_inverse_1n} and taking the $N\to\infty$ limit we get
	\begin{align}
		&\lim_{N \rightarrow \infty} \left[-\frac{a}{2} (\textbf{z}_0^2+\textbf{z}_N^2)+\frac{1}{2 a} \vec{J}^T F_{N-1}^{-1}\left(2+\frac{2b}{a}\right) \vec{J}\right]
		\nonumber \\ & \hskip0cm =
		\lim_{N \rightarrow \infty}\left[-\frac{a}{2} (\textbf{z}_0^2+\textbf{z}_N^2)\left(1- \left[F_{N-1}^{-1}\left(2+\frac{2b}{a}\right)\right]_{1,1} \right)+a \textbf{z}_0 \textbf{z}_N \left[F_{N-1}^{-1}\left(2+\frac{2b}{a}\right)\right]_{1,N-1}\right]
		\nonumber \\ & \hskip0cm =
		m \omega \left[\frac{\textbf{z}_0^2+\textbf{z}_N^2}{\tan \omega \Delta^+}-2 \frac{\textbf{z}_0 \textbf{z}_N}{\sin \omega \Delta^+}\right],
	\end{align}
	where we have used that
	\begin{align}\label{eq6:inverse_limit}
		\lim_{N\to\infty} N \left(1-\left[F_{N-1}\left( 2-\frac{x^2}{N^2} \right)\right]_{1,1}\right) = \frac{x}{\tan x}\ .
	\end{align}

	Therefore we obtain the well known solution of the harmonic oscillator path integral~\cite{Grosche:1998yu}:
	\begin{align}\label{eq6:path_int_1ho_sol}
		&\int_{\textbf{y}}^\textbf{x} \left[\mathcal{D} \textbf{z}\right] {\rm exp} \left\{ \int_{y^+}^{x^+}dz^+ \left[ m \dot{\textbf{z}}^2-\kappa \textbf{z}^2 \right]  \right\}
		\nonumber \\ & \hskip4cm =
		\frac{-m \omega}{\pi} \frac{1}{\sin \omega \Delta^+}
		{\rm exp} \left\{ m \omega \left[\frac{\textbf{z}_0^2+\textbf{z}_N^2}{\tan \omega \Delta^+}-2 \frac{\textbf{z}_0 \textbf{z}_N}{\sin \omega \Delta^+}\right] \right\}.
	\end{align}
	
	The second path integral that appears in Section~\ref{sec6:NE_target_averaging} is the one analogous to the one for two coupled harmonic oscillators of masses $m_1$ and $-m_2$ and coupling $\kappa$: 
	\begin{align}
		\int_{\textbf{y}}^\textbf{x} \left[\mathcal{D} \textbf{z}_1\right] \left[\mathcal{D} \textbf{z}_2\right] {\rm exp} \left\{ \int_{y^+}^{x^+}dz^+ \left[ m_1 \dot{\textbf{z}}_1^2-m_2 \dot{\textbf{z}}_2^2-\kappa (\textbf{z}_1-\textbf{z}_2)^2 \right]  \right\}.
	\end{align}
	
	In order to solve this integral it is convenient to make the change of variables\footnote{Note that a different definition $\textbf{B}=\frac{m_1 \textbf{z}_1-m_2 \textbf{z}_2}{m_1-m_2}$ would decouple the oscillators but then the limit $m_1\to m_2$ would become ill-defined. This is the reason why we choose the one in Eq.~\eqref{eq:var2osc}.}
	\begin{align}
	\label{eq:var2osc}
		\textbf{B}=\frac{m_1 \textbf{z}_1+m_2 \textbf{z}_2}{m_1+m_2},
		\qquad
		\textbf{R}=\textbf{z}_1-\textbf{z}_2,
	\end{align}
	in such a way that the integral can be rewritten as
	\begin{align}
		\int_{\textbf{y}}^\textbf{x} \left[\mathcal{D} \textbf{B}\right] \left[\mathcal{D} \textbf{R}\right] {\rm exp} \left\{ \int_{y^+}^{x^+}dz^+ \left[ \Delta m \bf{\dot{B}}^2-\frac{\Delta m \mu}{M} \bf{\dot{R}}^2+4 \mu \bf{\dot{B}} \cdot \bf{\dot{R}}-\kappa \textbf{R}^2 \right]  \right\},
	\end{align}
	where we have introduced
	\begin{align}
		\Delta m= m_1-m_2, \quad \mu=\frac{m_1 m_2 }{m_1+m_2} \quad {\rm and} \quad M=m_1+m_2.
	\end{align}
	
	In the discrete limit we can write the path integral as
	\begin{align}
		&\lim_{N \rightarrow \infty} 
		\int \left( \prod_{n=1}^{N-1} d^2 \textbf{r}_n d^2 \textbf{b}_n \right) 
		\left( \frac{-m_1 N}{\pi \Delta^+} \right)^N \left( \frac{m_2  N}{\pi \Delta^+} \right)^N
		 \\ & \hskip0cm \times 
		{\rm exp} \left\{ \sum_{n=1}^{N} \left[-\frac{a_1}{2} (\textbf{r}_n-\textbf{r}_{n-1})^2-\frac{a_2}{2} (\textbf{b}_n-\textbf{b}_{n-1})^2-\frac{a_3}{2} (\textbf{r}_n-\textbf{r}_{n-1}) \cdot (\textbf{b}_n-\textbf{b}_{n-1})-b \textbf{r}_n^2 \right]\right\},\nonumber
	\end{align}
	with
	\begin{align}
		a_1=\frac{2 \Delta m \mu N}{M \Delta^+}\ , \quad 
		a_2=-\frac{2 \Delta m N}{\Delta^+}
		\quad
		{\rm and}
		\quad
		a_3=-\frac{8 \mu N}{\Delta^+}\ .
	\end{align}
	
	The next step is to write, again, the path integral in a matrix form, i.e., as a multidimensional Gaussian integral. We therefore define $\vec{x}_{\bf{r}}=(\textbf{r}_1,\dots,\textbf{r}_{N-1})$ and $\vec{x}_{\bf{b}}=(\textbf{b}_1,\dots,\textbf{b}_{N-1})$ in order to make the following simplification:
	\begin{align}
		&\sum_{n=1}^{N} \left[-\frac{a_1}{2} (\textbf{r}_n-\textbf{r}_{n-1})^2-\frac{a_2}{2} (\textbf{b}_n-\textbf{b}_{n-1})^2-\frac{a_3}{2} (\textbf{r}_n-\textbf{r}_{n-1})(\textbf{b}_n-\textbf{b}_{n-1})-b \textbf{r}_n^2 \right]
		\nonumber \\ & \hskip0cm 
		=-\frac{a_1}{2} \vec{x}_{\bf{r}}^T \left(F_{N-1}(2) 
		+ \frac{2b}{a_1} \mathbb{I}\right) \vec{x}_{\bf{r}} 
		- \frac{a_2}{2} \textbf{x}_b^T F_{N-1}(2) \vec{x}_{\bf{b}} 
		\nonumber \\ & \hskip6cm 
		-\frac{a_3}{2} \left(\frac{1}{2} \vec{x}_{\bf{r}}^T F_{N-1}(2) \vec{x}_{\bf{b}} +\frac{1}{2} \vec{x}_{\bf{b}}^T F_{N-1}(2) \vec{x}_{\bf{r}} \right)
		\nonumber \\ & \hskip0cm 
		-\frac{a_1}{2} \left[ \textbf{r}_0^2+\textbf{r}_N^2-2\textbf{r}_0 \textbf{r}_1-2\textbf{r}_{N-1}\textbf{r}_N \right]-\frac{a_2}{2} \left[ \textbf{b}_0^2+\textbf{b}_N^2-2\textbf{b}_0 \textbf{b}_1-2\textbf{b}_{N-1}\textbf{b}_N \right]
		\nonumber \\ & \hskip0cm 
		-\frac{a_3}{2} \left[ \textbf{r}_0 \textbf{b}_0+\textbf{r}_N \textbf{b}_N-\textbf{r}_0 \textbf{b}_1- \textbf{r}_1 \textbf{b}_0-\textbf{r}_{N-1}\textbf{b}_N -\textbf{b}_{N-1} \textbf{r}_N \right] - b \textbf{r}_N^2.
	\end{align}
	
	Then the path integral can be simplified further by defining the matrix
	\begin{align}
		\mathcal{A}_{2(N-1)}=\begin{pmatrix}
			a_1 F_{N-1}\left(2+\frac{2b}{a_1}\right) & \frac{a_3}{2} F_{N-1}(2) \\
			\frac{a_3}{2} F_{N-1}(2) & a_2 F_{N-1}(2) 
		\end{pmatrix},
	\end{align}
	the vector
	\begin{align}
		J^i=&\left( a_1 \textbf{r}_0+\frac{a_3}{2} \textbf{b}_0 \right) \delta^{i,1}
		+ \left( a_1 \textbf{r}_N+\frac{a_3}{2} \textbf{b}_N \right) \delta^{i,N-1}
		\nonumber \\
		+ &\left( a_2 \textbf{b}_0+\frac{a_3}{2} \textbf{r}_0 \right) \delta^{i,N}
		+ \left( a_2 \textbf{b}_N+\frac{a_3}{2} \textbf{r}_N \right) \delta^{i,2(N-1)},
	\end{align}
	and $\vec{x}=(\textbf{r}_1,\dots,\textbf{r}_N,\textbf{b}_0,\dots,\textbf{b}_N)$. Thus we can write it as
	\begin{align}\label{eq6:path_int_aux3}
		&\lim_{N \rightarrow \infty} 
		\int d^{4(N-1)}\vec{x} 
		\left( \frac{-m_1 N}{\pi \Delta^+} \right)^N \left( \frac{m_2  N}{\pi \Delta^+} \right)^N
		{\rm exp} \left\{ -\frac{1}{2} \vec{x}^T \mathcal{A}_{2(N-1)} \vec{x} + \vec{J} \cdot \vec{x}  \right\}
		\nonumber \\ & \hskip2cm \times
		{\rm exp} \left\{ -\frac{a_1}{2} [\textbf{r}_0^2+\textbf{r}_N^2] -\frac{a_2}{2} [\textbf{b}_0^2+\textbf{b}_N^2]
		-\frac{a_3}{2} [\textbf{r}_0 \textbf{b}_0+\textbf{r}_N \textbf{b}_N] - b \textbf{r}_N^2 \right\}
		\nonumber \\ & \hskip0cm 
		=\lim_{N \rightarrow \infty} 
		\left( \frac{-m_1 N}{\pi \Delta^+} \right)^N \left( \frac{m_2  N}{\pi \Delta^+} \right)^N \frac{(2\pi)^{2(N-1)}}{\det \mathcal{A}_{2(N-1)}}
		{\rm exp} \left\{ \frac{1}{2} \vec{J}^T \mathcal{A}_{2(N-1)}^{-1} \vec{J}\right\}
		\nonumber \\ & \hskip2cm \times
		{\rm exp} \left\{ -\frac{a_1}{2} [\textbf{r}_0^2+\textbf{r}_N^2] -\frac{a_2}{2} [\textbf{b}_0^2+\textbf{b}_N^2]
		-\frac{a_3}{2} [\textbf{r}_0 \textbf{b}_0+\textbf{r}_N \textbf{b}_N] - b \textbf{r}_N^2 \right\},
	\end{align}
	where we have used \eqref{eq6:multi_gauss_int} in order to solve the integral over $\vec{x}$.
	
	The determinant of the matrix ${\cal A}$ can be evaluated by using the following property of block matrices:
	\begin{align}
		\det \begin{pmatrix}
			A & B \\
			C & D
		\end{pmatrix}
		= \det(A-B D^{-1} C) \det D.
	\end{align}
	By using this property we can write
	\begin{align}
		\det \mathcal{A}_{2(N-1)}=a_1^{N-1} a_2^{N-1} f^{N-1} \det F_{N-1}\left(2+\frac{2b}{ f a_1}\right) \det F_{N-1} \left(2\right),
	\end{align}
	where we have used the fact that $F_N(x)+b F_N(y) =(1+b) F_N\left(\frac{x+by}{1+b}\right)$ and defined
	\begin{align}
		f=1-\frac{a_3^2}{4 a_1 a_2} = \frac{M^2}{\Delta m^2}\ .
	\end{align}
	
	Therefore, using \eqref{eq6:determinant_limit} we can compute the limit $N\to\infty$ in the prefactor in \eqref{eq6:path_int_aux3} to read
		\begin{align}
		&\lim_{N \rightarrow \infty} 
		\left( \frac{-m_1 N}{\pi \Delta^+} \right)^N \left( \frac{m_2  N}{\pi \Delta^+} \right)^N \frac{(2\pi)^{2(N-1)}}{\det \mathcal{A}_{2(N-1)}} 
		\nonumber \\ & \hskip0cm  =
		- \frac{m_1 m_2}{\pi^2 (\Delta^+)^2}\lim_{N \rightarrow \infty} \frac{N}{\det F_{N-1} \left(2 + \frac{2b}{f a_1} \right)} \frac{N}{ \det F_{N-1} \left(2 \right)}
		=-\frac{- m_1 m_2}{\pi^2 {\Delta^+}} \frac{\omega_{-}}{\sin \omega_{-} \Delta^+}\ ,
	\end{align}
	where
	\begin{align}
		\omega_{-}^2=\frac{\kappa}{m_1}-\frac{\kappa}{m_2} =\omega_1^2 -\omega_2^2.
	\end{align}
	
	On the other hand, the limit $N\to\infty$ of the argument of the exponential in \eqref{eq6:path_int_aux3} can be evaluated by using \eqref{eq6:block_matrix_inv} and writing the inverse of the matrix ${\cal A}$ as
	\begin{align}
		\mathcal{A}_{2(N-1)}^{-1}=\begin{pmatrix}
			\frac{1}{f a_1} F_{N-1}\left(2+\frac{2 b}{f a_1}\right)^{-1}& \frac{-a_3}{2f a_1 a_2} F_{N-1}\left(2+\frac{2 b}{f a_1}\right)^{-1} \\
			\frac{-a_3}{2f a_1 a_2} F_{N-1}\left(2+\frac{2 b}{f a_1}\right)^{-1} & \frac{1}{a_2} F_{N-1}\left(2\right)^{-1}+\frac{a_3^2}{4f a_1 a_2^2} F_{N-1}\left(2+\frac{2 b}{f a_1}\right)^{-1}
		\end{pmatrix}.
	\end{align}
	Therefore we can write
	\begin{align}
		&\frac{1}{2} \vec{J}^T \mathcal{A}_{2(N-1)}^{-1} \vec{J} -\frac{a_1}{2} [\textbf{r}_0^2+\textbf{r}_N^2] -\frac{a_2}{2} [\textbf{b}_0^2+\textbf{b}_N^2]
		-\frac{a_3}{2} [\textbf{r}_0 \textbf{b}_0+\textbf{r}_N \textbf{b}_N] - b \textbf{r}_N^2
		\nonumber \\ & \hskip0cm  =
		(\bf{r}_0^2+\bf{r}_N^2)\left[ -\frac{a_1}{2} \left( 1-A_1 \right) +\frac{a_3^2}{8 a_2} \left[ \left(1-A_1\right) - \left(1-A_2\right) \right]\right]
		\nonumber \\ & \hskip0cm
		-\frac{a_2}{2}(\bf{b}_0^2+\bf{b}_N^2) \left( 1-A_2 \right) -\frac{a_3}{2} (\bf{r}_0\cdot \bf{b}_0+\bf{r}_N\cdot \bf{b}_N) \left(1-A_2\right)
		 \\ & \hskip0cm
		+\bf{r}_0 \cdot \bf{r}_N\left[ a_1 B_1 +\frac{a_3^2}{4 a_2} \left( B_2 -B_1 \right) \right]
		+ \frac{a_3}{2} B_2 (\bf{r}_0 \cdot \bf{b}_N+\bf{r}_N \cdot \bf{b}_0) +a_2 B_2 \bf{b}_0 \cdot \bf{b}_N - b \textbf{r}_N^2,\nonumber
	\end{align}
	where we have introduced for simplicity 
	\begin{align}
		A_1 &\equiv \left[F_{N-1}\left(2+\frac{2 b}{f a_1}\right)^{-1}\right]_{1,N-1},
		\qquad
		A_2 \equiv \left[F_{N-1}\left(2\right)^{-1}\right]_{1,N-1},
		\\
		B_1 &\equiv \left[F_{N-1}\left(2+\frac{2 b}{f a_1}\right)^{-1}\right]_{1,1},
		\qquad
		B_2 \equiv \left[F_{N-1}\left(2\right)^{-1}\right]_{1,1}.
	\end{align}
	
	Finally,  using~\eqref{eq6:determinant_limit} and \eqref{eq6:inverse_limit} and the definitions of the coefficients $a_i$ it is straightforward to see that
	\begin{align}
		&\lim_{N \rightarrow \infty} \left[ \frac{1}{2} \textbf{J}^T \mathcal{A}_{2(N-1)}^{-1} \textbf{J} -\frac{a_1}{2} [\textbf{r}_0^2+\textbf{r}_N^2] -\frac{a_2}{2} [\textbf{b}_0^2+\textbf{b}_N^2]
		-\frac{a_3}{2} [\textbf{r}_0 \textbf{b}_0+\textbf{r}_N \textbf{b}_N] - b \textbf{r}_N^2 \right]
		\nonumber \\
		&=(\textbf{r}_0^2+\textbf{r}_N^2) \frac{\mu}{\Delta^+}\left[ \frac{4 \mu}{\Delta m}-\frac{M \omega_{-} \Delta^+}{\Delta m} \frac{1}{\tan \omega_{-} \Delta^+} \right] - 2 \textbf{r}_0 \textbf{r}_N \frac{\mu}{\Delta^+}\left[ \frac{4 \mu}{\Delta m}-\frac{M \omega_{-} \Delta^+}{\Delta m} \frac{1}{\sin \omega_{-} \Delta^+} \right]
		\nonumber \\ & \hskip0cm
		+\frac{\Delta m}{\Delta^+}(\textbf{b}_0-\textbf{b}_N)^2 +\frac{4 \mu}{\Delta^+}(\textbf{r}_0-\textbf{r}_N)(\textbf{b}_0-\textbf{b}_N).
	\end{align}
	
	Thus we can write the solution of the double harmonic oscillator as
	\begin{align}
		&\int_{\textbf{y}}^\textbf{x} \mathcal{D} \textbf{z}_1 \mathcal{D} \textbf{z}_2 {\rm exp} \left\{ \int_{y^+}^{x^+}dz^+ \left[ m_1 \dot{\textbf{z}}_1^2-m_2 \dot{\textbf{z}}_2^2-\kappa (\textbf{z}_1-\textbf{z}_2)^2 \right]  \right\}
		 \\ & \hskip0cm
		=-\frac{- m_1 m_2 \omega_{-}}{\pi^2 {\Delta^+}} \frac{1}{\sin \omega_{-} \Delta^+} {\rm exp}\Bigg\{ (\textbf{r}_0^2+\textbf{r}_N^2) \frac{\mu}{\Delta^+}\left[ \frac{4 \mu}{\Delta m}-\frac{M \omega_{-} \Delta^+}{\Delta m} \frac{1}{\tan \omega_{-} \Delta^+} \right] 
		\nonumber \\ & \hskip-1cm
		- 2 \textbf{r}_0 \textbf{r}_N \frac{\mu}{\Delta^+}\left[ \frac{4 \mu}{\Delta m}-\frac{M \omega_{-} \Delta^+}{\Delta m} \frac{1}{\sin \omega_{-} \Delta^+} \right]
		+\frac{\Delta m}{\Delta^+}(\textbf{b}_0-\textbf{b}_N)^2 +\frac{4 \mu}{\Delta^+}(\textbf{r}_0-\textbf{r}_N)(\textbf{b}_0-\textbf{b}_N) \Bigg\} \nonumber
	\end{align}
	or, equivalently,
	\begin{align}\label{eq6:path_int_2ho_sol}
		&\int_{\textbf{y}}^\textbf{x} \mathcal{D} \textbf{z}_1 \mathcal{D} \textbf{z}_2 {\rm exp} \left\{ \int_{y^+}^{x^+}dz^+ \left[ \frac{\kappa}{\omega_1^2} \dot{\textbf{z}}_1^2-\frac{\kappa}{\omega_2^2} \dot{\textbf{z}}_2^2-\kappa (\textbf{z}_1-\textbf{z}_2)^2 \right]  \right\}
		\nonumber \\ & \hskip0cm
		=-\frac{\kappa^2 }{\pi^2 {\Delta^+} \omega_1^2 \omega_2^2} 
		\frac{\omega_{-}}{\sin \omega_{-} \Delta^+} {\rm exp}\Bigg\{ 
		\frac{\kappa}{\omega_{-}^2} \Bigg[
		(\textbf{r}_0^2+\textbf{r}_N^2)
		\frac{\omega_{-}}{\tan \omega_{-} \Delta^+}
		- 2 \bf{r}_0 \cdot \bf{r}_N \frac{\omega_{-}}{\sin \omega_{-} \Delta^+}
		\nonumber \\ & \hskip4cm
		- \frac{1}{\Delta^+ \omega_1^2 \omega_2^2 \omega_+^4}
		\Bigg(
		2 \omega_1^2 \omega_2^2(\bf{r}_N-\bf{r}_0)+\omega_+^2 \omega_{-}^2 (\bf{b}_0-\bf{b}_N)
		\Bigg)^2
		\Bigg] 
		\Bigg\},
	\end{align}
	where $\omega_{\pm}^2=\omega_1^2\pm \omega_2^2$. Note that this equation provides a well-defined limit for $m_1\to m_2\equiv \omega_1\to \omega_2$.

\section{Derivation of the terms in the target 2-point function}\label{sec:2point}
\label{app:2}

In this section we derive the explicit expressions of the functions appearing in Eq.~\eqref{eq6:2point_target}. We start by using the definition of the reduced amplitude given in Eq.~\eqref{eq6:reduced_amplitude} and the definitions of the eikonal and non-eikonal dipole functions introduced in Eqs.~\eqref{eq6:dipole_gg}, \eqref{eq6:dipole_gu} and \eqref{eq6:dipole_uu}. Thus, the 2-point correlator given in Eq.~\eqref{def_simp_lambda} reads
\begin{align}\label{eq6:2point_target_aux}
	&\Big \langle \Lambda_\alpha \Lambda_\beta \Big \rangle_T
	=\frac{\delta^{a_\alpha a_\beta} \delta^{b_\alpha b_\beta}}{N_c^2-1} 
	\int_{\textbf{y}_\alpha, \textbf{y}_\beta, \textbf{x}_\alpha, \textbf{x}_\beta }
	e^{ i (-1)^{\alpha+1} (\textbf{q}_\alpha \cdot \textbf{y}_\alpha-\textbf{k}_\alpha \cdot \textbf{x}_\alpha) +i (-1)^{\beta+1} (\textbf{q}_\beta \cdot \textbf{y}_\beta-\textbf{k}_\beta \cdot \textbf{x}_\beta)}
	\nonumber \\ & \hskip 1cm \times
	\Bigg \{ 
	2\frac{\bf{k}_\alpha^{\lambda_\alpha}\bf{k}_\beta^{\lambda_\beta}}{\bf{k}_\alpha^{2}\bf{k}_\beta^{2}}
	\delta^{(2)}(\bf{x}_\alpha-\bf{y}_\alpha) \delta^{(2)}(\bf{x}_\beta-\bf{y}_\beta)
	d^{(0)}(L^+,0|\bf{y}_\alpha,\bf{y}_\beta)
	\nonumber \\ & \hskip 1cm
	-
	4\frac{\bf{k}_\alpha^{\lambda_\alpha}\bf{q}_\beta^{\lambda_\beta}}{\bf{k}_\alpha^{2}\bf{q}_\beta^{2}}
	\delta^{(2)}(\bf{x}_\alpha-\bf{y}_\alpha)
	d^{(1)}(L^+,0|\bf{y}_\alpha;\bf{x}_\beta,\bf{y}_\beta,(-1)^{\beta+1}k_\beta^+)
	\nonumber \\ & \hskip 1cm
	+
	2\frac{\bf{q}_\alpha^{\lambda_\alpha}\bf{q}_\beta^{\lambda_\beta}}{\bf{q}_\alpha^{2}\bf{q}_\beta^{2}}
	d^{(2)}(L^+,0|\bf{x}_\alpha,\bf{y}_\alpha,(-1)^{\alpha+1}k_\alpha^+;\bf{x}_\beta,\bf{y}_\beta,(-1)^\beta k_\beta^+)
	\nonumber \\ & \hskip 1cm
	+
	2\frac{L^+}{k_\beta^+} \frac{\bf{k}_\alpha^{\lambda_\alpha}}{\bf{k}_\alpha^{2}}
	\delta^{(2)}(\bf{x}_\alpha-\bf{y}_\alpha)
	\int_0^1 df_\beta 
	\Big[ \partial_{\bf{y}_\alpha^{\lambda_\alpha}} d^{(1)}(L^+, f_\beta L^+ | \bf{y}_\alpha;\bf{x}_\beta,\bf{y}_\beta,(-1)^{\beta+1}k_\beta^+) \Big]
	\nonumber \\ & \hskip 10cm \times
	d^{(0)}(f_\beta L^+,0|\bf{y}_\alpha,\bf{y}_\beta)
	\nonumber \\ & \hskip 0.5cm
	-
	2\frac{L^+}{k_\beta^+} \frac{\bf{q}_\alpha^{\lambda_\alpha}}{\bf{q}_\alpha^{2}}
	\int_0^1 df_\beta  \int_{\bf{u}}
	\Big[ \partial_{\bf{y}_\beta^{\lambda_\beta}} d^{(2)}(L^+,f_\beta L^+|\bf{x}_\alpha,\bf{u},(-1)^{\alpha+1}k_\alpha^+;\bf{x}_\beta,\bf{y}_\beta,(-1)^\beta k_\beta^+) \Big]
	\nonumber \\ & \hskip 8cm \times
	d^{(1)}(f_\beta L^+, 0 | \bf{y}_\beta;\bf{u},\bf{y}_\alpha,(-1)^{\alpha+1}k_\alpha^+)
	\nonumber \\ & \hskip 0.5cm
	+
	\frac{L^+}{k_\alpha^+} \frac{L^+}{k_\beta^+} 
	\int_0^1 df_\beta \int_0^{f_\beta} df_\alpha  \int_{\bf{u}}
	\Big[ \partial_{\bf{y}_\beta^{\lambda_\beta}} d^{(2)}(L^+,f_\beta L^+|\bf{x}_\alpha,\bf{u},(-1)^{\alpha+1} k_\alpha^+;\bf{x}_\beta,\bf{y}_\beta,(-1)^\beta k_\beta^+) \Big]
	\nonumber \\ & \hskip 3cm
	\times
	\Big[ \partial_{\bf{y}_\alpha^{\lambda_\alpha}} d^{(1)}(f_\beta L^+, f_\alpha L^+ | \bf{y}_\beta;\bf{u},\bf{y}_\alpha,(-1)^{\alpha+1}k_\alpha^+) \Big]
	d^{(0)}(f_\alpha L^+,0|\bf{y}_\alpha,\bf{y}_\beta)
	\nonumber \\ & \hskip 10cm
	+(\alpha \leftrightarrow \beta)
	\Bigg\},
\end{align}
where we have introduced the notation $f_\alpha=y_\alpha^+/L^+$ (and analogous expression for $f_\beta$) for convenience. Note that the factors $(-1)^\alpha$ account for whether the object  $\Lambda_\alpha$ is sitting on the right or on the left side of the cut. It is also straightforward to realize that after setting $\alpha=1$ and $\beta=2$, this equation reduces to the single inclusive case computed in Eq.~\eqref{eq6:target_average}.
%
%

In Eq.~\eqref{eq6:2point_target_aux}, integrals over the transverse coordinates can be performed using the definitions of the non-eikonal dipole functions given in Eqs.~\eqref{eq6:gbw_local}, \eqref{eq6:gu_function} and \eqref{eq6:gg_function}, and also the solution of the Gaussian-like integral given in Eq.~\eqref{eq6:gauss_int}. In order to shorten the expressions, we use the notation $\pm k_{a} \equiv (-1)^{a+1} k_a$ introduced in Section~\ref{sec6:ne_multiparticle_production}. Thus, upon performing the integrations over the transverse coordinates and introducing this notation we arrive at Eq.~\eqref{eq6:2point_target}.

The first term in Eq.~\eqref{eq6:2point_target} corresponds to the case where both gluons are emitted after the interaction of the source with the target and it reads
\begin{align}
	\label{eq6:multi_afaf}
	I_{\rm aft-aft} \big(\bf{k}_\beta,\bf{q}_\beta \big) = 
	\frac{4 \pi}{Q_s^2} \exp{- \frac{(\bf{k}_\beta-\bf{q}_\beta)^2}{Q_s^2}}.
\end{align}
Note that this term is independent of the sign of the transverse momenta since its argument does not carry the $\pm$ notation introduced in Eq.~\eqref{eq6:2point_target}. Moreover, due to the appearance of the $\delta$-function in the first line of Eq.~\eqref{eq6:2point_target}, this term is invariant under the change $\alpha \leftrightarrow \beta$.

The second contribution in Eq.~\eqref{eq6:2point_target} corresponds to the case where the gluon represented by the index $\alpha$, which we refer to as gluon $\alpha$, is emitted after the interaction of the source with the target and the gluon $\beta$ is emitted before. On the other hand, the inverse situation is given by the mirror term $\alpha \leftrightarrow \beta$. Upon the performing the integrations, this term can be written as 
\begin{align}
	\label{eq6:multi_afbe}
	I_{\rm aft-bef} \Big(\bf{k}_\beta,\bf{q}_\beta,k_\beta^+ \Big) = 
	\frac{4 \pi}{Q_s^2}
	\frac{\epsilon_\beta }{\sin \epsilon_\beta} 
	\exp{- \frac{\epsilon_\beta}{Q_s^2 \sin \epsilon_\beta}
		\Big[
		(\bf{k}_\beta^2 +\bf{q}_\beta^2) \cos \epsilon_\beta
		- 2 \bf{k}_\beta \cdot \bf{q}_\beta
		\Big]},
\end{align}
where the $k^+$ dependence is given implicitly in the definition of $\epsilon$ (see Eq. ~\eqref{eq6:epsilon}). This contribution carries dependence only on $k_\beta^+$ (and not on $k_\alpha^+$) due to the fact that only the gluon $\beta$ traverses the target. 

The third contribution in Eq.~\eqref{eq6:2point_target} corresponds to the case in which both gluons are emitted before the target and it can be written as
\begin{align}
	\label{eq6:multi_bebe}
	&I_{\rm bef-bef} \Big(\bf{k}_\alpha,\bf{q}_\alpha,\bf{k}_\beta,\bf{q}_\beta; k_\alpha^+, k_\beta^+ \Big) 
	=
	\frac{4 \pi}{Q_s^2}
	\frac{\epsilon_{+}}{\sin \epsilon_{+}}  
	\nonumber \\ & \hskip1cm \times
	\exp{-\frac{1}{Q_s^2}
		\left[
		\frac{\epsilon_\alpha^2 \epsilon_\beta^2}{\epsilon_{+}^2} (\bf{k}_\alpha+\bf{k}_\beta)^2
		+\frac{\epsilon_{+}}{\tan \epsilon_{+}} (\bf{K^2}+\bf{Q}^2) 
		- 2 \frac{\epsilon_{+}}{\sin \epsilon_{+}} \bf{K}\cdot \bf{Q} 
		\right]},
\end{align}
where we have introduced the following notations for convenience: 
\begin{align}
	\epsilon_{+}^2=\epsilon_\alpha^2+\epsilon_\beta^2
\end{align}
and
\begin{align}
	\bf{K}=\frac{k_\beta^+ \bf{k}_\alpha - k_\alpha^+ \bf{k}_\beta}{k_\alpha^++k_\beta^+}, \qquad 
	\bf{Q}=\frac{k_\beta^+ \bf{q}_\alpha - k_\alpha^+ \bf{q}_\beta}{k_\alpha^++k_\beta^+}.
\end{align}
This contribution to the non-eikonal multigluon spectra is somewhat peculiar. In a situation where the  gluons have the same momenta but are sitting at different sides of the cut, which in our notation corresponds to  $\alpha=2m-1$ and $\beta=2m$ (or vice-versa) for any integer $m$, but $k_\beta^+ = k_\alpha^+$ and $\bf{k}_\beta =\bf{k}_\alpha$ due to the constraint introduced in Eq.~\eqref{eq6:constraint_momenta}. However, since the momenta on the right side of the cut are evaluated with a different sign in Eq.~\eqref{eq6:2point_target}, one would have $\epsilon_+ \to 0$, $\bf{K} \to \bf{k}_\alpha$ and\footnote{Because of the $\delta$-function in Eq.~\eqref{eq6:2point_target} one gets $\bf{q}_\beta=\bf{q}_\alpha$, but due to the constrains, it needs to be evaluated with different sign.} $\bf{Q} \to \bf{q}_\alpha$, so that the non-eikonal corrections vanish and it leads to the same contribution as in the case of single inclusive production given in Eq. \eqref{eq6:single_bebe}. On the other hand, Eq.~\eqref{eq6:multi_bebe} is symmetric under the exchange $\alpha \leftrightarrow \beta$ as expected. 


The fourth contribution in Eq.~\eqref{eq6:2point_target} corresponds to the case where the gluon $\alpha$ is emitted after the interaction of the source with the target and the gluon $\beta$ is emitted inside the target, with the opposite situation given by the mirror term $\alpha \leftrightarrow \beta$. The expression of this contribution reads
\begin{align}
	\label{eq6:multi_afin}
	&I^{\lambda_\beta}_{\rm aft-in} \Big(\bf{k}_\beta,\bf{q}_\beta,k_\beta^+ ; f_\beta \Big)
	=
	-\frac{4 \pi i}{Q_s^2 \left[1 + \epsilon_\beta f_\beta \cot(\tilde{f}_\beta \epsilon_\beta)\right]^2}\frac{\epsilon_\beta }{\sin (\tilde{f}_\beta \epsilon_\beta)}
	\left[\bf{q}_\beta^{\lambda_\beta} 
	+  \frac{f_\beta \epsilon_\beta }{\sin (\tilde{f}_\beta \epsilon_\beta)} \bf{k}_\beta^{\lambda_\beta}
	\right]
	\nonumber \\ & \hskip1cm	\times
	\exp{- \frac{\epsilon_\beta}{Q_s^2 \left[1 + f_\beta \epsilon_\beta \cot(\tilde{f}_\beta \epsilon_\beta)\right]}
		\left[
		\frac{\bf{k}_\beta^2+\bf{q}_\beta^2}{\tan(\tilde{f}_\beta \epsilon_\beta)}  
		-
		2\frac{\bf{k}_\beta \cdot \bf{q}_\beta}{\sin(\tilde{f}_\beta \epsilon_\beta)}
		-\epsilon_\beta f_\beta \bf{k}_\beta^2 
		\right]	
	},
\end{align}
where we have defined $\tilde{f}_\beta=1-f_\beta=\frac{L^+-y_\beta^+}{L^+}$ that corresponds to the fraction of the medium traversed by gluon $\beta$.
This result is identical to the one evaluated for the single inclusive case given in Eq.~\eqref{eq6:single_afin}, which can be explained as follows. The main difference between the multigluon and single gluon calculations relies on the form of the $2^{\rm nd}$ order NE dipole function. For the multigluon production presented in this section, the $2^{\rm nd}$ order NE dipole functions, in general, are evaluated at different longitudinal momenta whereas in the single inclusive case they are evaluated at the same longitudinal momenta. However, some of the contributions to the multigluon production, namely aft-aft, aft-bef and aft-in contributions (given in Eqs.~\eqref{eq6:multi_afaf}, \eqref{eq6:multi_afbe} and \eqref{eq6:multi_afin} respectively), contain only one longitudinal momentum and therefore their expressions are almost the same as the ones in the single inclusive case up to some overall factors.  


The fifth contribution in Eq.~\eqref{eq6:2point_target} corresponds to the case where the gluon $\alpha$ is emitted before the interaction of the source with the target and the gluon $\beta$ is emitted inside the target, with the opposite situation given by its mirror term 
$\alpha \leftrightarrow \beta$. We can write this contribution as
\begin{align}
	\label{eq6:multi_bein}
	&I^{\lambda_\beta}_{\rm in-bef} \Big(\bf{k}_\alpha, \bf{q}_\alpha,\bf{k}_\beta,\bf{q}_\beta; k_\alpha^+, k_\beta^+ ; f_\beta \Big)
	=
	-\frac{4 \pi i \epsilon_\alpha
		\csc(\tilde{f}_\beta \epsilon_{+})  \csc( f_\beta \epsilon_\alpha)}{ \epsilon_{+} Q_s^2 \left[\epsilon_\alpha \cot(f_\beta \epsilon_\alpha) + \epsilon_{+} \cot(\tilde{f}_\beta \epsilon_{+}) \right]^2}
	\nonumber \\ & \hskip0cm \times
	\Bigg[
	\Bigg(
	\epsilon_{+}^2
	+
	\frac{\epsilon_\beta^2}{\cos (f_\beta \epsilon_\alpha)}
	+\frac{\epsilon_\alpha \epsilon_{+} \tan(f_\beta \epsilon_\alpha)}{\tan(\tilde{f}_\beta \epsilon_{+})}
	\Bigg)
	\epsilon_\alpha 
	\Delta \bf{k}^{\lambda_\beta}
	-\frac{\epsilon_\alpha \epsilon_{+}^2}{\cos(f_\beta \epsilon_\alpha)}
	\bf{Q}^{\lambda_\beta}
	-\frac{\epsilon_{+}^3 \tan(f_\beta \epsilon_\alpha)}{\sin(\tilde{f}_\beta \epsilon_{+})}
	\bf{K}^{\lambda_\beta}
	\Bigg]
	\nonumber \\ & \hskip0cm \times
	{\rm exp} \Bigg\{ 
	-\frac{\cot(\tilde{f}_\beta \epsilon_{+}) }{\epsilon_{+} Q_s^2 \left[ 
		\epsilon_\alpha \cot(f_\beta \epsilon_\alpha) + \epsilon_{+} \cot(\tilde{f}_\beta \epsilon_{+})
		\right]}
	\Bigg[ 
	\epsilon_{+}^2
	\left(
	\frac{\epsilon_\alpha
	}{\tan(f_\beta \epsilon_\alpha)}
	-
	\epsilon_{+}
	\tan(\tilde{f}_\beta \epsilon_{+})
	\right)
	\bf{K}^2
	\nonumber \\ & \hskip-1cm
	-
	\epsilon_\alpha^2 \epsilon_{+}^2
	\left(
	\frac{\cot(f_\beta \epsilon_\alpha)}{\epsilon_\alpha}
	-
	\frac{\tan(\tilde{f}_\beta \epsilon_{+})}{\epsilon_{+}}
	\right)
	\bf{q}_\alpha^2
	+
	\frac{\epsilon_\alpha \epsilon_\beta^2}{\epsilon_{+}^2}
	\left(
	\left(
	\frac{\epsilon_\beta^2}{\tan (f_\beta \epsilon_\alpha)}
	+ \epsilon_\alpha \epsilon_{+}^2 \tilde{f}_\beta
	\right)
	+\frac{\epsilon_\alpha^2 \epsilon_{+} \tilde{f}_\beta
		\tan(\tilde{f}_\beta \epsilon_{+})}{\tan(f_\beta \epsilon_\alpha)}
	\right)
	\Delta \bf{k}^2
	\nonumber \\ & \hskip1cm 
	+
	2 \frac{\epsilon_\alpha \epsilon_\beta^2 
	}{\sin(f_\beta \epsilon_\alpha)}
	\Delta \bf{k} \cdot \bf{q}_\alpha
	-
	2 \frac{\epsilon_\alpha \epsilon_{+}^2 \sec(\tilde{f}_\beta \epsilon_{+})}{\sin(f_\beta \epsilon_\alpha)}
	\bf{K} \cdot \bf{q}_\alpha
	-
	2 \frac{\epsilon_\alpha \epsilon_\beta^2 \sec(\tilde{f}_\beta \epsilon_{+})}{\tan(f_\beta \epsilon_\alpha)}
	\Delta \bf{k} \cdot \bf{K}
	\Bigg]
	\Bigg\},
\end{align}
where we have introduced $\Delta \bf{k} = \bf{k}_\alpha + \bf{k}_\beta$. Since in this contribution we are including the non-eikonal effects that stem from the difference between the two longitudinal momenta, in the fifth term of Eq.~\eqref{eq6:2point_target_aux} we performed the derivative  with respect to $\bf{y}^{\lambda_\beta}$ and integrated over an extra transverse position $\bf{u}$. We would like to note that in the case of having the two momenta equal, i.e., when $\Delta \bf{k} \to 0$, $\epsilon_+ \to 0$, $\bf{K} \to \bf{k}_\alpha$ and $\bf{Q}\to \bf{q}_\alpha$, this expression simplifies to the one presented for the single inclusive case, Eq.~\eqref{eq6:single_bein}, up to some prefactors.


Finally, the last contribution to Eq.~\eqref{eq6:2point_target},  $I^{\lambda_\alpha \lambda_\beta}_{\rm in-in} \Big(\bf{k}_\alpha,\bf{k}_\beta, \bf{q}_\beta; k_\alpha^+, k_\beta^+ ; f_\alpha, f_\beta \Big)$ corresponds to the case when both gluons are emitted inside the target. This expression involves an extra derivative with respect to $\bf{y}^{\lambda_\alpha}$ compared to the in-bef contribution given in Eq.~\eqref{eq6:multi_bein}. It can be written as
\begin{align}
	&(2\pi)^2 \delta^{(2)}[(\bf{k}_\alpha-\bf{q}_\alpha)+(\bf{k}_\beta-\bf{q}_\beta)]I^{\lambda_\alpha \lambda_\beta}_{\rm in-in} \Big(\bf{k}_\alpha,\bf{k}_\beta,\bf{q}_\beta; k_\alpha^+, k_\beta^+ ; f_\alpha, f_\beta \Big)
	\nonumber \\ & \hskip1cm
	=
	\frac{Q_s^6}{(4 \pi)^3 \epsilon_\alpha \sin(\epsilon_\alpha (f_\beta-f_\alpha))} 
	\frac{\epsilon_{-}}{(1-f_\beta) \epsilon_\alpha^2 \epsilon_\beta^2 \sin (\epsilon_{-}(1-f_\beta))}
	\nonumber \\ & \hskip1cm \times
	\int_{\bf{y}_\alpha, \bf{y}_\beta, \bf{x}_\alpha, \bf{x}_\beta, \bf{u}} e^{i(\bf{q}_\alpha \cdot \bf{y}_\alpha-\bf{k}_\alpha \cdot \bf{x}_\alpha)+i(\bf{q}_\beta \cdot \bf{y}_\beta-\bf{k}_\beta \cdot \bf{x}_\beta)}
	e^{-f_\alpha \frac{Q_s^2}{4}(\bf{y}_\alpha-\bf{y}_\beta)^2}
	\nonumber \\ & \hskip1cm \times
	\frac{\partial}{\partial \bf{y}_\alpha^{\lambda_\alpha}}\Bigg[
	\exp{\frac{Q_s^2}{4\epsilon_\alpha} \left[ \frac{(\bf{y}_\alpha- \bf{y}_\beta)^2+(\bf{u}-\bf{y}_\beta)^2}{\tan(\epsilon_\alpha (f_\beta-f_\alpha))} -2 \frac{ (\bf{y}_\alpha- \bf{y}_\beta) \cdot (\bf{u}-\bf{y}_\beta)}{\sin(\epsilon_\alpha (f_\beta-f_\alpha))} \right]}
	\Bigg]
	\nonumber \\ & \hskip1cm \times
	\frac{\partial}{\partial \bf{y}_\beta^{\lambda_\beta}}\Bigg[
	{\rm exp} \Bigg\{\frac{Q_s^2}{4 \epsilon_{-}^2} \Bigg(
	\frac{\epsilon_{-} [(\bf{u}-\bf{y}_\beta)^2+(\bf{x}_\alpha-\bf{x}_\beta)^2]}{\tan(\epsilon_{-}(1-f_\beta))}
	-2\frac{\epsilon_{-} (\bf{u}-\bf{y}_\beta) \cdot (\bf{x}_\alpha-\bf{x}_\beta)}{\sin(\epsilon_{-}(1-f_\beta))}
	\nonumber \\ & \hskip3cm
	+
	\frac{1}{(1-f_\beta) \epsilon_\alpha^2 \epsilon_\beta^2 \epsilon_{+}^4} \Big[
	2\epsilon_\alpha^2\epsilon_\beta^2 (\bf{x}_\alpha+\bf{y}_\beta-\bf{x}_\beta-\bf{u})
	\nonumber \\ & \hskip4cm
	-\epsilon_{+}^2 \epsilon_{-}^2
	\left(\frac{k_\alpha^+ (\bf{x}_\alpha -\bf{u})+ k_\beta^+ (\bf{x}_\beta - \bf{y}_\beta)}{k_\alpha^++k_\beta^+} \right)
	(\bf{b}_N-\bf{b}_0)
	\Big]^2
	\Bigg)	
	\Bigg\}
	\Bigg],
\end{align}
where $\epsilon_{\pm}=\epsilon_\alpha^2 \pm \epsilon_\beta^2$. The five 2-dimensional integrals in this expression are of the form given in Eq.~\eqref{eq6:gauss_int} and therefore they can be solved analytically. However, since its final form is too long we do not write it explicitly in this manuscript.


\begin{thebibliography}{10}

\bibitem{Jeon:2016uym}
S.~Jeon and U.~Heinz, \emph{{Introduction to Hydrodynamics}},  in
  \emph{{Quark-Gluon Plasma 5}}, pp.~131--187 (2016),
  \href{https://doi.org/10.1142/9789814663717\_0003}{DOI}.

\bibitem{Romatschke:2017ejr}
P.~Romatschke and U.~Romatschke, \emph{{Relativistic Fluid Dynamics In and Out
  of Equilibrium}}, Cambridge Monographs on Mathematical Physics, Cambridge
  University Press (5, 2019),
  \href{https://doi.org/10.1017/9781108651998}{10.1017/9781108651998},
  [\href{https://arxiv.org/abs/1712.05815}{{\ttfamily 1712.05815}}].

\bibitem{Schlichting:2016sqo}
S.~Schlichting and P.~Tribedy, \emph{{Collectivity in Small Collision Systems:
  An Initial-State Perspective}},
  \href{https://doi.org/10.1155/2016/8460349}{\emph{Adv.\ High Energy Phys.}
  {\bfseries 2016} (2016) 8460349}
  [\href{https://arxiv.org/abs/1611.00329}{{\ttfamily 1611.00329}}].

\bibitem{Schenke:2017bog}
B.~Schenke, \emph{{Origins of collectivity in small systems}},
  \href{https://doi.org/10.1016/j.nuclphysa.2017.05.017}{\emph{Nucl.\ Phys.\ A}
  {\bfseries 967} (2017) 105}
  [\href{https://arxiv.org/abs/1704.03914}{{\ttfamily 1704.03914}}].

\bibitem{Loizides:2016tew}
C.~Loizides, \emph{{Experimental overview on small collision systems at the
  LHC}}, \href{https://doi.org/10.1016/j.nuclphysa.2016.04.022}{\emph{Nucl.
  Phys. A} {\bfseries 956} (2016) 200}
  [\href{https://arxiv.org/abs/1602.09138}{{\ttfamily 1602.09138}}].

\bibitem{Citron:2018lsq}
Z.~Citron et~al., \emph{{Report from Working Group 5}: {Future physics
  opportunities for high-density QCD at the LHC with heavy-ion and proton
  beams}}, \href{https://doi.org/10.23731/CYRM-2019-007.1159}{\emph{CERN Yellow
  Rep. Monogr.} {\bfseries 7} (2019) 1159}
  [\href{https://arxiv.org/abs/1812.06772}{{\ttfamily 1812.06772}}].

\bibitem{Nagle:2018nvi}
J.L.~Nagle and W.A.~Zajc, \emph{{Small System Collectivity in Relativistic
  Hadronic and Nuclear Collisions}},
  \href{https://doi.org/10.1146/annurev-nucl-101916-123209}{\emph{Ann. Rev.
  Nucl. Part. Sci.} {\bfseries 68} (2018) 211}
  [\href{https://arxiv.org/abs/1801.03477}{{\ttfamily 1801.03477}}].

\bibitem{Gelis:2010nm}
F.~Gelis, E.~Iancu, J.~Jalilian-Marian and R.~Venugopalan, \emph{{The Color
  Glass Condensate}},
  \href{https://doi.org/10.1146/annurev.nucl.010909.083629}{\emph{Ann. Rev.
  Nucl. Part. Sci.} {\bfseries 60} (2010) 463}
  [\href{https://arxiv.org/abs/1002.0333}{{\ttfamily 1002.0333}}].

\bibitem{Kovchegov:2012mbw}
Y.V.~Kovchegov and E.~Levin, \emph{{Quantum chromodynamics at high energy}},
  vol.~33, Cambridge University Press (8, 2012),
  \href{https://doi.org/10.1017/CBO9781139022187}{10.1017/CBO9781139022187}.

\bibitem{Lappi:2006fp}
T.~Lappi and L.~McLerran, \emph{{Some features of the glasma}},
  \href{https://doi.org/10.1016/j.nuclphysa.2006.04.001}{\emph{Nucl. Phys. A}
  {\bfseries 772} (2006) 200}
  [\href{https://arxiv.org/abs/hep-ph/0602189}{{\ttfamily hep-ph/0602189}}].

\bibitem{Altinoluk:2020wpf}
T.~Altinoluk and N.~Armesto, \emph{{Particle correlations from the initial
  state}}, \href{https://doi.org/10.1140/epja/s10050-020-00225-6}{\emph{Eur.
  Phys. J. A} {\bfseries 56} (2020) 215}
  [\href{https://arxiv.org/abs/2004.08185}{{\ttfamily 2004.08185}}].

\bibitem{Dumitru:2008wn}
A.~Dumitru, F.~Gelis, L.~McLerran and R.~Venugopalan, \emph{{Glasma flux tubes
  and the near side ridge phenomenon at RHIC}},
  \href{https://doi.org/10.1016/j.nuclphysa.2008.06.012}{\emph{Nucl. Phys.}
  {\bfseries A810} (2008) 91}
  [\href{https://arxiv.org/abs/0804.3858}{{\ttfamily 0804.3858}}].

\bibitem{Dumitru:2010iy}
A.~Dumitru, K.~Dusling, F.~Gelis, J.~Jalilian-Marian, T.~Lappi and
  R.~Venugopalan, \emph{{The Ridge in proton-proton collisions at the LHC}},
  \href{https://doi.org/10.1016/j.physletb.2011.01.024}{\emph{Phys. Lett.}
  {\bfseries B697} (2011) 21}
  [\href{https://arxiv.org/abs/1009.5295}{{\ttfamily 1009.5295}}].

\bibitem{Kovchegov:2012nd}
Y.V.~Kovchegov and D.E.~Wertepny, \emph{{Long-Range Rapidity Correlations in
  Heavy-Light Ion Collisions}},
  \href{https://doi.org/10.1016/j.nuclphysa.2013.03.006}{\emph{Nucl. Phys. A}
  {\bfseries 906} (2013) 50} [\href{https://arxiv.org/abs/1212.1195}{{\ttfamily
  1212.1195}}].

\bibitem{Kovchegov:2013ewa}
Y.V.~Kovchegov and D.E.~Wertepny, \emph{{Two-Gluon Correlations in Heavy-Light
  Ion Collisions: Energy and Geometry Dependence, IR Divergences, and
  $k_T$-Factorization}},
  \href{https://doi.org/10.1016/j.nuclphysa.2014.02.021}{\emph{Nucl. Phys.}
  {\bfseries A925} (2014) 254}
  [\href{https://arxiv.org/abs/1310.6701}{{\ttfamily 1310.6701}}].

\bibitem{Altinoluk:2015uaa}
T.~Altinoluk, N.~Armesto, G.~Beuf, A.~Kovner and M.~Lublinsky, \emph{{Bose
  enhancement and the ridge}},
  \href{https://doi.org/10.1016/j.physletb.2015.10.072}{\emph{Phys. Lett. B}
  {\bfseries 751} (2015) 448}
  [\href{https://arxiv.org/abs/1503.07126}{{\ttfamily 1503.07126}}].

\bibitem{Altinoluk:2015eka}
T.~Altinoluk, N.~Armesto, G.~Beuf, A.~Kovner and M.~Lublinsky,
  \emph{{Hanbury?Brown?Twiss measurements at large rapidity separations, or can
  we measure the proton radius in p-A collisions?}},
  \href{https://doi.org/10.1016/j.physletb.2015.11.033}{\emph{Phys. Lett.}
  {\bfseries B752} (2016) 113}
  [\href{https://arxiv.org/abs/1509.03223}{{\ttfamily 1509.03223}}].

\bibitem{Altinoluk:2016vax}
T.~Altinoluk, N.~Armesto, G.~Beuf, A.~Kovner and M.~Lublinsky, \emph{{Quark
  correlations in the Color Glass Condensate: Pauli blocking and the ridge}},
  \href{https://doi.org/10.1103/PhysRevD.95.034025}{\emph{Phys. Rev. D}
  {\bfseries 95} (2017) 034025}
  [\href{https://arxiv.org/abs/1610.03020}{{\ttfamily 1610.03020}}].

\bibitem{Li:2021zmf}
M.~Li and V.V.~Skokov, \emph{{First saturation correction in high energy
  proton-nucleus collisions. Part I. Time evolution of classical Yang-Mills
  fields beyond leading order}},
  \href{https://doi.org/10.1007/JHEP06(2021)140}{\emph{JHEP} {\bfseries 06}
  (2021) 140} [\href{https://arxiv.org/abs/2102.01594}{{\ttfamily
  2102.01594}}].

\bibitem{Li:2021yiv}
M.~Li and V.V.~Skokov, \emph{{First saturation correction in high energy
  proton-nucleus collisions. Part II. Single inclusive semi-hard gluon
  production}}, \href{https://doi.org/10.1007/JHEP06(2021)141}{\emph{JHEP}
  {\bfseries 06} (2021) 141}
  [\href{https://arxiv.org/abs/2104.01879}{{\ttfamily 2104.01879}}].

\bibitem{Li:2021ntt}
M.~Li and V.V.~Skokov, \emph{{First saturation correction in high energy
  proton-nucleus collisions. Part III. Ensemble averaging}},
  \href{https://doi.org/10.1007/JHEP01(2022)160}{\emph{JHEP} {\bfseries 01}
  (2022) 160} [\href{https://arxiv.org/abs/2111.05304}{{\ttfamily
  2111.05304}}].

\bibitem{Kovner:2018fxj}
A.~Kovner and V.V.~Skokov, \emph{{Does shape matter? $v_2$ vs eccentricity in
  small x gluon production}},
  \href{https://doi.org/10.1016/j.physletb.2018.09.001}{\emph{Phys. Lett. B}
  {\bfseries 785} (2018) 372}
  [\href{https://arxiv.org/abs/1805.09297}{{\ttfamily 1805.09297}}].

\bibitem{Altinoluk:2020psk}
T.~Altinoluk, N.~Armesto, A.~Kovner, M.~Lublinsky and V.V.~Skokov,
  \emph{{Angular correlations in pA collisions from CGC: multiplicity and mean
  transverse momentum dependence of $v_2$}},
  \href{https://arxiv.org/abs/2012.01810}{{\ttfamily 2012.01810}}.

\bibitem{Kovner:2016jfp}
A.~Kovner, M.~Lublinsky and V.~Skokov, \emph{{Exploring correlations in the CGC
  wave function: odd azimuthal anisotropy}},
  \href{https://doi.org/10.1103/PhysRevD.96.016010}{\emph{Phys. Rev. D}
  {\bfseries 96} (2017) 016010}
  [\href{https://arxiv.org/abs/1612.07790}{{\ttfamily 1612.07790}}].

\bibitem{Kovchegov:2018jun}
Y.V.~Kovchegov and V.V.~Skokov, \emph{{How classical gluon fields generate odd
  azimuthal harmonics for the two-gluon correlation function in high-energy
  collisions}}, \href{https://doi.org/10.1103/PhysRevD.97.094021}{\emph{Phys.
  Rev. D} {\bfseries 97} (2018) 094021}
  [\href{https://arxiv.org/abs/1802.08166}{{\ttfamily 1802.08166}}].

\bibitem{Dumitru:2014vka}
A.~Dumitru and V.~Skokov, \emph{{Anisotropy of the semiclassical gluon field of
  a large nucleus at high energy}},
  \href{https://doi.org/10.1103/PhysRevD.91.074006}{\emph{Phys. Rev. D}
  {\bfseries 91} (2015) 074006}
  [\href{https://arxiv.org/abs/1411.6630}{{\ttfamily 1411.6630}}].

\bibitem{Dumitru:2014dra}
A.~Dumitru and A.V.~Giannini, \emph{{Initial state angular asymmetries in high
  energy p+A collisions: spontaneous breaking of rotational symmetry by a color
  electric field and C-odd fluctuations}},
  \href{https://doi.org/10.1016/j.nuclphysa.2014.10.037}{\emph{Nucl. Phys. A}
  {\bfseries 933} (2015) 212}
  [\href{https://arxiv.org/abs/1406.5781}{{\ttfamily 1406.5781}}].

\bibitem{Agostini:2019avp}
P.~Agostini, T.~Altinoluk and N.~Armesto, \emph{{Non-eikonal corrections to
  multi-particle production in the Color Glass Condensate}},
  \href{https://doi.org/10.1140/epjc/s10052-019-7097-5}{\emph{Eur. Phys. J. C}
  {\bfseries 79} (2019) 600}
  [\href{https://arxiv.org/abs/1902.04483}{{\ttfamily 1902.04483}}].

\bibitem{Agostini:2019hkj}
P.~Agostini, T.~Altinoluk and N.~Armesto, \emph{{Effect of non-eikonal
  corrections on azimuthal asymmetries in the Color Glass Condensate}},
  \href{https://doi.org/10.1140/epjc/s10052-019-7315-1}{\emph{Eur. Phys. J. C}
  {\bfseries 79} (2019) 790}
  [\href{https://arxiv.org/abs/1907.03668}{{\ttfamily 1907.03668}}].

\bibitem{Accardi:2012qut}
A.~Accardi et~al., \emph{{Electron Ion Collider: The Next QCD Frontier}:
  {Understanding the glue that binds us all}},
  \href{https://doi.org/10.1140/epja/i2016-16268-9}{\emph{Eur. Phys. J. A}
  {\bfseries 52} (2016) 268} [\href{https://arxiv.org/abs/1212.1701}{{\ttfamily
  1212.1701}}].

\bibitem{AbdulKhalek:2021gbh}
R.~Abdul~Khalek et~al., \emph{{Science Requirements and Detector Concepts for
  the Electron-Ion Collider: EIC Yellow Report}},
  \href{https://arxiv.org/abs/2103.05419}{{\ttfamily 2103.05419}}.

\bibitem{Altinoluk:2014oxa}
T.~Altinoluk, N.~Armesto, G.~Beuf, M.~Mart\'\i{}nez and C.A.~Salgado,
  \emph{{Next-to-eikonal corrections in the CGC: gluon production and spin
  asymmetries in pA collisions}},
  \href{https://doi.org/10.1007/JHEP07(2014)068}{\emph{JHEP} {\bfseries 07}
  (2014) 068} [\href{https://arxiv.org/abs/1404.2219}{{\ttfamily 1404.2219}}].

\bibitem{Altinoluk:2015gia}
T.~Altinoluk, N.~Armesto, G.~Beuf and A.~Moscoso,
  \emph{{Next-to-next-to-eikonal corrections in the CGC}},
  \href{https://doi.org/10.1007/JHEP01(2016)114}{\emph{JHEP} {\bfseries 01}
  (2016) 114} [\href{https://arxiv.org/abs/1505.01400}{{\ttfamily
  1505.01400}}].

\bibitem{Altinoluk:2015xuy}
T.~Altinoluk and A.~Dumitru, \emph{{Particle production in high-energy
  collisions beyond the shockwave limit}},
  \href{https://doi.org/10.1103/PhysRevD.94.074032}{\emph{Phys. Rev. D}
  {\bfseries 94} (2016) 074032}
  [\href{https://arxiv.org/abs/1512.00279}{{\ttfamily 1512.00279}}].

\bibitem{Chirilli:2018kkw}
G.A.~Chirilli, \emph{{Sub-eikonal corrections to scattering amplitudes at high
  energy}}, \href{https://doi.org/10.1007/JHEP01(2019)118}{\emph{JHEP}
  {\bfseries 01} (2019) 118}
  [\href{https://arxiv.org/abs/1807.11435}{{\ttfamily 1807.11435}}].

\bibitem{Chirilli:2021lif}
G.A.~Chirilli, \emph{{High-energy operator product expansion at sub-eikonal
  level}}, \href{https://doi.org/10.1007/JHEP06(2021)096}{\emph{JHEP}
  {\bfseries 06} (2021) 096}
  [\href{https://arxiv.org/abs/2101.12744}{{\ttfamily 2101.12744}}].

\bibitem{Altinoluk:2020oyd}
T.~Altinoluk, G.~Beuf, A.~Czajka and A.~Tymowska, \emph{{Quarks at
  next-to-eikonal accuracy in the CGC: Forward quark-nucleus scattering}},
  \href{https://doi.org/10.1103/PhysRevD.104.014019}{\emph{Phys. Rev. D}
  {\bfseries 104} (2021) 014019}
  [\href{https://arxiv.org/abs/2012.03886}{{\ttfamily 2012.03886}}].

\bibitem{Altinoluk:2021lvu}
T.~Altinoluk and G.~Beuf, \emph{{Quark and scalar propagators at
  next-to-eikonal accuracy in the CGC through a dynamical background gluon
  field}}, \href{https://doi.org/10.1103/PhysRevD.105.074026}{\emph{Phys. Rev.
  D} {\bfseries 105} (2022) 074026}
  [\href{https://arxiv.org/abs/2109.01620}{{\ttfamily 2109.01620}}].

\bibitem{Kovchegov:2018znm}
Y.V.~Kovchegov and M.D.~Sievert, \emph{{Small-$x$ Helicity Evolution: an
  Operator Treatment}},
  \href{https://doi.org/10.1103/PhysRevD.99.054032}{\emph{Phys. Rev. D}
  {\bfseries 99} (2019) 054032}
  [\href{https://arxiv.org/abs/1808.09010}{{\ttfamily 1808.09010}}].

\bibitem{Cougoulic:2019aja}
F.~Cougoulic and Y.V.~Kovchegov, \emph{{Helicity-dependent generalization of
  the JIMWLK evolution}},
  \href{https://doi.org/10.1103/PhysRevD.100.114020}{\emph{Phys. Rev. D}
  {\bfseries 100} (2019) 114020}
  [\href{https://arxiv.org/abs/1910.04268}{{\ttfamily 1910.04268}}].

\bibitem{Kovchegov:2015pbl}
Y.V.~Kovchegov, D.~Pitonyak and M.D.~Sievert, \emph{{Helicity Evolution at
  Small-x}}, \href{https://doi.org/10.1007/JHEP01(2016)072}{\emph{JHEP}
  {\bfseries 01} (2016) 072}
  [\href{https://arxiv.org/abs/1511.06737}{{\ttfamily 1511.06737}}].

\bibitem{Kovchegov:2016zex}
Y.V.~Kovchegov, D.~Pitonyak and M.D.~Sievert, \emph{{Helicity Evolution at
  Small $x$: Flavor Singlet and Non-Singlet Observables}},
  \href{https://doi.org/10.1103/PhysRevD.95.014033}{\emph{Phys. Rev. D}
  {\bfseries 95} (2017) 014033}
  [\href{https://arxiv.org/abs/1610.06197}{{\ttfamily 1610.06197}}].

\bibitem{Kovchegov:2021iyc}
Y.V.~Kovchegov and M.G.~Santiago, \emph{{Quark sivers function at small $x$:
  spin-dependent odderon and the sub-eikonal evolution}},
  \href{https://doi.org/10.1007/JHEP11(2021)200}{\emph{JHEP} {\bfseries 11}
  (2021) 200} [\href{https://arxiv.org/abs/2108.03667}{{\ttfamily
  2108.03667}}].

\bibitem{Cougoulic:2020tbc}
F.~Cougoulic and Y.V.~Kovchegov, \emph{{Helicity-dependent extension of the
  McLerran\textendash{}Venugopalan model}},
  \href{https://doi.org/10.1016/j.nuclphysa.2020.122051}{\emph{Nucl. Phys. A}
  {\bfseries 1004} (2020) 122051}
  [\href{https://arxiv.org/abs/2005.14688}{{\ttfamily 2005.14688}}].

\bibitem{Cougoulic:2022gbk}
F.~Cougoulic, Y.V.~Kovchegov, A.~Tarasov and Y.~Tawabutr, \emph{{Quark and
  gluon helicity evolution at small x: revised and updated}},
  \href{https://doi.org/10.1007/JHEP07(2022)095}{\emph{JHEP} {\bfseries 07}
  (2022) 095} [\href{https://arxiv.org/abs/2204.11898}{{\ttfamily
  2204.11898}}].

\bibitem{Adamiak:2021ppq}
{\scshape Jefferson Lab Angular Momentum} collaboration, \emph{{First analysis
  of world polarized DIS data with small-x helicity evolution}},
  \href{https://doi.org/10.1103/PhysRevD.104.L031501}{\emph{Phys. Rev. D}
  {\bfseries 104} (2021) L031501}
  [\href{https://arxiv.org/abs/2102.06159}{{\ttfamily 2102.06159}}].

\bibitem{Kovchegov:2017jxc}
Y.V.~Kovchegov, D.~Pitonyak and M.D.~Sievert, \emph{{Small-$x$ Asymptotics of
  the Quark Helicity Distribution: Analytic Results}},
  \href{https://doi.org/10.1016/j.physletb.2017.06.032}{\emph{Phys. Lett. B}
  {\bfseries 772} (2017) 136}
  [\href{https://arxiv.org/abs/1703.05809}{{\ttfamily 1703.05809}}].

\bibitem{Jalilian-Marian:2017ttv}
J.~Jalilian-Marian, \emph{{Elastic scattering of a quark from a color field:
  longitudinal momentum exchange}},
  \href{https://doi.org/10.1103/PhysRevD.96.074020}{\emph{Phys. Rev. D}
  {\bfseries 96} (2017) 074020}
  [\href{https://arxiv.org/abs/1708.07533}{{\ttfamily 1708.07533}}].

\bibitem{Jalilian-Marian:2018iui}
J.~Jalilian-Marian, \emph{{Quark jets scattering from a gluon field: from
  saturation to high $p_t$}},
  \href{https://doi.org/10.1103/PhysRevD.99.014043}{\emph{Phys. Rev. D}
  {\bfseries 99} (2019) 014043}
  [\href{https://arxiv.org/abs/1809.04625}{{\ttfamily 1809.04625}}].

\bibitem{Jalilian-Marian:2019kaf}
J.~Jalilian-Marian, \emph{{Rapidity loss, spin, and angular asymmetries in the
  scattering of a quark from the color field of a proton or nucleus}},
  \href{https://doi.org/10.1103/PhysRevD.102.014008}{\emph{Phys. Rev. D}
  {\bfseries 102} (2020) 014008}
  [\href{https://arxiv.org/abs/1912.08878}{{\ttfamily 1912.08878}}].

\bibitem{Sadofyev:2021ohn}
A.V.~Sadofyev, M.D.~Sievert and I.~Vitev, \emph{{Ab~initio coupling of jets to
  collective flow in the opacity expansion approach}},
  \href{https://doi.org/10.1103/PhysRevD.104.094044}{\emph{Phys. Rev. D}
  {\bfseries 104} (2021) 094044}
  [\href{https://arxiv.org/abs/2104.09513}{{\ttfamily 2104.09513}}].

\bibitem{Andres:2022ndd}
C.~Andres, F.~Dominguez, A.V.~Sadofyev and C.A.~Salgado, \emph{{Jet Broadening
  in Flowing Matter -- Resummation}},
  \href{https://arxiv.org/abs/2207.07141}{{\ttfamily 2207.07141}}.

\bibitem{Casalderrey-Solana:2007knd}
J.~Casalderrey-Solana and C.A.~Salgado, \emph{{Introductory lectures on jet
  quenching in heavy ion collisions}}, {\emph{Acta Phys. Polon. B} {\bfseries
  38} (2007) 3731} [\href{https://arxiv.org/abs/0712.3443}{{\ttfamily
  0712.3443}}].

\bibitem{Mehtar-Tani:2013pia}
Y.~Mehtar-Tani, J.G.~Milhano and K.~Tywoniuk, \emph{{Jet physics in heavy-ion
  collisions}}, \href{https://doi.org/10.1142/S0217751X13400137}{\emph{Int. J.
  Mod. Phys. A} {\bfseries 28} (2013) 1340013}
  [\href{https://arxiv.org/abs/1302.2579}{{\ttfamily 1302.2579}}].

\bibitem{Blaizot:2015lma}
J.-P.~Blaizot and Y.~Mehtar-Tani, \emph{{Jet Structure in Heavy Ion
  Collisions}}, \href{https://doi.org/10.1142/S021830131530012X}{\emph{Int. J.
  Mod. Phys. E} {\bfseries 24} (2015) 1530012}
  [\href{https://arxiv.org/abs/1503.05958}{{\ttfamily 1503.05958}}].

\bibitem{Agostini:2021xca}
P.~Agostini, T.~Altinoluk and N.~Armesto, \emph{{Multi-particle production in
  proton\textendash{}nucleus collisions in the color glass condensate}},
  \href{https://doi.org/10.1140/epjc/s10052-021-09475-0}{\emph{Eur. Phys. J. C}
  {\bfseries 81} (2021) 760}
  [\href{https://arxiv.org/abs/2103.08485}{{\ttfamily 2103.08485}}].

\bibitem{Kovner:2017ssr}
A.~Kovner and A.H.~Rezaeian, \emph{{Double parton scattering in the CGC: Double
  quark production and effects of quantum statistics}},
  \href{https://doi.org/10.1103/PhysRevD.96.074018}{\emph{Phys. Rev.}
  {\bfseries D96} (2017) 074018}
  [\href{https://arxiv.org/abs/1707.06985}{{\ttfamily 1707.06985}}].

\bibitem{Kovner:2018vec}
A.~Kovner and A.H.~Rezaeian, \emph{{Multiquark production in $p+A$ collisions:
  Quantum interference effects}},
  \href{https://doi.org/10.1103/PhysRevD.97.074008}{\emph{Phys. Rev.}
  {\bfseries D97} (2018) 074008}
  [\href{https://arxiv.org/abs/1801.04875}{{\ttfamily 1801.04875}}].

\bibitem{Altinoluk:2018ogz}
T.~Altinoluk, N.~Armesto, A.~Kovner and M.~Lublinsky, \emph{{Double and triple
  inclusive gluon production at mid rapidity: quantum interference in p-A
  scattering}},
  \href{https://doi.org/10.1140/epjc/s10052-018-6186-1}{\emph{Eur. Phys. J. C}
  {\bfseries 78} (2018) 702}
  [\href{https://arxiv.org/abs/1805.07739}{{\ttfamily 1805.07739}}].

\bibitem{Mehtar-Tani:2006vpj}
Y.~Mehtar-Tani, \emph{{Relating the description of gluon production in pA
  collisions and parton energy loss in AA collisions}},
  \href{https://doi.org/10.1103/PhysRevC.75.034908}{\emph{Phys. Rev. C}
  {\bfseries 75} (2007) 034908}
  [\href{https://arxiv.org/abs/hep-ph/0606236}{{\ttfamily hep-ph/0606236}}].

\bibitem{Armesto:2006bv}
N.~Armesto, L.~McLerran and C.~Pajares, \emph{{Long Range Forward-Backward
  Correlations and the Color Glass Condensate}},
  \href{https://doi.org/10.1016/j.nuclphysa.2006.10.074}{\emph{Nucl. Phys. A}
  {\bfseries 781} (2007) 201}
  [\href{https://arxiv.org/abs/hep-ph/0607345}{{\ttfamily hep-ph/0607345}}].

\bibitem{Ozonder:2014sra}
c.~\"Ozonder, \emph{{Triple-gluon and quadruple-gluon azimuthal correlations
  from glasma and higher-dimensional ridges}},
  \href{https://doi.org/10.1103/PhysRevD.91.034005}{\emph{Phys. Rev. D}
  {\bfseries 91} (2015) 034005}
  [\href{https://arxiv.org/abs/1409.6347}{{\ttfamily 1409.6347}}].

\bibitem{Ozonder:2017wmh}
c.~\"Oz\"onder, \emph{{Predictions on three-particle azimuthal correlations in
  proton-proton collisions}},
  \href{https://doi.org/10.3906/fiz-1710-6}{\emph{Turk. J. Phys.} {\bfseries
  42} (2018) 78} [\href{https://arxiv.org/abs/1712.05571}{{\ttfamily
  1712.05571}}].

\bibitem{Altinoluk:2018hcu}
T.~Altinoluk, N.~Armesto and D.E.~Wertepny, \emph{{Correlations and the ridge
  in the Color Glass Condensate beyond the glasma graph approximation}},
  \href{https://doi.org/10.1007/JHEP05(2018)207}{\emph{JHEP} {\bfseries 05}
  (2018) 207} [\href{https://arxiv.org/abs/1804.02910}{{\ttfamily
  1804.02910}}].

\bibitem{McLerran:1993ni}
L.D.~McLerran and R.~Venugopalan, \emph{{Computing quark and gluon distribution
  functions for very large nuclei}},
  \href{https://doi.org/10.1103/PhysRevD.49.2233}{\emph{Phys. Rev. D}
  {\bfseries 49} (1994) 2233}
  [\href{https://arxiv.org/abs/hep-ph/9309289}{{\ttfamily hep-ph/9309289}}].

\bibitem{McLerran:1994vd}
L.D.~McLerran and R.~Venugopalan, \emph{{Green's functions in the color field
  of a large nucleus}},
  \href{https://doi.org/10.1103/PhysRevD.50.2225}{\emph{Phys. Rev. D}
  {\bfseries 50} (1994) 2225}
  [\href{https://arxiv.org/abs/hep-ph/9402335}{{\ttfamily hep-ph/9402335}}].

\bibitem{GolecBiernat:1998js}
K.J.~Golec-Biernat and M.~Wusthoff, \emph{{Saturation effects in deep inelastic
  scattering at low Q**2 and its implications on diffraction}},
  \href{https://doi.org/10.1103/PhysRevD.59.014017}{\emph{Phys. Rev. D}
  {\bfseries 59} (1998) 014017}
  [\href{https://arxiv.org/abs/hep-ph/9807513}{{\ttfamily hep-ph/9807513}}].

\bibitem{GolecBiernat:1999qd}
K.J.~Golec-Biernat and M.~Wusthoff, \emph{{Saturation in diffractive deep
  inelastic scattering}},
  \href{https://doi.org/10.1103/PhysRevD.60.114023}{\emph{Phys. Rev. D}
  {\bfseries 60} (1999) 114023}
  [\href{https://arxiv.org/abs/hep-ph/9903358}{{\ttfamily hep-ph/9903358}}].

\bibitem{Zakharov:1998sv}
B.G.~Zakharov, \emph{{Light cone path integral approach to the
  Landau-Pomeranchuk-Migdal effect}}, {\emph{Phys. Atom. Nucl.} {\bfseries 61}
  (1998) 838} [\href{https://arxiv.org/abs/hep-ph/9807540}{{\ttfamily
  hep-ph/9807540}}].

\bibitem{Blaizot:2012fh}
J.-P.~Blaizot, F.~Dominguez, E.~Iancu and Y.~Mehtar-Tani, \emph{{Medium-induced
  gluon branching}}, \href{https://doi.org/10.1007/JHEP01(2013)143}{\emph{JHEP}
  {\bfseries 01} (2013) 143} [\href{https://arxiv.org/abs/1209.4585}{{\ttfamily
  1209.4585}}].

\bibitem{Apolinario:2014csa}
L.~Apolin\'ario, N.~Armesto, J.G.~Milhano and C.A.~Salgado,
  \emph{{Medium-induced gluon radiation and colour decoherence beyond the soft
  approximation}}, \href{https://doi.org/10.1007/JHEP02(2015)119}{\emph{JHEP}
  {\bfseries 02} (2015) 119} [\href{https://arxiv.org/abs/1407.0599}{{\ttfamily
  1407.0599}}].

\bibitem{Baier:1996sk}
R.~Baier, Y.L.~Dokshitzer, A.H.~Mueller, S.~Peigne and D.~Schiff,
  \emph{{Radiative energy loss and p(T) broadening of high-energy partons in
  nuclei}}, \href{https://doi.org/10.1016/S0550-3213(96)00581-0}{\emph{Nucl.
  Phys. B} {\bfseries 484} (1997) 265}
  [\href{https://arxiv.org/abs/hep-ph/9608322}{{\ttfamily hep-ph/9608322}}].

\bibitem{Baier:1996kr}
R.~Baier, Y.L.~Dokshitzer, A.H.~Mueller, S.~Peigne and D.~Schiff,
  \emph{{Radiative energy loss of high-energy quarks and gluons in a finite
  volume quark - gluon plasma}},
  \href{https://doi.org/10.1016/S0550-3213(96)00553-6}{\emph{Nucl. Phys. B}
  {\bfseries 483} (1997) 291}
  [\href{https://arxiv.org/abs/hep-ph/9607355}{{\ttfamily hep-ph/9607355}}].

\bibitem{Zakharov:1997uu}
B.G.~Zakharov, \emph{{Radiative energy loss of high-energy quarks in finite
  size nuclear matter and quark - gluon plasma}},
  \href{https://doi.org/10.1134/1.567389}{\emph{JETP Lett.} {\bfseries 65}
  (1997) 615} [\href{https://arxiv.org/abs/hep-ph/9704255}{{\ttfamily
  hep-ph/9704255}}].

\bibitem{Wiedemann:2000za}
U.A.~Wiedemann, \emph{{Gluon radiation off hard quarks in a nuclear
  environment: Opacity expansion}},
  \href{https://doi.org/10.1016/S0550-3213(00)00457-0}{\emph{Nucl. Phys. B}
  {\bfseries 588} (2000) 303}
  [\href{https://arxiv.org/abs/hep-ph/0005129}{{\ttfamily hep-ph/0005129}}].

\bibitem{Grosche:1998yu}
C.~Grosche and F.~Steiner, \emph{{Handbook of Feynman Path Integrals}},
  vol.~145 (1998).

\end{thebibliography}

\providecommand{\href}[2]{#2}\begingroup\raggedright\endgroup

\end{document}